\documentclass[twocolumn,floatfix]{revtex4}
\pdfoutput=1
\usepackage{graphicx}
\usepackage{dcolumn}
\usepackage{longtable}
\usepackage{amsmath}
\usepackage{amssymb}
\usepackage{color}
\usepackage{isotope}

\begin{document}
\title{Robustness of the proxy-SU(3) symmetry in atomic nuclei and the role of the next highest weight irreducible representation}

\author
{Dennis Bonatsos$^1$, Andriana Martinou$^1$,  S.K. Peroulis$^1$, D. Petrellis$^2$, P. Vasileiou$^3$, T.J. Mertzimekis$^3$, and N. Minkov$^4$ }

\affiliation
{$^1$Institute of Nuclear and Particle Physics, National Centre for Scientific Research ``Demokritos'', GR-15310 Aghia Paraskevi, Attiki, Greece}

\affiliation
{$^2$ Nuclear Physics Institute, Czech Academy of Sciences, CZ-250 68 \v{R}e\v{z} near Prague, Czech Republic}

\affiliation
{$^3$  Department of Physics, National and Kapodistrian University of Athens, Zografou Campus, GR-15784 Athens, Greece}

\affiliation
{$^4$Institute of Nuclear Research and Nuclear Energy, Bulgarian Academy of Sciences, 72 Tzarigrad Road, 1784 Sofia, Bulgaria}

\begin{abstract}

The proxy-SU(3) symmetry predicts, in a parameter-free way, the collective deformation variables $\beta$ and $\gamma$ in even-even atomic nuclei away from closed shells based on the highest weight irreducible representations (irreps) of SU(3) in the relevant proton and neutron shells, which are the most symmetric irreps allowed by the Pauli principle and the short-range nature of the nucleon-nucleon interactions. The special cases in which the use of the next highest weight irrep of SU(3) becomes necessary are pointed out and numerical results are given for several regions of the nuclear chart, which can be used as input for irrep-mixing calculations.
 
\end{abstract}

\maketitle

\section{Introduction}

The SU(3) symmetry has been playing an important role in nuclear structure since its infancy \cite{Kota2020}. It has been present already in 1949 in the introduction \cite{Mayer1948,Mayer1949,Haxel1949,Mayer1955} of the shell model \cite{Heyde1990,Talmi1993}, which is based on the three-dimensional isotropic harmonic oscillator (3D-HO) \cite{Wybourne1974,Moshinsky1996,Iachello2006}, the shells of which are characterized by unitary symmetries possessing SU(3) subalgebras \cite{Bonatsos1986}. Elliott 
\cite{Elliott1958a,Elliott1958b,Elliott1963,Elliott1968,Harvey1968} in 1958 managed to bridge the gap between the shell model and the collective model of Bohr and Mottelson \cite{Bohr1952,Bohr1953,Bohr1998a,Bohr1998b}, introduced in 1952, by proving that nuclear quadrupole deformation occurs in the shell model framework, its size depending on the second order Casimir operator \cite{Wybourne1974,Iachello2006} of SU(3). Since the SU(3) symmetry is broken beyond the $sd$ nuclear shell, because of the presence of the spin-orbit interaction \cite{Mayer1948,Mayer1949,Haxel1949,Mayer1955}, several approximation schemes have been introduced, aiming at reestablishing the SU(3) symmetry in higher nuclear shells. The pseudo-SU(3) symmetry \cite{Arima1969,Hecht1969,RatnaRaju1973,Draayer1982,Draayer1983,Draayer1984,Bahri1992,Ginocchio1997} has been introduced in 1973, the quasi-SU(3) symmetry \cite{Zuker1995,Zuker2015} appeared in 1995, while the proxy-SU(3) symmetry \cite{Bonatsos2017a,Bonatsos2017b,Bonatsos2023} occurred in 2017. In parallel, the symplectic model \cite{Rosensteel1977,Rosensteel1980,Rowe1985}, which is a generalization of Elliott's SU(3) model, taking advantage of an overall non-compact algebra Sp(6,R) possessing an SU(3) subalgebra, has been introduced in 1977. The SU(3) subalgrebra of the symplectic symmetry has played a crucial role in the recent development of the symmetry-adapted no-core shell model (SANCSM) \cite{Dytrych2008,Launey2015,Launey2016,Launey2020,Launey2021}, applicable to light nuclei because of the large scale computational effort demanded. In all the models mentioned so far, fermions (protons and neutrons) are used. Along an alternative path, initiated in 1975 through the introduction of the Interacting Boson Model (IBM) \cite{Arima1975,Iachello1987,Iachello1991,Casten1993,Frank2005}, bosons have been used as approximations to correlated fermion pairs. The SU(3) symmetry is related to nuclear deformation also in the IBM framework \cite{Arima1978,Casten1988,Casten1990}, as well as in the framework of the Vector Boson Model (VBM) \cite{Minkov1997}.   

In the present work, attention is focused on the proxy-SU(3) approximation to the shell model \cite{Bonatsos2017a,Bonatsos2017b,Bonatsos2023}. Its introduction has been based on a unitary transformation \cite{Martinou2020} allowing the replacement of the intruder orbitals, invading a nuclear shell from above due to the action of the spin-orbit interaction breaking the SU(3) symmetry of the 3D-HO, by the orbitals which have deserted this nuclear shell, sinking into the shell below. This replacement of the intruder orbitals by the deserting orbitals, acting as their proxies, re-establishes the SU(3) symmetry of the 3D-HO \cite{Wybourne1974,Moshinsky1996,Iachello2006} and allows its use for the description of medium-mass and heavy nuclei, the approximation becoming more accurate at heavier shells \cite{Bonatsos2017a}. The next major ingredient of the proxy-SU(3) symmetry is the dominance of the highest weight (hw) irreducible representations (irreps) of SU(3) \cite{Martinou2021b}, which lie lowest in energy because of the Pauli principle characterizing the nucleons within the shell and the short-range nature of the nucleon-nucleon interaction \cite{Ring1980,Casten1990}. It has been proved \cite{Martinou2021b} that the hw irrep, which lies lowest in energy, is the most symmetric irrep allowed by the Pauli principle and the short-range nature of the nucleon-nucleon interaction. 

The use of the hw irreps has led to several interesting results within the proxy-SU(3) scheme, like the explanation of the dominance \cite{Bonatsos2017b,Bonatsos2017c} of the prolate over oblate shapes in the ground states of even-even nuclei and the appearance of shape coexistence only within certain regions of the nuclear chart, called the islands of shape coexistence \cite{Martinou2021,Martinou2023,Bonatsos2023c}. It has also allowed for parameter-free predictions of the collective deformation variables 
$\beta$ (describing deviation from sphericity) and $\gamma$ (describing deviation from triaxiality), which are in satisfactory agreement to available data, as well as to predictions of various theoretical models using free parameters \cite{Bonatsos2017b,Martinou2017}. However, some discrepancies occur, which are more evident in nuclei in which the hw irrep of the valence protons and/or of the valence neutrons happen to acquire the completely symmetric form, which is analogous to the completely symmetric form occurring for bosons \cite{Arima1978,Iachello1987}. 

It is the aim of the present work to pave the way for amending these discrepancies. As it will be shown, this can be done by involving the next highest weight (nhw) irrep, which is the irrep having the second highest weight among all irreps allowed by the Pauli principle and the short-range nature of the nucleon-nucleon interaction.  

\section{Highest weight irreducible representations}\label{hw}

The highest weight (hw) irreducible representations (irreps) are the most symmetric irreps allowed for a given number of identical nucleons (protons or neutrons) within a given shell, allowed by the Pauli principle and the short-range nature of the nucleon-nucleon interaction, which pushes them to lie lowest in energy \cite{Martinou2021b}. 

The shells of the 3D-HO are known to possess the unitary symmetry U$((n+1)(n+2)/2)$, where $n$ is the shell number \cite{Wybourne1974,Moshinsky1996,Iachello2006}. In more detail, the $p$, $sd$, $pf$, $sdg$, $pfh$, $sdgi$ shells possess the U(3), U(6), U(10), U(15), U(21), U(28) symmetry respectively. All these symmetries possess SU(3) subalgebras \cite{Bonatsos1986}. 

The hw irrep for each particle number within each shell has been calculated using the code UNTOU3 \cite{Draayer1989a}, while a newer code \cite{Langr2019}, as well as an analytic formula for the hw irrep also exist \cite{Kota2018}. The results are shown in Table I, in which the next highest weight (nhw) irrep, which is the one lying in energy immediately above the hw irrep, is also shown, calculated using the code UNTOU3 \cite{Draayer1989a}. The Elliott notation $(\lambda,\mu)$ \cite{Elliott1958a} is used for the SU(3) irreps, with $\lambda=f_1-f_2$ and $\mu=f_2$, where $f_1$ and $f_2$ are the number of boxes in the first and second row of the relevant Young diagram \cite{Wybourne1974}, respectively.  We remark that the value $\mu=0$, which corresponds to a fully symmetric (boson-like) irrep \cite{Arima1978,Iachello1987},  occurs very rarely in the hw irreps.   

In order to find the hw irrep corresponding to the ground state of an even-even nucleus, the hw irreps $(\lambda_p,\mu_p)$ corresponding to the number of valence protons,
and the hw irreps $(\lambda_n,\mu_n)$ corresponding to the number of valence neutrons, are coupled into the most stretched irrep 
\begin{equation}\label{str}
 (\lambda,\mu) = (\lambda_p+\lambda_n,\mu_p+\mu_n). 
\end{equation}
The resulting hw irreps for nuclei in 5 different regions of the nuclear chart, depicted in Fig. 1, are listed in Tables II-VI. We remark that the value of $\mu=0$ occurs very rarely, while the value $\mu=2$ occurs only once, in \isotope[124][54]{Xe}$_{70}$. 

It is known \cite{Elliott1958a,Iachello1987} that the $(\lambda,\mu)$ SU(3) irrep contains bands characterized by $K=\mu, \mu-2, \mu-4, \dots 0$ or 1, where $K$ is the Elliott quantum number, which is the missing quantum number in the decomposition SU(3)$\supset$SO(3). Bands with a given $K\neq 0$ contain levels with angular momenta
\begin{equation}
 L=K, K+1, K+2, \ldots, K+\max\{\lambda,\mu\}, 
\end{equation}
while in the case of $K=0$ one has 
\begin{equation}
 L=\max\{\lambda,\mu\},  \max\{\lambda,\mu\}-2, \ldots, 1 \quad  {\rm or} \quad 0.
\end{equation}
The missing quantum number $K$ serves for distinguishing levels with the same $L$ occurring within the same irrep.

From the physics point of view, the quantum number $K$, corresponding to the projection of the angular momentum $L$ on the symmetry axis (taken as the $z$-axis) in the intrinsic nuclear frame is used. The ground state band (gsb) of even-even nuclei always has $K=0$, therefore possessing the levels $L=0$, 2, 4, 6, \dots, while higher $K=0$ bands  can also exist, usually referred to as the $\beta_1$-band, $\beta_2$ band, and so on \cite{ensdf}. The lowest $K=2$ band, usually referred to as the $\gamma_1$-band, contains the levels $L=2$, 3, 4, 5,\dots, while higher $K=2$ bands can also exist \cite{ensdf}. The lowest $K=4$ band (sometimes referred to as a
two-phonon $\gamma\gamma$ band \cite{Borner1991,Wu1994} or a hexadecapole band \cite{Burke1994,Garrett2005}) contains the levels $L=4$, 5, 6, 7, 8, \dots, the lowest $K=6$ band contains the levels $L=6$, 7, 8, 9, 10, \dots, and so on.  

From the mathematical point of view, it is known that electromagnetic transitions can occur only between states belonging to the same irrep \cite{Iachello1987}. Much experimental information exists on electric transitions connecting the $\gamma_1$ band and the gsb, having substantial $B(E2)$ transition rates \cite{ensdf}. In addition, the gsb and the $\gamma_1$ band exhibit certain structural similarities, as, for example, similar moments of inertia \cite{Jolos2006,Jolos2007,Bonatsos2021}. It is therefore imperative for the gsb and the $\gamma_1$ band to belong to the same irrep \cite{Draayer1984}. 
This is impossible if the hw irrep has $\mu=0$, since such an irrep would accommodate only the gsb, while the $\gamma_1$ band would be pushed to the nhw irrep. 
Therefore we expect serious discrepancies from the data to occur in nuclei in which the hw irrep has $\mu=0$. For these nuclei it seems necessary to involve the nhw irrep in their description, an issue to be considered later on in section \ref{nhw}.  

\section{Parameter-free predictions for the $\beta$ and $\gamma$ collective variables from the hw irreps} \label{bg}

A mapping between the deformation variables $\beta$ and $\gamma$ of the collective model and the Elliot SU(3) quantum numbers $\lambda$ and $\mu$ can be achieved 
\cite{Castanos1988,Draayer1989} by making a mapping between the invariant quantities of the two theories, which are $\beta^2$ and $\beta^3 \cos 3\gamma$ for the former \cite{Bohr1998b}, while for the latter are the second order Casimir operator, $C_2$, having eigenvalues \cite{Kota2020}
  \begin{equation}\label{C2} 
 C_2(\lambda,\mu)= {2 \over 3} (\lambda^2+\lambda \mu + \mu^2+ 3\lambda +3 \mu), 
\end{equation}
and the third order Casimir operator, $C_3$, possessing eigenvalues \cite{Kota2020}
  \begin{equation}\label{C3} 
 C_3(\lambda,\mu)= {1 \over 9} (2\lambda^3-2 \mu^3 + 3\lambda \mu(\lambda-\mu) +18\lambda^2 + 9\lambda\mu +36 \lambda  +18 \mu).  
\end{equation}
This mapping results in the relations  \cite{Castanos1988,Draayer1989}
\begin{equation}\label{g1}
\gamma = \arctan \left( {\sqrt{3} (\mu+1) \over 2\lambda+\mu+3}  \right),
\end{equation}
and \cite{Castanos1988,Draayer1989}
\begin{equation}\label{b1}
	\beta^2= {4\pi \over 5} {1\over (A \bar{r^2})^2} (\lambda^2+\lambda \mu + \mu^2+ 3\lambda +3 \mu +3), 
\end{equation}
where $A$ is the mass number of the nucleus, while $\bar{r^2}$ is related to the dimensionless mean square radius \cite{Ring1980}, $\sqrt{\bar{r^2}}= r_0 A^{1/6}$. The dimensionless mean square radius is obtained by dividing the mean square radius, which grows as $A^{1/3}$, by the oscillator length, which is proportional to $A^{1/6}$ \cite{Ring1980}. The constant $r_0$ is found from a fit over a wide range of nuclei \cite{DeVries1987,Stone2014} to have the value $r_0=0.87$. We remark that $\beta^2$ is proportional to $C_2+3$. Taking into account that only the valence shells have been taken into account, the values of $\beta$ should be multiplied by a scaling factor $A/(S_p+S_n)$, where $S_p$ ($S_n$) is the size of the proton (neutron) valence shell \cite{Bonatsos2017b}. For example, in the case of the rare earth region, in which the valence protons lie in the 50-82 shell and the valence neutrons lie in the 82-126 shell, one has $S_p=32$ and $S_n=44$, thus the scaling factor is $A/76$.  

Empirical values for $\beta$ can be obtained from the experimental $B(E2)$ transition rates connecting the $0^+$ ground state to the first excited $2^+$ state, and are listed in Ref. \cite{Pritychenko2016}. Empirical values for $\gamma$ can be obtained from the energy ratio 
\begin{equation}
R = {E(2_2^+) \over E(2_1^+)}, 
\end{equation}
of the second excited $2^+$ state (which is the bandhead of the lowest lying  $K=2$ band, usually referred to as the $\gamma_1$ band) and the first excited $2^+$ state (which is a member of the gsb, having $K=0$) through the relation
 \cite{Casten1990}
\begin{equation}\label{R}
\sin 3\gamma= {3\over 2\sqrt{2}} \sqrt{1-\left({R-1\over R+1}  \right)^2}, 
\end{equation} 
 which is obtained within the Davydov model \cite{Davydov1958,Davydov1959} for triaxial nuclei. 

The proxy-SU(3) predictions for $\beta$ and $\gamma$, obtained through the use of the hw irreps given in Tables II-VI and listed in these tables, have been compared in the past \cite{Bonatsos2017b,Martinou2017,Martinou2017b} to the above mentioned empirical values, as well as to parameter-dependent predictions of various theoretical approaches. The following remarks apply

a) The proxy-SU(3) values of $\beta$  have in general been found to be in good agreement with the empirical values, as well as with the predictions of several theoretical models \cite{Bonatsos2017b,Martinou2017,Martinou2017b}. 

b) The proxy-SU(3) values of $\gamma$ are in general in good agreement with the empirical values, except in cases in which hw irreps for protons or neutrons with $\mu=0$ are involved, in which case the proxy-SU(3) values are too low \cite{Bonatsos2017b,Martinou2017,Martinou2017b}. 

An example can be seen in Fig. 2, in which the empirical values of $\gamma$ coming from Eq. (\ref{R}) and the proxy-SU(3) values obtained with the hw irrep, are shown for several $N=94$, 96, 98 isotones in the rare earth region. We remark that the proxy-SU(3) values are too low for $Z=62$ and $Z=70$, corresponding respectively to 12 and 20 valence protons within the 50-82 shell, i.e., within the proxy-SU(3) $sdg$ shell having the U(15) symmetry, as seen in Table I. In addition we remark that   the proxy-SU(3) values are systematically too low for $N=94$,  which corresponds to 12 valence neutrons within the 82-126 shell, i.e., within the proxy-SU(3) $pfh$ shell having the U(21) symmetry. 

In Table I we observe that hw irreps with $\mu=0$ occur at $M=2$, 6, 12, 20, 30, 42, 56 particles. Therefore these are the protons and/or neutron numbers in which too low values of $\gamma$ are expected to be obtained from the hw proxy-SU(3) irreps. 

\section{Next highest weight irreducible representations}

As already remarked in section \ref{hw}, in cases in which the hw irrep has $\mu=0$, the next highest weight irrep (nhw) would have to be involved. The nhw irreps for a given number of valence nucleons in a given shell, calculated using the code UNTOU3 \cite{Draayer1989a},   are listed in Table I. The following observations can be made.

i) The  nhw irreps with $\mu=0$ occur only for $M=8$ and 10 nucleons, thus they never coincide with hw irreps having $\mu=0$. 

ii) The hw irreps with $\mu=0$ occurring  at $M=2$, 6, 12, 20, 30, 42 particles are followed by nhw irreps having $\mu=2$, 6, 10, 14, 18, 22, respectively, which can accommodate bands with $K>0$.  

iii) The hw irreps $(\lambda,\mu)$ with $\mu\neq 0$ (occurring  at $M\neq 2$, 6, 12, 20, 30, 42 particles) are followed by nhw irreps of the form $(\lambda+2,\mu-4)$. 

In view of the above, the nhw for a given nucleus can be obtained using Eq. (\ref{str}) in the following way. 

1) If the proton (neutron) hw irrep has $\mu=0$, the nhw proton (neutron) irrep will enter the nhw of the whole nucleus. 

2) In the very rare case in which both the proton hw irrep and the neutron hw irrep have $\mu=0$, the nhw of both protons and neutrons will enter the nhw of the whole nucleus. 

3) In most nuclei, both the hw proton irrep and the hw neutron irrep have $\mu \neq 0$. In this case, if the hw irrep of the whole nucleus is $(\lambda,\mu)$, the nhw irrep of the whole nucleus will be $(\lambda+2,\mu-4)$, since the combination of the hw proton irrep with the nhw neutron irrep, and the inverse combination of the  nhw proton irrep with the hw neutron irrep give the same result when inserted in Eq. (\ref{str}). 

Based on Table I and the above comments, we list in Tables II-VI the nhw irreps obtained for nuclei in five different regions of the nuclear chart, depicted in Fig. 1.  The following observations can be made. 

a) The nhw irreps in general have $\mu \neq 0$, the only two exceptions occurring in \isotope[170][80]{Hg}$_{90}$ and   \isotope[172][80]{Hg}$_{92}$.

b) In series of isotopes in which the hw irrep for protons has $\mu\neq 0$, the hw and nhw irreps are connected in the following way: to the hw irrep $(\lambda,\mu)$, corresponds the nhw irrep $(\lambda+2,\mu-4)$, if $\mu \geq 4$. If $\mu < 4$, obviously this rule does not apply. 

c) In series of isotopes in which the hw irrep for protons has $\mu=0$, the hw and nhw irreps are radically different from each other for any neutron number $N$. 

d) From Eq. (\ref{b1}) it is seen that $\beta^2$ is proportional to the quantity $\lambda^2+\lambda \mu + \mu^2+ 3\lambda +3 \mu +3$. A simple calculation shows that when this quantity is calculated for the nhw $(\lambda+2,\mu-4)$, instead for the hw irrep  $(\lambda,\mu)$, a reduction of $6(\mu-1)$ occurs. Since usually $\lambda$ is much higher than $\mu$, we conclude that the value of $\beta$ will be only slightly reduced when using the nhw irrep instead of the hw irrep in the same nucleus. This observation proves the robustness of hw proxy-SU(3) predictions for the $\beta$ deformation variable. It should be remembered that this observation holds only for the nuclei satisfying the condition 3) given above, which, however, represent the vast majority of nuclei. 

e) In contrast, in Eq. (\ref{g1}) one sees that while the denominator $2\lambda+\mu+3$ remains unchanged when passing from the hw irrep to the nhw irrep, the numerator is reduced by 4 units. Therefore the value of $\gamma$ will be substantially changed when passing from the hw irrep to the nhw irrep. In other words, the hw proxy-SU(3) predictions for the collective variable $\gamma$ are not expected to be as robust as these for the collective vabiable $\beta$. Discrepancies may get larger in cases in which the above mentioned condition 3) is violated, as seen, for example, in the Yb isotopes included in Fig. 2 and further discussed in the beginning of Section VI.

f) The picture appearing in b), with the hw irrep being $(\lambda,\mu)$  with $\mu \geq 4$,  and the nhw irrep being $(\lambda+2,\mu-4)$, should be compared to the picture appearing in the SU(3) limit of IBM-1 \cite{Arima1978,Iachello1987}, in which the lowest lying irrep for a nucleus corresponding to  ${\cal N}$ bosons is $(2 {\cal N}, 0)$, while the next lowest lying irrep is $(2 {\cal N}-4, 2)$. In the IBM-1 framework it is clear that the gsb will belong to the $(2 {\cal N}, 0)$ irrep, while the next $K=0$ band, usually called the $\beta_1$ band \cite{ensdf}, and the first $K=2$ band, usually called the $\gamma_1$ band \cite{ensdf}, will belong to the $(2 {\cal N}-4, 2)$ irrep. As a consequence, no $B(E2)$ transitions can connect the $\gamma_1$ band and the gsb, since they belong to different irreps, against experimental evidence \cite{ensdf} suggesting that these transitions are strong, thus leading to the need to break the SU(3) symmetry in IBM-1. In contrast, in the proxy-SU(3) framework, the lowest-lying $(\lambda,\mu)$  irrep can accommodate at least the ground state band (having $K=0$), the first $K=2$ band (the $\gamma_1$ band), and the first $K=4$ band, since $\mu \geq 4$, while only the second $K=0$ band (the $\beta_1$ band) is forced to go into the next lowest-lying irrep,  $(\lambda+2,\mu-4)$. As a consequence, non-vanishing $B(E2)$s connecting the $\gamma_1$ band to the gsb are allowed, and no need to break the SU(3) symmetry occurs, thus resolving the problem mentioned at the end of section \ref{hw}.  
 
g) In a similar way, in the framework of the Vector Boson Model (VBM) \cite{Minkov1997} the ground state band and the first excited $K=2$ band, as well as the first excited  $K=4$ band, always belong to the same irrep \cite{Minkov1999,Minkov2000}.  

The regions shown in Fig. 1 cover all nuclei above $Z=28$ and $N=28$ away from magic numbers, up to the end of the rare earth region. Lighter nuclei can be successfully considered in a variety of methods, thus they are not of particular interest in the proxy-SU(3) framework. In addition, since in these nuclei valence protons and valence neutrons occupy the same shell, the recently developed proxy-SU(4) symmetry \cite{Kota2024} should be employed for them. Earlier work \cite{Sarantopoulou2017,Martinou2017b} in the actinides and in superheavy nuclei suggests that the method is applicable in these heavier regions, which are reserved for possible further work, due also to their large size. In this respect, it should be noted that the proxy-SU(3) approximation becomes more accurate in heavier shells \cite{Bonatsos2017a}.  
 
\section{Parameter-free predictions for the $\beta$ and $\gamma$ collective variables from the nhw irreps} \label{nhw} 
 
The predictions for the collective variables $\beta$ and $\gamma$ corresponding to the nhw irreps, obtained from Eqs. (\ref{b1}) and (\ref{g1}) respectively, are listed in Tables II-VI and compared to the predictions of the hw irreps in Figs. 3-14.  

In the region $Z=70$-80, $N=90$-122, we observe in Fig. 4 large discrepancies between the hw and nhw values of $\gamma$ at $N=94$, 102, 112. Deviations also appear at these $N$ values in Fig. 3 between the hw and nhw predictions for $\beta$, but they are much smaller in size. In addition, in Fig. 4 we observe that the hw $\gamma$ predictions for Yb ($Z=70$) are systematically much lower than the nhw predictions for the other series of isotopes. 

A similar picture is obtained in the region  $Z=58$-68, $N=88$-100. In Fig. 6, large discrepancies between the hw and nhw values of $\gamma$ are seen at $N=88$ and 94. Deviations also appear at these $N$ values in Fig. 5 between the hw and nhw predictions for $\beta$, but they are much smaller in size. In addition, in Fig. 6 we observe that the hw $\gamma$ predictions for Sm ($Z=62$) are systematically much lower than the nhw predictions for the other series of isotopes. This difference is also visible for the $\beta$ values, shown in Fig. 5. 

For corroboration of the above observations, which have been made by plotting the collective parameters $\beta$ and $\gamma$ for different series of isotopes vs. the neutron number $N$, we plot in Figs. 7 and 8 the collective parameters vs. the proton number $Z$ for different series of isotones, focusing on the region $Z=58$-80, $N=90$-100. The observations made are fully compatible with the remarks made above.   Large discrepancies between the hw and nhw values of $\gamma$ are observed in Fig. 8 at $Z=62$, 70, 80, while for $N=94$ the hw predictions for $\gamma$ are systematically lower than the nhw predictions. These discrepancies are also seen in Fig. 7 for the $\beta$ values, although their size is much smaller. 

As a consequence of the compatibility of observations made in plots vs. $N$ and vs. $Z$, in the rest of the regions the study is confined to plots vs. $N$. 

In the region $Z=54$-64, $N=64$-78, we observe in Fig. 10 large discrepancies between the hw and nhw values of $\gamma$ at $N=70$. Deviations also appear at this $N$ value in Fig. 9 between the hw and nhw predictions for $\beta$, but they are much smaller in size. In addition, in Fig. 10 we observe that the hw $\gamma$ predictions for Ba ($Z=56$) and Sm ($Z=62$) are systematically much lower than the nhw predictions for the other series of isotopes. 

In the region $Z=36$-46, $N=54$-72, we observe in Fig. 12 large discrepancies between the hw and nhw values of $\gamma$ at $N=56$, 62, 70. Deviations also appear at these $N$ values in Fig. 11 between the hw and nhw predictions for $\beta$, but they are much smaller in size. In addition, in Fig. 12 we observe that the hw $\gamma$ predictions for Zr ($Z=40$) are systematically much lower than the nhw predictions for the other series of isotopes. This difference is also visible for the $\beta$ values, shown in Fig. 11. 

In the region $Z=36$-46, $N=36$-46, we observe in Fig. 14 large discrepancies between the hw and nhw values of $\gamma$ at $N=40$. Deviations also appear at this $N$ value in Fig. 13 between the hw and nhw predictions for $\beta$, but they are much smaller in size. In addition, in Fig. 14 we observe that the hw $\gamma$ predictions for Zr ($Z=40$) are systematically much lower than the nhw predictions for the other series of isotopes. 

Summarizing the above results, we see that large discrepancies between the hw and nhw predictions occur for nucleon numbers 40, 56, 62, 70, 88, 94, 102, 112, which can be written as 28+12, 50+6, 50+12, 50+20, 82+6, 82+12, 82+20, 82+30, respectively, showing that discrepancies occur at 6, 12, 20, 30 nucleons above the magic numbers, corresponding to 6, 12, 20, 30 valence nucleons in Table I, for which hw irreps with $\mu=0$ occur. In other words, all discrepancies are rooted in the hw irreps with 
$\mu=0$. Therefore, these are the cases in which the hw irrep for a nucleus does not suffice for its description and, as a consequence, the nhw irrep has to be taken into account. 

\section{Combining the hw and nhw predictions} \label{mixing}

The question now surfaces on how to take the hw and nhw irreps simultaneously into account. As the simplest approximation, we consider that the two irreps will contribute to the final result on equal footing. Returning to Fig. 2, we replace the hw predictions for $\gamma$ by the average of the hw and nhw predictions given in Table II. In the $N=96$, 98 series of isotones, the replacement is made for $Z=70$, which corresponds to 20 valence protons in the 50-82 shell. In the $N=94$ series of isotones, 
all values are replaced, since $N=94$ corresponds to 12 valence neutrons in the 82-126 shell. We see that these replacements dramatically improve the agreement of the proxy-SU(3) predictions to the empirical values.

Some comments regarding further work are now in place.

a) The idea of an average SU(3) representation has been considered by Bhatt \textit{et al.} \cite{Bhatt2000} in a different context, while trying to estimate the contribution of the intruder orbitals to deformation \cite{Bhatt1992,Bhatt1994}. It has been based on the observation \cite{Kahane1997} that the distribution of angular momenta within the intrinsic state of the intruder nucleons, which does not have any SU(3) symmetry, is very similar to the distribution of angular momenta contained in an SU(3) intrinsic state with the same average value of the angular momentum. In earlier work \cite{Raman1988}, schematic $pf$, $sdg$, $pfh$, and $sdgi$ shells with SU(3) symmetry have been proposed as approximations to the 28-50, 50-82, 82-126, and 126-184 nucleon shells, in agreement with the proxy-SU(3) proposal, while for the $pf$ shell it has been remarked that it can be appropriate also for the 20-40 shell of the 3D-HO.   

b) From the mathematics point of view, it is known that when coupling two SU(3) irreps, a large number of irreps is obtained. Confining ourselves within the ground state, in order to have angular momentum quantum numbers equal to zero, the following equation occurs \cite{Alex2011}
\begin{equation}
(\lambda_2, \mu_2)\otimes (\lambda_1, \mu_1) = \sum_{\lambda,\mu,\rho} ( (\lambda_1, \mu_1, (\lambda_2, \mu_2)) |\rho \lambda \mu\rangle, \label{zeros}
\end{equation}
where  $c_{\lambda,\mu,\rho}= ( (\lambda_1, \mu_1, (\lambda_2, \mu_2))$ stands for an SU(3) to SO(3) Clebsch-Gordan coefficient with all angular momentum quantum numbers equal to zero, and $\rho$ is a multiplicity index. The resulting irreps can be readily calculated using the code \cite{vonDelft2010} accompanying Ref. \cite{Alex2011}, which lists the resulting irreps in order of increasing weight, thus the hw irrep, which is the most stretched irrep of Eq. (\ref{str}) and always has multiplicity equal  to unity, appears at the bottom of the list.  
The coefficients $c_{\lambda,\mu,\rho}$ can be readily calculated using a recent code of Dytrych et al. \cite{Dytrych2021}, which can be used up to $\lambda+\mu+L \leq 268$, while the older code of  Bahri et al. \cite{Bahri2004} was limited to lower values. 
It has been verified that the quantities  $c_{\lambda,\mu,\rho}^2$ represent the probability for the irrep $|\rho \lambda \mu\rangle$ to appear, if all possible irreps are taken into account. Although mathematically correct, the above set of irreps and relative probabilities ignores the nature of the nucleon-nucleon interaction, which dictates that the hw irrep alone dominates the collective properties of the nucleus. Under the special circumstances mentioned above, in the cases in which fully symmetric hw irreps with $\mu=0$ occur, the reasonable thing to assume is that the nhw will also be involved, the rest of the irreps remaining practically inactive.   

c) Since the early days of nuclear physics, it has been realized that nuclear structure is shaped up through the competition \cite{Ring1980} between the pairing \cite{Brink2005} and the quadrupole-quadrupole interaction. Pairing is known to be associated with the O(5) symmetry, allowing seniority to be a good quantum number \cite{Rakavy1957,Talmi1993}.  O(5) underlies both the vibrational [U(5)] and $\gamma$-unstable [O(6)] symmetries \cite{Iachello1987}, but not the rotational [SU(3)] symmetry. Therefore pairing is known to break SU(3). In other words, the presence of pairing will bring in additional SU(3) irreps for the description of a nucleus deviating from pure SU(3) behavior. The question comes on how to obtain the additional irreps called in by pairing, as well as on how to estimate their relevant importance. 
In the framework of pseudo-SU(3), the way the pairing interaction is coupling different irreps of SU(3) has been studied in detail in 
\cite{Troltenier1995a,Troltenier1995b,Bahri1994,Bahri1995,Escher1998}. The SU(3) tensor decomposition of the pairing interaction in the $pf$ shell is given in Table 7.1 of \cite{Bahri1994}, as well as in Table 1 of \cite{Bahri1995}, with the usual Elliott basis used, paving the way for analogous calculations in the proxy-SU(3) framework, in which pairing is expected to moderate the values of the deformation parameter $\beta$ near the beginning of the shells, where they appear to exceed the data \cite{Bonatsos2017b,Martinou2017}. 

\section{Conclusions} 

The proxy-SU(3) symmetry predicts, in a parameter-free way, the collective deformation variables $\beta$ and $\gamma$ in even-even atomic nuclei away from closed shells based on the highest weight (hw) irreducible representations (irreps) of SU(3) in the relevant proton and neutron shells, which are the most symmetric irreps allowed by the Pauli principle and the short-range nature of the nucleon-nucleon interactions. Within the proxy-SU(3) scheme, the hw irrep can accommodate the ground state band, the lowest $K=2$ band (usually referred to as the $\gamma_1$ band), and the lowest $K=4$ band.

However, substantial deviations from the empirical values occur when the hw irrep happens to be fully symmetric, i.e., boson-like, since in this case the hw irrep can accommodate only the ground state band, while the lowest $K=2$ and $K=4$ bands are pushed into the next lower-lying in energy irrep, which is the next highest weight (nhw) irrep. 

The hw and nhw irreps, along with their predictions for the collective variables $\beta$ and $\gamma$, calculated in five different regions of the nuclear chart, have been tabulated. A comparison between the two sets reveals that in most cases the hw and nhw predictions are very similar, providing evidence for the robustness of the proxy-SU(3) approach. 

As expected, substantial deviations occur in the above mentioned case, in which the hw irrep happens to be fully symmetric. The deviations are more dramatic in the case of the collective variable $\gamma$. It is seen that taking the average of the hw and nhw values in these special cases, results in predictions very close to the empirical values, thus eliminating the problematic  deviations. The proxy-SU(3) predictions taking into account the nhw irreps can also be compared to the results of other models, a first step in this direction taken in Ref. \cite{Bonatsos2024}.  

The calculated $\beta$ and $\gamma$ values could be used as input for more elaborate irrep mixing calculations. In addition, it would be interesting to examine to which extent the nhw irreps are useful in the framework of the pseudo-SU(3) symmetry, in which the use of the hw irreps has led \cite{Bonatsos2020} to predictions similar to these of the proxy-SU(3) scheme, despite the different assumptions made and different unitary transformations \cite{Martinou2020,Castanos1992a,Castanos1992b,Castanos1994} used in the two models.      

Finally, the extension of the present calculations to the actinides and to superheavy nuclei \cite{Martinou2017b,Sarantopoulou2017} is a straightforward task, 
expected to provide good results, since the proxy-SU(3) approximation becomes more accurate in heavier shells \cite{Bonatsos2017a}, while the extension to lighter 
nuclei, in which the valence protons and the valence neutrons occupy the same shells would require the use of the recently introduced proxy-SU(4) symmetry \cite{Kota2024}.
 
\section*{Acknowledgements} 

Support by the Bulgarian National Science Fund (BNSF) under Contract No. KP-06-N48/1  is gratefully acknowledged.



\begin{table*}

\caption{Highest weight (hw) irreducible representations of SU(3) and next highest weight (nhw) irreps of SU(3) for $M$ nucleons within the proxy-SU(3) scheme in the $sd$, $pf$, $sdg$, $pfh$, $sdgi$ shells having the overall symmetry  U(6), U(10), U(15), U(21), and U(28) respectively, calculated using the code UNTOU3 \cite{Draayer1989a}. The Elliott \cite{Elliott1958a} notation $(\lambda,\mu)$ is used for the SU(3) irreps.  
The corresponding shells of the shell model, within the proxy-SU(3) scheme \cite{Martinou2020} are also shown. Adapted from Ref. \cite{Bonatsos2024}.  
 See Section \ref{hw} for further discussion. 
}
\begin{tabular}{ r r r r r r r r r r r  }
\hline
$M$  & U(6) & U(6) & U(10) & U(10) & U(15) & U(15) & U(21) & U(21) & U(28) & U(28) \\
     & $sd$ & $sd$ & $pf$  & $pf$  & $sdg$ & $sdg$ & $pfh$ & $pfh$ & $sdgi$ & $sdgi$ \\
     & 8-20 & 8-20 & 28-50 & 28-50 & 50-82 & 50-82 & 82-126 & 82-126 & 126-128 & 126-184 \\
     & hw   & nhw  & hw    & nhw   & hw    & nhw   & hw    & nhw   & hw    & nhw   \\

      \hline
 2 & 4,0 & 0,2 &  6,0 &  2,2 &  8,0 &   4,2 &  10,0 &   6,2 &  12,0 &   8,2 \\ 
 4 & 4,2 & 0,4 &  8,2 &  4,4 & 12,2 &   8,4 &  16,2 &  12,4 &  20,2 &  16,4 \\
 6 & 6,0 & 0,6 & 12.0 &  6,6 & 18,0 &  12,6 &  24,0 &  18,6 &  30,0 &  24,6 \\
 8 & 2,4 & 4,0 & 10,4 & 12,0 & 18,4 &  20,0 &  26,4 &  28,0 &  34,4 &  36,0 \\
10 & 0,4 & 2,0 & 10,4 & 12,0 & 20,4 &  22,0 &  30,4 &  32,0 &  40,4 &  42,0 \\
12 & 0,0 &     & 12,0 & 4,10 & 24,0 & 16,10 &  36,0 & 28,10 &  48,0 & 40,10 \\
14 &     &     &  6,6 &  8,2 & 20,6 &  22,2 &  34,6 &  36,2 &  48,6 &  50,2 \\
16 &     &     &  2,8 &  4,4 & 18,8 &  20,4 &  34,8 &  36,4 &  50,8 &  52,4 \\
18 &     &     &  0,6 &  2,2 & 18,6 &  20,2 &  36,6 &  38,2 &  54,6 &  56,2 \\
20 &     &     &  0,0 &      & 20,0 & 10,14 &  40,0 & 30,14 &  60,0 & 50,14 \\
22 &     &     &      &      & 12,8 &  14,4 &  34,8 &  36,4 &  56,8 &  58,4 \\
24 &     &     &      &      & 6,12 &   8,8 & 30,12 &  32,8 & 54,12 &  56,8 \\
26 &     &     &      &      & 2,12 &   4,8 & 28,12 &  30,8 & 54,12 &  56,8 \\
28 &     &     &      &      &  0,8 &   2,4 &  28,8 &  30,4 &  56,8 &  58,4 \\
30 &     &     &      &      &  0,0 &       &  30,0 & 18,18 &  60,0 & 48,18 \\
32 &     &     &      &      &      &       & 20,10 &  22,6 & 52,10 &  54,6 \\
34 &     &     &      &      &      &       & 12,16 & 14,12 & 46,16 & 48,12 \\ 
36 &     &     &      &      &      &       &  6,18 &  8,14 & 42,18 & 44,14 \\
38 &     &     &      &      &      &       &  2,16 &  4,12 & 40,16 & 42,12 \\
40 &     &     &      &      &      &       &  0,10 &   2,6 & 40,10 &  42,6 \\
42 &     &     &      &      &      &       &   0,0 &       &  42,0 & 28,22 \\
44 &     &     &      &      &      &       &       &       & 30,12 &  32,8 \\
46 &     &     &      &      &      &       &       &       & 20,20 & 22,16 \\
48 &     &     &      &      &      &       &       &       & 12,24 & 14,20 \\
50 &     &     &      &      &      &       &       &       &  6,24 &  8,20 \\
52 &     &     &      &      &      &       &       &       &  2,20 &  4,16 \\
54 &     &     &      &      &      &       &       &       &  0,12 &   2,8 \\
56 &     &     &      &      &      &       &       &       &   0,0 &       \\

\hline
\end{tabular}

\end{table*}


\begin{table}

\caption{Highest weight (hw) irreducible representations of SU(3) and next highest weight (nhw) irreps of SU(3) within the proxy-SU(3) scheme for nuclei in the $Z=70$-80, $N=90$-122 region. The Elliott \cite{Elliott1958a} notation $(\lambda,\mu)$ is used for the SU(3) irreps. Parameter-independent predictions for the collective variables $\beta$ and $\gamma$ within each irrep, calculated from Eqs. (\ref{b1}) and (\ref{g1}) are also given, labeled by hw and nhw respectively. See Section \ref{nhw} for further discussion. 
}
\begin{tabular}{ r   r  r r c c r  }
\hline
nucleus  & hw & nhw & $\beta_{hw}$ & $\beta_{nhw}$ & $\gamma_{hw}$ & $\gamma_{nhw}$  \\ 
        
\hline

 \isotope[160][70]{Yb}$_{90}$ & 46,  4 & 38, 14 & 0.252 & 0.245 &  5.00 & 15.61 \\
 \isotope[162][70]{Yb}$_{92}$ & 50,  4 & 42, 14 & 0.271 & 0.263 &  4.63 & 14.43 \\
 \isotope[164][70]{Yb}$_{94}$ & 56,  0 & 38, 24 & 0.290 & 0.282 &  0.86 & 22.80 \\
 \isotope[166][70]{Yb}$_{96}$ & 54,  6 & 46, 16 & 0.295 & 0.288 &  5.92 & 14.86 \\
 \isotope[168][70]{Yb}$_{98}$ & 54,  8 & 46, 18 & 0.300 & 0.294 &  7.46 & 16.24 \\
\isotope[170][70]{Yb}$_{100}$ & 56,  6 & 48, 16 & 0.302 & 0.295 &  5.72 & 14.36 \\
\isotope[172][70]{Yb}$_{102}$ & 60,  0 & 40, 28 & 0.305 & 0.302 &  0.81 & 24.35 \\
\isotope[174][70]{Yb}$_{104}$ & 54,  8 & 46, 18 & 0.296 & 0.290 &  7.46 & 16.24 \\
\isotope[176][70]{Yb}$_{106}$ & 50, 12 & 42, 22 & 0.288 & 0.285 & 11.08 & 20.08 \\
\isotope[178][70]{Yb}$_{108}$ & 48, 12 & 40, 22 & 0.277 & 0.275 & 11.47 & 20.78 \\
\isotope[180][70]{Yb}$_{110}$ & 48,  8 & 40, 18 & 0.264 & 0.259 &  8.29 & 18.05 \\
\isotope[182][70]{Yb}$_{112}$ & 50,  0 & 28, 32 & 0.251 & 0.262 &  0.96 & 32.13 \\
\isotope[184][70]{Yb}$_{114}$ & 40, 10 & 32, 20 & 0.230 & 0.228 & 11.58 & 22.69 \\
\isotope[186][70]{Yb}$_{116}$ & 32, 16 & 24, 26 & 0.213 & 0.218 & 19.53 & 31.27 \\
\isotope[188][70]{Yb}$_{118}$ & 26, 18 & 18, 28 & 0.193 & 0.202 & 24.27 & 36.86 \\
\isotope[190][70]{Yb}$_{120}$ & 22, 16 & 14, 26 & 0.167 & 0.177 & 25.05 & 39.37 \\
\isotope[192][70]{Yb}$_{122}$ & 20, 10 & 12, 20 & 0.135 & 0.142 & 19.77 & 37.74 \\

 \isotope[162][72]{Hf}$_{90}$ & 38, 12 & 40,  8 & 0.237 & 0.233 & 13.90 &  9.72 \\
 \isotope[164][72]{Hf}$_{92}$ & 42, 12 & 44,  8 & 0.256 & 0.253 & 12.81 &  8.95 \\
 \isotope[166][72]{Hf}$_{94}$ & 48,  8 & 40, 18 & 0.271 & 0.266 &  8.29 & 18.05 \\
 \isotope[168][72]{Hf}$_{96}$ & 46, 14 & 48, 10 & 0.280 & 0.276 & 13.41 &  9.91 \\
 \isotope[170][72]{Hf}$_{98}$ & 46, 16 & 48, 12 & 0.286 & 0.282 & 14.86 & 11.47 \\
\isotope[172][72]{Hf}$_{100}$ & 48, 14 & 50, 10 & 0.287 & 0.284 & 12.95 &  9.57 \\
\isotope[174][72]{Hf}$_{102}$ & 52,  8 & 42, 22 & 0.286 & 0.286 &  7.72 & 20.08 \\
\isotope[176][72]{Hf}$_{104}$ & 46, 16 & 48, 12 & 0.282 & 0.278 & 14.86 & 11.47 \\
\isotope[178][72]{Hf}$_{106}$ & 42, 20 & 44, 16 & 0.277 & 0.272 & 18.77 & 15.39 \\
\isotope[180][72]{Hf}$_{108}$ & 40, 20 & 42, 16 & 0.267 & 0.262 & 19.45 & 15.95 \\
\isotope[182][72]{Hf}$_{110}$ & 40, 16 & 42, 12 & 0.251 & 0.247 & 16.56 & 12.81 \\
\isotope[184][72]{Hf}$_{112}$ & 42,  8 & 30, 26 & 0.233 & 0.243 &  9.32 & 27.72 \\
\isotope[186][72]{Hf}$_{114}$ & 32, 18 & 34, 14 & 0.220 & 0.215 & 21.16 & 17.00 \\
\isotope[188][72]{Hf}$_{116}$ & 24, 24 & 26, 20 & 0.208 & 0.200 & 30.00 & 25.87 \\
\isotope[190][72]{Hf}$_{118}$ & 18, 26 & 20, 22 & 0.192 & 0.183 & 35.73 & 31.50 \\
\isotope[192][72]{Hf}$_{120}$ & 14, 24 & 16, 20 & 0.167 & 0.157 & 38.21 & 33.48 \\
\isotope[194][72]{Hf}$_{122}$ & 12, 18 & 14, 14 & 0.133 & 0.124 & 36.18 & 30.00 \\

 \isotope[164][74]{W}$_{90}$ & 32, 16 & 34, 12 & 0.222 & 0.217 & 19.53 & 15.18 \\
 \isotope[166][74]{W}$_{92}$ & 36, 16 & 38, 12 & 0.240 & 0.235 & 17.93 & 13.90 \\
 \isotope[168][74]{W}$_{94}$ & 42, 12 & 34, 22 & 0.254 & 0.253 & 12.81 & 23.19 \\
 \isotope[170][74]{W}$_{96}$ & 40, 18 & 42, 14 & 0.264 & 0.259 & 18.05 & 14.43 \\
 \isotope[172][74]{W}$_{98}$ & 40, 20 & 42, 16 & 0.271 & 0.266 & 19.45 & 15.95 \\
\isotope[174][74]{W}$_{100}$ & 42, 18 & 44, 14 & 0.272 & 0.267 & 17.40 & 13.90 \\
\isotope[176][74]{W}$_{102}$ & 46, 12 & 36, 26 & 0.269 & 0.274 & 11.88 & 24.85 \\
\isotope[178][74]{W}$_{104}$ & 40, 20 & 42, 16 & 0.268 & 0.263 & 19.45 & 15.95 \\
\isotope[180][74]{W}$_{106}$ & 36, 24 & 38, 20 & 0.264 & 0.258 & 23.62 & 20.17 \\
\isotope[182][74]{W}$_{108}$ & 34, 24 & 36, 20 & 0.254 & 0.247 & 24.50 & 20.95 \\
\isotope[184][74]{W}$_{110}$ & 34, 20 & 36, 16 & 0.237 & 0.231 & 21.79 & 17.93 \\
\isotope[186][74]{W}$_{112}$ & 36, 12 & 24, 30 & 0.217 & 0.235 & 14.51 & 33.54 \\
\isotope[188][74]{W}$_{114}$ & 26, 22 & 28, 18 & 0.209 & 0.202 & 27.36 & 23.14 \\
\isotope[190][74]{W}$_{116}$ & 18, 28 & 20, 24 & 0.201 & 0.191 & 36.86 & 32.87 \\
\isotope[192][74]{W}$_{118}$ & 12, 30 & 14, 26 & 0.187 & 0.176 & 43.29 & 39.37 \\
\isotope[194][74]{W}$_{120}$ &  8, 28 & 10, 24 & 0.164 & 0.152 & 46.90 & 42.65 \\
\isotope[196][74]{W}$_{122}$ &  6, 22 &  8, 18 & 0.129 & 0.117 & 47.11 & 41.65 \\


\hline
\end{tabular}

\end{table}

\setcounter{table}{1}

\begin{table}

\caption{(continued)
}
\begin{tabular}{ r   r  r r c c r  }
\hline
nucleus  & hw & nhw & $\beta_{hw}$ & $\beta_{nhw}$ & $\gamma_{hw}$ & $\gamma_{nhw}$  \\ 
        
\hline

 \isotope[166][76]{Os}$_{90}$ & 28, 16 & 30, 12 & 0.202 & 0.196 & 21.43 & 16.71 \\
 \isotope[168][76]{Os}$_{92}$ & 32, 16 & 34, 12 & 0.220 & 0.215 & 19.53 & 15.18 \\
 \isotope[170][76]{Os}$_{94}$ & 38, 12 & 30, 22 & 0.233 & 0.233 & 13.90 & 25.11 \\
 \isotope[172][76]{Os}$_{96}$ & 36, 18 & 38, 14 & 0.244 & 0.239 & 19.49 & 15.61 \\
 \isotope[174][76]{Os}$_{98}$ & 36, 20 & 38, 16 & 0.251 & 0.245 & 20.95 & 17.22 \\
\isotope[176][76]{Os}$_{100}$ & 38, 18 & 40, 14 & 0.252 & 0.247 & 18.74 & 14.99 \\
\isotope[178][76]{Os}$_{102}$ & 42, 12 & 32, 26 & 0.249 & 0.255 & 12.81 & 26.70 \\
\isotope[180][76]{Os}$_{104}$ & 36, 20 & 38, 16 & 0.248 & 0.242 & 20.95 & 17.22 \\
\isotope[182][76]{Os}$_{106}$ & 32, 24 & 34, 20 & 0.245 & 0.238 & 25.45 & 21.79 \\
\isotope[184][76]{Os}$_{108}$ & 30, 24 & 32, 20 & 0.235 & 0.228 & 26.46 & 22.69 \\
\isotope[186][76]{Os}$_{110}$ & 30, 20 & 32, 16 & 0.219 & 0.213 & 23.66 & 19.53 \\
\isotope[188][76]{Os}$_{112}$ & 32, 12 & 20, 30 & 0.198 & 0.218 & 15.91 & 36.34 \\
\isotope[190][76]{Os}$_{114}$ & 22, 22 & 24, 18 & 0.191 & 0.183 & 30.00 & 25.50 \\
\isotope[192][76]{Os}$_{116}$ & 14, 28 & 16, 24 & 0.185 & 0.175 & 40.41 & 36.28 \\
\isotope[194][76]{Os}$_{118}$ &  8, 30 & 10, 26 & 0.173 & 0.161 & 47.62 & 43.66 \\
\isotope[196][76]{Os}$_{120}$ &  4, 28 &  6, 24 & 0.151 & 0.138 & 51.17 & 47.99 \\
\isotope[198][76]{Os}$_{122}$ &  2, 22 &  4, 18 & 0.116 & 0.103 & 53.95 & 48.61 \\

 \isotope[168][78]{Pt}$_{90}$ & 26, 12 & 28,  8 & 0.177 & 0.172 & 18.58 & 13.10 \\
 \isotope[170][78]{Pt}$_{92}$ & 30, 12 & 32,  8 & 0.195 & 0.191 & 16.71 & 11.74 \\
 \isotope[172][78]{Pt}$_{94}$ & 36,  8 & 28, 18 & 0.209 & 0.207 & 10.64 & 23.14 \\
 \isotope[174][78]{Pt}$_{96}$ & 34, 14 & 36, 10 & 0.219 & 0.215 & 17.00 & 12.63 \\
 \isotope[176][78]{Pt}$_{98}$ & 34, 16 & 36, 12 & 0.226 & 0.221 & 18.70 & 14.51 \\
\isotope[178][78]{Pt}$_{100}$ & 36, 14 & 38, 10 & 0.227 & 0.223 & 16.27 & 12.08 \\
\isotope[180][78]{Pt}$_{102}$ & 40,  8 & 30, 22 & 0.225 & 0.229 &  9.72 & 25.11 \\
\isotope[182][78]{Pt}$_{104}$ & 34, 16 & 36, 12 & 0.223 & 0.218 & 18.70 & 14.51 \\
\isotope[184][78]{Pt}$_{106}$ & 30, 20 & 32, 16 & 0.220 & 0.214 & 23.66 & 19.53 \\
\isotope[186][78]{Pt}$_{108}$ & 28, 20 & 30, 16 & 0.210 & 0.204 & 24.72 & 20.44 \\
\isotope[188][78]{Pt}$_{110}$ & 28, 16 & 30, 12 & 0.194 & 0.189 & 21.43 & 16.71 \\
\isotope[190][78]{Pt}$_{112}$ & 30,  8 & 18, 26 & 0.174 & 0.192 & 12.38 & 35.73 \\
\isotope[192][78]{Pt}$_{114}$ & 20, 18 & 22, 14 & 0.166 & 0.159 & 28.35 & 23.07 \\
\isotope[194][78]{Pt}$_{116}$ & 12, 24 & 14, 20 & 0.159 & 0.149 & 40.33 & 35.50 \\
\isotope[196][78]{Pt}$_{118}$ &  6, 26 &  8, 22 & 0.148 & 0.136 & 48.76 & 44.18 \\
\isotope[198][78]{Pt}$_{120}$ &  2, 24 &  4, 20 & 0.126 & 0.113 & 54.40 & 49.56 \\
\isotope[200][78]{Pt}$_{122}$ &  0, 18 &  2, 14 & 0.092 & 0.079 & 57.46 & 51.05 \\


 \isotope[170][80]{Hg}$_{90}$ & 26,  4 & 28,  0 & 0.148 & 0.146 &  8.35 &  1.68 \\
 \isotope[172][80]{Hg}$_{92}$ & 30,  4 & 32,  0 & 0.167 & 0.166 &  7.37 &  1.48 \\
 \isotope[174][80]{Hg}$_{94}$ & 36,  0 & 28, 10 & 0.185 & 0.177 &  1.32 & 15.44 \\
 \isotope[176][80]{Hg}$_{96}$ & 34,  6 & 36,  2 & 0.192 & 0.190 &  8.95 &  3.86 \\
 \isotope[178][80]{Hg}$_{98}$ & 34,  8 & 36,  4 & 0.197 & 0.194 & 11.16 &  6.26 \\
\isotope[180][80]{Hg}$_{100}$ & 36,  6 & 38,  2 & 0.200 & 0.198 &  8.51 &  3.67 \\
\isotope[182][80]{Hg}$_{102}$ & 40,  0 & 30, 14 & 0.202 & 0.198 &  1.20 & 18.65 \\
\isotope[184][80]{Hg}$_{104}$ & 34,  8 & 36,  4 & 0.195 & 0.192 & 11.16 &  6.26 \\
\isotope[186][80]{Hg}$_{106}$ & 30, 12 & 32,  8 & 0.189 & 0.185 & 16.71 & 11.74 \\
\isotope[188][80]{Hg}$_{108}$ & 28, 12 & 30,  8 & 0.179 & 0.175 & 17.60 & 12.38 \\
\isotope[190][80]{Hg}$_{110}$ & 28,  8 & 30,  4 & 0.165 & 0.162 & 13.10 &  7.37 \\
\isotope[192][80]{Hg}$_{112}$ & 30,  0 & 18, 18 & 0.151 & 0.158 &  1.57 & 30.00 \\
\isotope[194][80]{Hg}$_{114}$ & 20, 10 & 22,  6 & 0.134 & 0.129 & 19.77 & 12.89 \\
\isotope[196][80]{Hg}$_{116}$ & 12, 16 & 14, 12 & 0.124 & 0.115 & 34.40 & 27.64 \\
\isotope[198][80]{Hg}$_{118}$ &  6, 18 &  8, 14 & 0.110 & 0.099 & 44.92 & 38.21 \\
\isotope[200][80]{Hg}$_{120}$ &  2, 16 &  4, 12 & 0.088 & 0.076 & 52.01 & 44.39 \\
\isotope[202][80]{Hg}$_{122}$ &  0, 10 &  2,  6 & 0.054 & 0.042 & 55.69 & 43.00 \\

\hline
\end{tabular}

\end{table}


\begin{table}

\caption{Highest weight (hw) irreducible representations of SU(3) and next highest weight (nhw) irreps of SU(3) within the proxy-SU(3) scheme for nuclei in the $Z=58$-68, $N=88$-100 region. The Elliott \cite{Elliott1958a} notation $(\lambda,\mu)$ is used for the SU(3) irreps. Parameter-independent predictions for the collective variables $\beta$ and $\gamma$ within each irrep, calculated from Eqs. (\ref{b1}) and (\ref{g1}) are also given, labeled by hw and nhw respectively. See Section \ref{nhw} for further discussion.}

\begin{tabular}{ r   r  r r c c r  }
\hline
nucleus  & hw & nhw & $\beta_{hw}$ & $\beta_{nhw}$ & $\gamma_{hw}$ & $\gamma_{nhw}$  \\ 
        
\hline

 \isotope[146][58]{Ce}$_{88}$ & 42,  4 & 36, 10 & 0.239 & 0.228 &  5.44 & 12.63 \\
 \isotope[148][58]{Ce}$_{90}$ & 44,  8 & 46,  4 & 0.261 & 0.259 &  8.95 &  5.00 \\
 \isotope[150][58]{Ce}$_{92}$ & 48,  8 & 50,  4 & 0.280 & 0.278 &  8.29 &  4.63 \\
 \isotope[152][58]{Ce}$_{94}$ & 54,  4 & 46, 14 & 0.298 & 0.290 &  4.31 & 13.41 \\
 \isotope[154][58]{Ce}$_{96}$ & 52, 10 & 54,  6 & 0.305 & 0.303 &  9.25 &  5.92 \\
 \isotope[156][58]{Ce}$_{98}$ & 52, 12 & 54,  8 & 0.310 & 0.307 & 10.71 &  7.46 \\
\isotope[158][58]{Ce}$_{100}$ & 54, 10 & 56,  6 & 0.312 & 0.310 &  8.95 &  5.72 \\


 \isotope[148][60]{Nd}$_{88}$ & 44,  4 & 38, 10 & 0.249 & 0.238 &  5.21 & 12.08 \\
 \isotope[150][60]{Nd}$_{90}$ & 46,  8 & 48,  4 & 0.270 & 0.268 &  8.61 &  4.81 \\
 \isotope[152][60]{Nd}$_{92}$ & 50,  8 & 52,  4 & 0.289 & 0.287 &  7.99 &  4.46 \\
 \isotope[154][60]{Nd}$_{94}$ & 56,  4 & 48, 14 & 0.307 & 0.298 &  4.16 & 12.95 \\
 \isotope[156][60]{Nd}$_{96}$ & 54, 10 & 56,  6 & 0.314 & 0.312 &  8.95 &  5.72 \\
 \isotope[158][60]{Nd}$_{98}$ & 54, 12 & 56,  8 & 0.319 & 0.316 & 10.37 &  7.22 \\
\isotope[160][60]{Nd}$_{100}$ & 56, 10 & 58,  6 & 0.321 & 0.319 &  8.67 &  5.54 \\

 \isotope[150][62]{Sm}$_{88}$ & 48,  0 & 34, 16 & 0.257 & 0.238 &  1.00 & 18.70 \\
 \isotope[152][62]{Sm}$_{90}$ & 50,  4 & 44, 10 & 0.277 & 0.265 &  4.63 & 10.68 \\
 \isotope[154][62]{Sm}$_{92}$ & 54,  4 & 48, 10 & 0.296 & 0.284 &  4.31 &  9.91 \\
 \isotope[156][62]{Sm}$_{94}$ & 60,  0 & 44, 20 & 0.315 & 0.299 &  0.81 & 18.14 \\
 \isotope[158][62]{Sm}$_{96}$ & 58,  6 & 52, 12 & 0.320 & 0.309 &  5.54 & 10.71 \\
 \isotope[160][62]{Sm}$_{98}$ & 58,  8 & 52, 14 & 0.325 & 0.314 &  7.00 & 12.12 \\
\isotope[162][62]{Sm}$_{100}$ & 60,  6 & 54, 12 & 0.328 & 0.317 &  5.37 & 10.37 \\

 \isotope[152][64]{Gd}$_{88}$ & 44,  6 & 38, 12 & 0.252 & 0.242 &  7.12 & 13.90 \\
 \isotope[154][64]{Gd}$_{90}$ & 46, 10 & 48,  6 & 0.274 & 0.271 & 10.28 &  6.59 \\
 \isotope[156][64]{Gd}$_{92}$ & 50, 10 & 52,  6 & 0.293 & 0.291 &  9.57 &  6.12 \\
 \isotope[158][64]{Gd}$_{94}$ & 56,  6 & 48, 16 & 0.310 & 0.303 &  5.72 & 14.36 \\
 \isotope[160][64]{Gd}$_{96}$ & 54, 12 & 56,  8 & 0.317 & 0.314 & 10.37 &  7.22 \\
 \isotope[162][64]{Gd}$_{98}$ & 54, 14 & 56, 10 & 0.323 & 0.320 & 11.74 &  8.67 \\
\isotope[164][64]{Gd}$_{100}$ & 56, 12 & 58,  8 & 0.325 & 0.322 & 10.05 &  7.00 \\

 \isotope[154][66]{Dy}$_{88}$ & 42,  8 & 36, 14 & 0.247 & 0.238 &  9.32 & 16.27 \\
 \isotope[156][66]{Dy}$_{90}$ & 44, 12 & 46,  8 & 0.270 & 0.267 & 12.33 &  8.61 \\
 \isotope[158][66]{Dy}$_{92}$ & 48, 12 & 50,  8 & 0.289 & 0.286 & 11.47 &  7.99 \\
 \isotope[160][66]{Dy}$_{94}$ & 54,  8 & 46, 18 & 0.305 & 0.299 &  7.46 & 16.24 \\
 \isotope[162][66]{Dy}$_{96}$ & 52, 14 & 54, 10 & 0.313 & 0.310 & 12.12 &  8.95 \\
 \isotope[164][66]{Dy}$_{98}$ & 52, 16 & 54, 12 & 0.318 & 0.314 & 13.46 & 10.37 \\
\isotope[166][66]{Dy}$_{100}$ & 54, 14 & 56, 10 & 0.320 & 0.317 & 11.74 &  8.67 \\


 \isotope[156][66]{Er}$_{88}$ & 42,  6 & 36, 12 & 0.240 & 0.230 &  7.43 & 14.51 \\
 \isotope[158][66]{Er}$_{90}$ & 44, 10 & 46,  6 & 0.262 & 0.259 & 10.68 &  6.85 \\
 \isotope[160][66]{Er}$_{92}$ & 48, 10 & 50,  6 & 0.281 & 0.279 &  9.91 &  6.35 \\
 \isotope[162][66]{Er}$_{94}$ & 54,  6 & 46, 16 & 0.297 & 0.290 &  5.92 & 14.86 \\
 \isotope[164][66]{Er}$_{96}$ & 52, 12 & 54,  8 & 0.305 & 0.302 & 10.71 &  7.46 \\
 \isotope[166][66]{Er}$_{98}$ & 52, 14 & 54, 10 & 0.310 & 0.307 & 12.12 &  8.95 \\
\isotope[168][66]{Er}$_{100}$ & 54, 12 & 56,  8 & 0.312 & 0.309 & 10.37 &  7.22 \\

\hline
\end{tabular}

\end{table}


\begin{table}

\caption{Highest weight (hw) irreducible representations of SU(3) and next highest weight (nhw) irreps of SU(3) within the proxy-SU(3) scheme for nuclei in the $Z=54$-64, $N=64$-78 region. The Elliott \cite{Elliott1958a} notation $(\lambda,\mu)$ is used for the SU(3) irreps. Parameter-independent predictions for the collective variables $\beta$ and $\gamma$ within each irrep, calculated from Eqs. (\ref{b1}) and (\ref{g1}) are also given, labeled by hw and nhw respectively. See Section \ref{nhw} for further discussion.
}
\begin{tabular}{ r   r  r r c c r  }
\hline
nucleus  & hw & nhw & $\beta_{hw}$ & $\beta_{nhw}$ & $\gamma_{hw}$ & $\gamma_{nhw}$  \\ 
        
\hline

\isotope[118][54]{Xe}$_{64}$ & 32,  8 & 34,  4 & 0.256 & 0.252 & 11.74 &  6.59 \\
\isotope[120][54]{Xe}$_{66}$ & 30, 10 & 32,  6 & 0.250 & 0.245 & 14.63 &  9.43 \\
\isotope[122][54]{Xe}$_{68}$ & 30,  8 & 32,  4 & 0.240 & 0.236 & 12.38 &  6.95 \\
\isotope[124][54]{Xe}$_{70}$ & 32,  2 & 22, 16 & 0.227 & 0.228 &  4.31 & 25.05 \\
\isotope[126][54]{Xe}$_{72}$ & 24, 10 & 26,  6 & 0.209 & 0.203 & 17.35 & 11.24 \\
\isotope[128][54]{Xe}$_{74}$ & 18, 14 & 20, 10 & 0.192 & 0.183 & 26.11 & 19.77 \\
\isotope[130][54]{Xe}$_{76}$ & 14, 14 & 16, 10 & 0.168 & 0.158 & 30.00 & 22.95 \\
\isotope[132][54]{Xe}$_{78}$ & 12, 10 & 14,  6 & 0.134 & 0.117 & 27.25 & 18.14 \\


\isotope[120][56]{Ba}$_{64}$ & 38,  6 & 34,  8 & 0.285 & 0.267 &  8.12 & 11.16 \\
\isotope[122][56]{Ba}$_{66}$ & 36,  8 & 32, 10 & 0.279 & 0.262 & 10.64 & 13.90 \\
\isotope[124][56]{Ba}$_{68}$ & 36,  6 & 32,  8 & 0.269 & 0.252 &  8.51 & 11.74 \\
\isotope[126][56]{Ba}$_{70}$ & 38,  0 & 22, 20 & 0.258 & 0.249 &  1.26 & 28.50 \\
\isotope[128][56]{Ba}$_{72}$ & 30,  8 & 26, 10 & 0.236 & 0.220 & 12.38 & 16.34 \\
\isotope[130][56]{Ba}$_{74}$ & 24, 12 & 20, 14 & 0.216 & 0.202 & 19.67 & 24.50 \\
\isotope[132][56]{Ba}$_{76}$ & 20, 12 & 16, 14 & 0.191 & 0.178 & 22.26 & 27.93 \\
\isotope[134][56]{Ba}$_{78}$ & 18,  8 & 14, 10 & 0.158 & 0.144 & 18.35 & 24.92 \\


\isotope[122][58]{Ce}$_{64}$ & 38, 10 & 40,  6 & 0.300 & 0.296 & 12.08 &  7.76 \\
\isotope[124][58]{Ce}$_{66}$ & 36, 12 & 38,  8 & 0.295 & 0.290 & 14.51 & 10.16 \\
\isotope[126][58]{Ce}$_{68}$ & 36, 10 & 38,  6 & 0.284 & 0.276 & 12.63 &  8.12 \\
\isotope[128][58]{Ce}$_{70}$ & 38,  4 & 28, 18 & 0.271 & 0.272 &  5.96 & 23.14 \\
\isotope[130][58]{Ce}$_{72}$ & 30, 12 & 32,  8 & 0.253 & 0.247 & 16.71 & 11.74 \\
\isotope[132][58]{Ce}$_{74}$ & 24, 16 & 26, 12 & 0.235 & 0.227 & 23.72 & 18.58 \\
\isotope[134][58]{Ce}$_{76}$ & 20, 16 & 22, 12 & 0.211 & 0.202 & 26.52 & 20.89 \\
\isotope[136][58]{Ce}$_{78}$ & 18, 12 & 20,  8 & 0.177 & 0.169 & 23.82 & 17.00 \\


\isotope[124][60]{Nd}$_{64}$ & 40, 10 & 42,  6 & 0.312 & 0.308 & 11.58 &  7.43 \\
\isotope[126][60]{Nd}$_{66}$ & 38, 12 & 40,  8 & 0.306 & 0.301 & 13.90 &  9.72 \\
\isotope[128][60]{Nd}$_{68}$ & 38, 10 & 40,  6 & 0.296 & 0.292 & 12.08 &  7.76 \\
\isotope[130][60]{Nd}$_{70}$ & 40,  4 & 30, 18 & 0.282 & 0.282 &  5.68 & 22.11 \\
\isotope[132][60]{Nd}$_{72}$ & 32, 12 & 34,  8 & 0.264 & 0.259 & 15.91 & 11.16 \\
\isotope[134][60]{Nd}$_{74}$ & 26, 16 & 28, 12 & 0.246 & 0.238 & 22.52 & 17.60 \\
\isotope[136][60]{Nd}$_{76}$ & 22, 16 & 24, 12 & 0.221 & 0.213 & 25.05 & 19.67 \\
\isotope[138][60]{Nd}$_{78}$ & 20, 12 & 22,  8 & 0.188 & 0.174 & 22.26 & 15.82 \\

\isotope[126][62]{Sm}$_{64}$ & 44,  6 & 38, 12 & 0.319 & 0.306 &  7.12 & 13.90 \\
\isotope[128][62]{Sm}$_{66}$ & 42,  8 & 36, 14 & 0.313 & 0.301 &  9.32 & 16.27 \\
\isotope[130][62]{Sm}$_{68}$ & 42,  6 & 36, 12 & 0.303 & 0.290 &  7.43 & 14.51 \\
\isotope[132][62]{Sm}$_{70}$ & 44,  0 & 26, 24 & 0.293 & 0.290 &  1.09 & 28.73 \\
\isotope[134][62]{Sm}$_{72}$ & 36,  8 & 30, 14 & 0.270 & 0.260 & 10.64 & 18.65 \\
\isotope[136][62]{Sm}$_{74}$ & 30, 12 & 24, 18 & 0.249 & 0.243 & 16.71 & 25.50 \\
\isotope[138][62]{Sm}$_{76}$ & 26, 12 & 20, 18 & 0.224 & 0.220 & 18.58 & 28.35 \\
\isotope[140][62]{Sm}$_{78}$ & 24,  8 & 18, 14 & 0.192 & 0.186 & 14.80 & 26.11 \\

\isotope[128][64]{Gd}$_{64}$ & 40, 12 & 42,  8 & 0.317 & 0.313 & 13.33 &  9.32 \\
\isotope[130][64]{Gd}$_{66}$ & 38, 14 & 40, 10 & 0.312 & 0.307 & 15.61 & 11.58 \\
\isotope[132][64]{Gd}$_{68}$ & 38, 12 & 40,  8 & 0.301 & 0.296 & 13.90 &  9.72 \\
\isotope[134][64]{Gd}$_{70}$ & 40,  6 & 30, 20 & 0.287 & 0.290 &  7.76 & 23.66 \\
\isotope[136][64]{Gd}$_{72}$ & 32, 14 & 34, 10 & 0.271 & 0.265 & 17.78 & 13.24 \\
\isotope[138][64]{Gd}$_{74}$ & 26, 18 & 28, 14 & 0.254 & 0.246 & 24.27 & 19.59 \\
\isotope[140][64]{Gd}$_{76}$ & 22, 18 & 24, 14 & 0.230 & 0.221 & 26.85 & 21.79 \\
\isotope[142][64]{Gd}$_{78}$ & 20, 14 & 22, 10 & 0.196 & 0.188 & 24.50 & 18.48 \\

\hline
\end{tabular}

\end{table}


\begin{table}

\caption{Highest weight (hw) irreducible representations of SU(3) and next highest weight (nhw) irreps of SU(3) within the proxy-SU(3) scheme for nuclei in the $Z=36$-46, $N=62$-72 region. The Elliott \cite{Elliott1958a} notation $(\lambda,\mu)$ is used for the SU(3) irreps. Parameter-independent predictions for the collective variables $\beta$ and $\gamma$ within each irrep, calculated from Eqs. (\ref{b1}) and (\ref{g1}) are also given, labeled by hw and nhw respectively. See Section \ref{nhw} for further discussion.
}
\begin{tabular}{ r   r  r r c c r  }
\hline
nucleus  & hw & nhw & $\beta_{hw}$ & $\beta_{nhw}$ & $\gamma_{hw}$ & $\gamma_{nhw}$  \\ 
        
\hline

 \isotope[90][36]{Kr}$_{54}$ & 22,  6 & 18,  8 & 0.235 & 0.214 & 12.89 & 18.35 \\
 \isotope[92][36]{Kr}$_{56}$ & 28,  4 & 22, 10 & 0.273 & 0.258 &  7.83 & 18.48 \\
 \isotope[94][36]{Kr}$_{58}$ & 28,  8 & 30,  4 & 0.293 & 0.288 & 13.10 &  7.37 \\
 \isotope[96][36]{Kr}$_{60}$ & 30,  8 & 32,  4 & 0.308 & 0.303 & 12.38 &  6.95 \\
 
 \isotope[98][36]{Kr}$_{62}$ & 34,  4 & 26, 10 & 0.318 & 0.285 &  6.59 & 16.34 \\
\isotope[100][36]{Kr}$_{64}$ & 30, 10 & 32,  6 & 0.315 & 0.309 & 14.63 &  9.43 \\
\isotope[102][36]{Kr}$_{66}$ & 28, 12 & 30,  8 & 0.309 & 0.302 & 17.60 & 12.38 \\
\isotope[104][36]{Kr}$_{68}$ & 28, 10 & 30,  6 & 0.295 & 0.289 & 15.44 &  9.97 \\
\isotope[106][36]{Kr}$_{70}$ & 30,  4 & 20, 18 & 0.277 & 0.284 &  7.37 & 28.35 \\
\isotope[108][36]{Kr}$_{72}$ & 22, 12 & 24,  8 & 0.257 & 0.248 & 20.89 & 14.80 \\


 \isotope[92][38]{Sr}$_{54}$ & 22,  6 & 18,  8 & 0.234 & 0.213 & 12.89 & 18.35 \\
 \isotope[94][38]{Sr}$_{56}$ & 28,  4 & 22, 10 & 0.271 & 0.256 &  7.83 & 18.48 \\
 \isotope[96][38]{Sr}$_{58}$ & 28,  8 & 30,  4 & 0.291 & 0.286 & 13.10 &  7.37 \\
 \isotope[98][38]{Sr}$_{60}$ & 30,  8 & 32,  4 & 0.306 & 0.301 & 12.38 &  6.95 \\
 
\isotope[100][38]{Sr}$_{62}$ & 34,  4 & 26, 10 & 0.315 & 0.283 &  6.59 & 16.34 \\
\isotope[102][38]{Sr}$_{64}$ & 30, 10 & 32,  6 & 0.313 & 0.307 & 14.63 &  9.43 \\
\isotope[104][38]{Sr}$_{66}$ & 28, 12 & 30,  8 & 0.307 & 0.300 & 17.60 & 12.38 \\
\isotope[106][38]{Sr}$_{68}$ & 28, 10 & 30,  6 & 0.293 & 0.287 & 15.44 &  9.97 \\
\isotope[108][38]{Sr}$_{70}$ & 30,  4 & 20, 18 & 0.275 & 0.282 &  7.37 & 28.35 \\
\isotope[110][38]{Sr}$_{72}$ & 22, 12 & 24,  8 & 0.256 & 0.247 & 20.89 & 14.80 \\
 
  \isotope[94][40]{Zr}$_{54}$ & 24,  2 & 12, 14 & 0.227 & 0.207 &  5.60 & 32.36 \\
  \isotope[96][40]{Zr}$_{56}$ & 30,  0 & 16, 16 & 0.267 & 0.249 &  1.57 & 30.00 \\
  \isotope[98][40]{Zr}$_{58}$ & 30,  4 & 24, 10 & 0.284 & 0.269 &  7.37 & 17.35 \\
 \isotope[100][40]{Zr}$_{60}$ & 32,  4 & 26, 10 & 0.299 & 0.283 &  6.95 & 16.34 \\
 
\isotope[102][40]{Zr}$_{62}$ & 36,  0 & 20, 20 & 0.311 & 0.302 &  1.32 & 30.00 \\
\isotope[104][40]{Zr}$_{64}$ & 32,  6 & 26, 12 & 0.305 & 0.291 &  9.43 & 18.58 \\
\isotope[106][40]{Zr}$_{66}$ & 30,  8 & 24, 14 & 0.298 & 0.287 & 12.38 & 21.79 \\
\isotope[108][40]{Zr}$_{68}$ & 30,  6 & 24, 12 & 0.285 & 0.272 &  9.97 & 19.67 \\
\isotope[110][40]{Zr}$_{70}$ & 32,  0 & 14, 24 & 0.271 & 0.283 &  1.48 & 38.21 \\
\isotope[112][40]{Zr}$_{72}$ & 24,  8 & 18, 14 & 0.246 & 0.238 & 14.80 & 26.11 \\
 
\hline
\end{tabular}

\end{table}

\setcounter{table}{4}

\begin{table}

\caption{(continued) 
}
\begin{tabular}{ r   r  r r c c r  }
\hline
nucleus  & hw & nhw & $\beta_{hw}$ & $\beta_{nhw}$ & $\gamma_{hw}$ & $\gamma_{nhw}$  \\ 
        
\hline

  \isotope[96][42]{Mo}$_{54}$ & 18,  8 & 14, 10 & 0.210 & 0.192 & 18.35 & 24.92 \\
  \isotope[98][42]{Mo}$_{56}$ & 24,  6 & 18, 12 & 0.245 & 0.234 & 12.01 & 23.82 \\
 \isotope[100][42]{Mo}$_{58}$ & 24, 10 & 26,  6 & 0.267 & 0.260 & 17.35 & 11.24 \\
 \isotope[102][42]{Mo}$_{60}$ & 26, 10 & 28,  6 & 0.281 & 0.274 & 16.34 & 10.57 \\

\isotope[104][42]{Mo}$_{62}$ & 30,  6 & 22, 16 & 0.289 & 0.287 &  9.97 & 25.05 \\
\isotope[106][42]{Mo}$_{64}$ & 26, 12 & 28,  8 & 0.290 & 0.282 & 18.58 & 13.10 \\
\isotope[108][42]{Mo}$_{66}$ & 24, 14 & 26, 10 & 0.285 & 0.276 & 21.79 & 16.34 \\
\isotope[110][42]{Mo}$_{68}$ & 24, 12 & 26,  8 & 0.271 & 0.263 & 19.67 & 13.90 \\
\isotope[112][42]{Mo}$_{70}$ & 26,  6 & 16, 20 & 0.250 & 0.265 & 11.24 & 33.48 \\
\isotope[114][42]{Mo}$_{72}$ & 18, 14 & 20, 10 & 0.236 & 0.225 & 26.11 & 19.77 \\ 
 
  \isotope[98][44]{Ru}$_{54}$ & 14, 10 & 10, 12 & 0.190 & 0.175 & 24.92 & 32.75 \\
 \isotope[100][44]{Ru}$_{56}$ & 20,  8 & 14, 14 & 0.223 & 0.212 & 17.00 & 30.00 \\
 \isotope[102][44]{Ru}$_{58}$ & 20, 12 & 22,  8 & 0.247 & 0.229 & 22.26 & 15.82 \\
 \isotope[104][44]{Ru}$_{60}$ & 22, 12 & 24,  8 & 0.260 & 0.243 & 20.89 & 14.80 \\
 
\isotope[106][44]{Ru}$_{62}$ & 26,  8 & 18, 18 & 0.266 & 0.270 & 13.90 & 30.00 \\
\isotope[108][44]{Ru}$_{64}$ & 22, 14 & 24, 10 & 0.270 & 0.260 & 23.07 & 17.35 \\
\isotope[110][44]{Ru}$_{66}$ & 20, 16 & 22, 12 & 0.267 & 0.256 & 26.52 & 20.89 \\
\isotope[112][44]{Ru}$_{68}$ & 20, 14 & 22, 10 & 0.252 & 0.242 & 24.50 & 18.48 \\
\isotope[114][44]{Ru}$_{70}$ & 22,  8 & 12, 22 & 0.229 & 0.253 & 15.82 & 39.11 \\
\isotope[116][44]{Ru}$_{72}$ & 14, 16 & 16, 12 & 0.221 & 0.208 & 32.07 & 25.13 \\ 
 
 \isotope[100][46]{Pd}$_{54}$ & 12,  8 &  8, 10 & 0.160 & 0.145 & 24.01 & 33.30 \\
 \isotope[102][46]{Pd}$_{56}$ & 18,  6 & 12, 12 & 0.193 & 0.186 & 15.08 & 30.00 \\
 \isotope[104][46]{Pd}$_{58}$ & 18, 10 & 20,  6 & 0.217 & 0.208 & 21.25 & 13.90 \\
 \isotope[106][46]{Pd}$_{60}$ & 20, 10 & 22,  6 & 0.231 & 0.230 & 19.77 & 12.89 \\

\isotope[108][46]{Pd}$_{62}$ & 24,  6 & 16, 16 & 0.237 & 0.239 & 12.01 & 30.00 \\
\isotope[110][46]{Pd}$_{64}$ & 20, 12 & 22,  8 & 0.241 & 0.232 & 22.26 & 15.82 \\
\isotope[112][46]{Pd}$_{66}$ & 18, 14 & 20, 10 & 0.237 & 0.226 & 26.11 & 19.77 \\
\isotope[114][46]{Pd}$_{68}$ & 18, 12 & 20,  8 & 0.223 & 0.213 & 23.82 & 17.00 \\
\isotope[116][46]{Pd}$_{70}$ & 20,  6 & 10, 20 & 0.201 & 0.224 & 13.90 & 40.23 \\
\isotope[118][46]{Pd}$_{72}$ & 12, 14 & 14, 10 & 0.192 & 0.179 & 32.36 & 24.92 \\ 

\hline
\end{tabular}

\end{table}


\begin{table}

\caption{Highest weight (hw) irreducible representations of SU(3) and next highest weight (nhw) irreps of SU(3) within the proxy-SU(3) scheme for nuclei in the $Z=36$-46, $N=36$-46 region. The Elliott \cite{Elliott1958a} notation $(\lambda,\mu)$ is used for the SU(3) irreps. Parameter-independent predictions for the collective variables $\beta$ and $\gamma$ within each irrep, calculated from Eqs. (\ref{b1}) and (\ref{g1}) are also given, labeled by hw and nhw respectively. See Section \ref{nhw} for further discussion.
}
\begin{tabular}{ r   r  r r c c r  }
\hline
nucleus  & hw & nhw & $\beta_{hw}$ & $\beta_{nhw}$ & $\gamma_{hw}$ & $\gamma_{nhw}$  \\ 
        
\hline

\isotope[72][36]{Kr}$_{36}$ & 20,  8 & 22,  4 & 0.305 & 0.296 & 17.00 &  9.64 \\
\isotope[74][36]{Kr}$_{38}$ & 20,  8 & 22,  4 & 0.302 & 0.293 & 17.00 &  9.64 \\
\isotope[76][36]{Kr}$_{40}$ & 22,  4 & 14, 14 & 0.291 & 0.292 &  9.64 & 30.00 \\
\isotope[78][36]{Kr}$_{42}$ & 16, 10 & 18,  6 & 0.272 & 0.259 & 22.95 & 15.08 \\ 
\isotope[80][36]{Kr}$_{44}$ & 12, 12 & 14,  8 & 0.249 & 0.232 & 30.00 & 21.79 \\
\isotope[82][36]{Kr}$_{46}$ & 10, 10 & 12,  6 & 0.209 & 0.193 & 30.00 & 20.17 \\


\isotope[74][38]{Sr}$_{36}$ & 20,  8 & 22,  4 & 0.302 & 0.293 & 17.00 &  9.64 \\
\isotope[76][38]{Sr}$_{38}$ & 20,  8 & 22,  4 & 0.300 & 0.291 & 17.00 &  9.64 \\
\isotope[78][38]{Sr}$_{40}$ & 22,  4 & 14, 14 & 0.288 & 0.289 &  9.64 & 30.00 \\
\isotope[80][38]{Sr}$_{42}$ & 16, 10 & 18,  6 & 0.270 & 0.257 & 22.95 & 15.08 \\ 
\isotope[82][38]{Sr}$_{44}$ & 12, 12 & 14,  8 & 0.247 & 0.230 & 30.00 & 21.79 \\
\isotope[84][36]{Sr}$_{46}$ & 10, 10 & 12,  6 & 0.207 & 0.191 & 30.00 & 20.17 \\

\isotope[76][40]{Zr}$_{36}$ & 22,  4 & 16, 10 & 0.291 & 0.275 &  9.64 & 22.95 \\
\isotope[78][40]{Zr}$_{38}$ & 22,  4 & 16, 10 & 0.288 & 0.272 &  9.64 & 22.95 \\
\isotope[80][40]{Zr}$_{40}$ & 24,  0 &  8, 20 & 0.282 & 0.295 &  1.95 & 43.00 \\
\isotope[82][40]{Zr}$_{42}$ & 18,  6 & 18, 12 & 0.255 & 0.246 & 15.08 & 30.00 \\ 
\isotope[84][40]{Zr}$_{44}$ & 14,  8 & 14, 14 & 0.228 & 0.228 & 21.79 & 38.21 \\
\isotope[86][40]{Zr}$_{46}$ & 12,  6 & 12, 12 & 0.190 & 0.190 & 20.17 & 39.83 \\

\isotope[78][42]{Mo}$_{36}$ & 16, 10 & 18,  6 & 0.272 & 0.259 & 22.95 & 15.08 \\
\isotope[80][42]{Mo}$_{38}$ & 16, 10 & 18,  6 & 0.270 & 0.257 & 22.95 & 15.08 \\
\isotope[82][42]{Mo}$_{40}$ & 18,  6 & 10, 16 & 0.255 & 0.267 & 15.08 & 37.05 \\
\isotope[84][42]{Mo}$_{42}$ & 12, 12 & 14,  8 & 0.245 & 0.228 & 30.00 & 21.79 \\ 
\isotope[86][42]{Mo}$_{44}$ &  8, 14 & 10, 10 & 0.226 & 0.205 & 38.21 & 30.00 \\
\isotope[88][42]{Mo}$_{46}$ &  6, 12 &  8,  8 & 0.188 & 0.167 & 39.83 & 30.00 \\

\isotope[80][44]{Ru}$_{36}$ & 12, 12 & 14,  8 & 0.249 & 0.232 & 30.00 & 21.79 \\
\isotope[82][44]{Ru}$_{38}$ & 12, 12 & 14,  8 & 0.247 & 0.230 & 30.00 & 21.79 \\
\isotope[84][44]{Ru}$_{40}$ & 14,  8 &  6, 18 & 0.228 & 0.253 & 21.79 & 44.92 \\
\isotope[86][44]{Ru}$_{42}$ &  8, 14 & 10, 10 & 0.226 & 0.205 & 38.21 & 30.00 \\ 
\isotope[88][44]{Ru}$_{44}$ &  4, 16 &  6, 12 & 0.214 & 0.188 & 47.48 & 39.83 \\
\isotope[90][44]{Ru}$_{46}$ &  2, 14 &  4, 10 & 0.177 & 0.150 & 51.05 & 42.22 \\

\isotope[82][46]{Pd}$_{36}$ & 10, 10 & 12,  6 & 0.209 & 0.193 & 30.00 & 20.17 \\
\isotope[84][46]{Pd}$_{38}$ & 10, 10 & 12,  8 & 0.207 & 0.191 & 30.00 & 20.17 \\
\isotope[86][46]{Pd}$_{40}$ & 12,  6 &  4, 16 & 0.190 & 0.216 & 20.17 & 47.48 \\
\isotope[88][46]{Pd}$_{42}$ &  6, 12 &  8,  8 & 0.188 & 0.167 & 39.83 & 30.00 \\ 
\isotope[90][46]{Pd}$_{44}$ &  2, 14 &  4, 10 & 0.177 & 0.150 & 51.05 & 42.22 \\
\isotope[92][46]{Pd}$_{46}$ &  0, 12 &  2,  8 & 0.143 & 0.114 & 56.33 & 46.10 \\

\hline
\end{tabular}

\end{table}


\begin{figure*} [htb]

    \includegraphics[width=150mm]{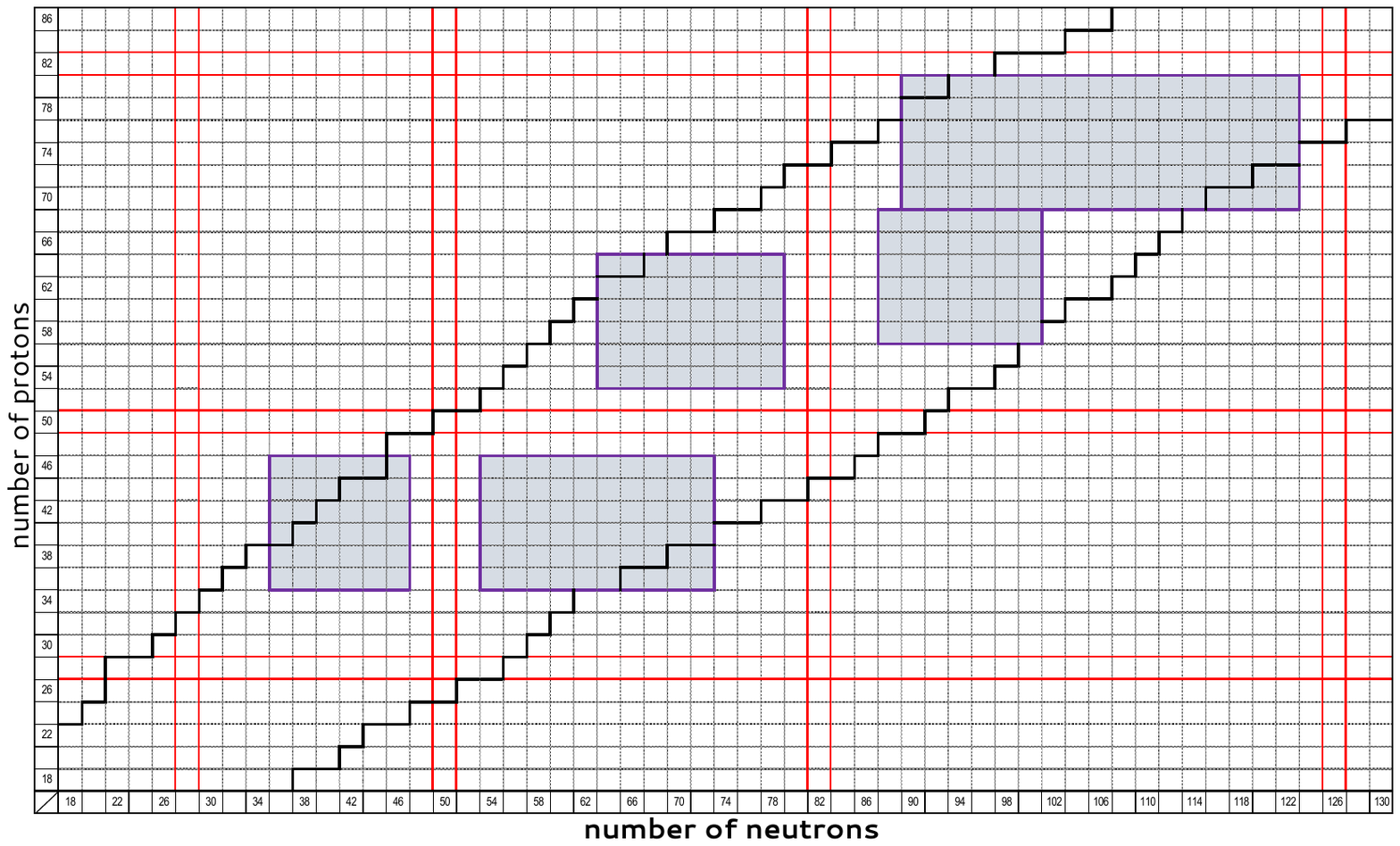}

    \caption{Regions of nuclei included in the present study. } 
    
\end{figure*}


\begin{figure*} [htb]

    {\includegraphics[width=75mm]{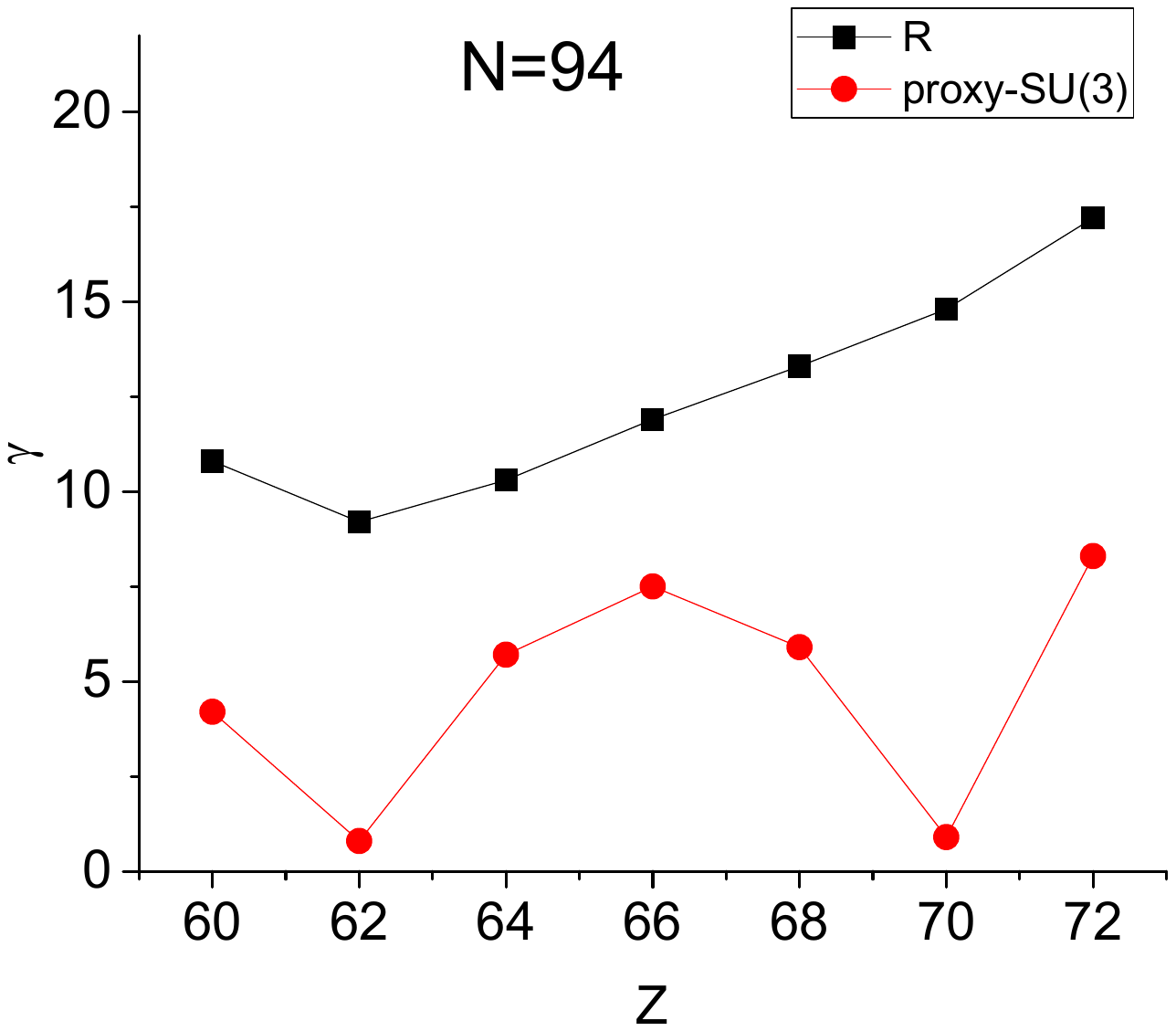} \hspace{5mm}   \includegraphics[width=75mm]{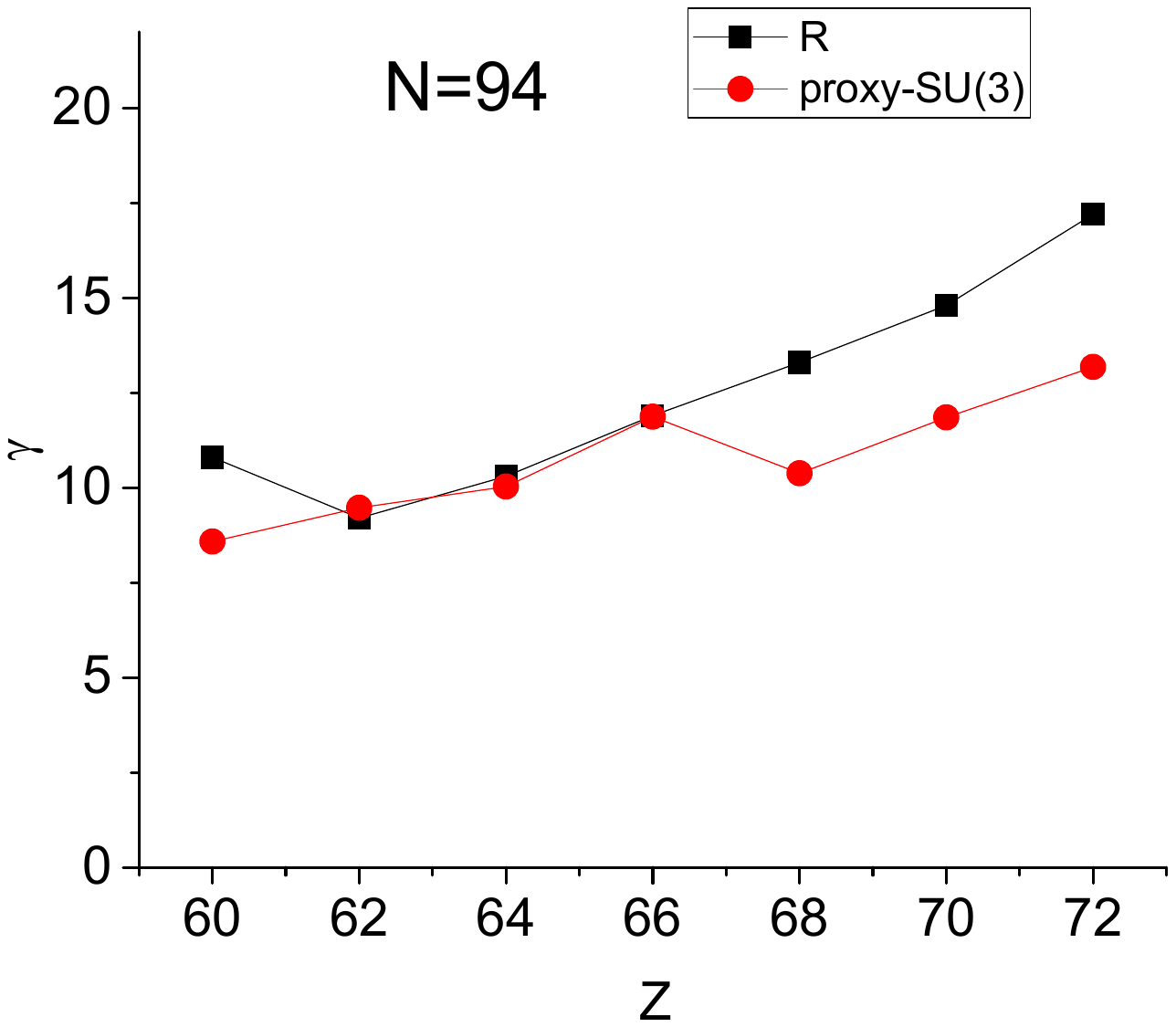} }
    {\includegraphics[width=75mm]{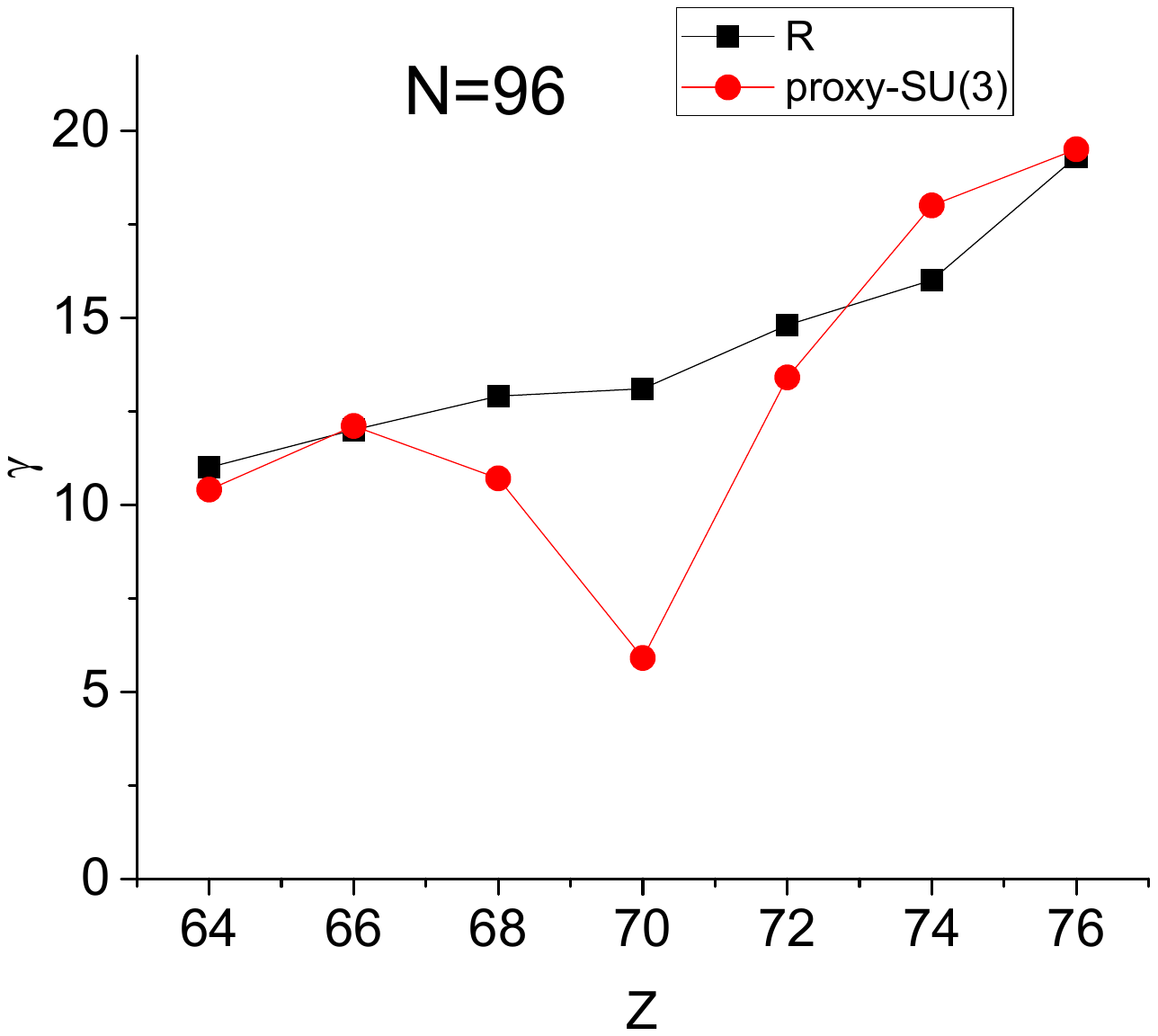} \hspace{5mm}    \includegraphics[width=75mm]{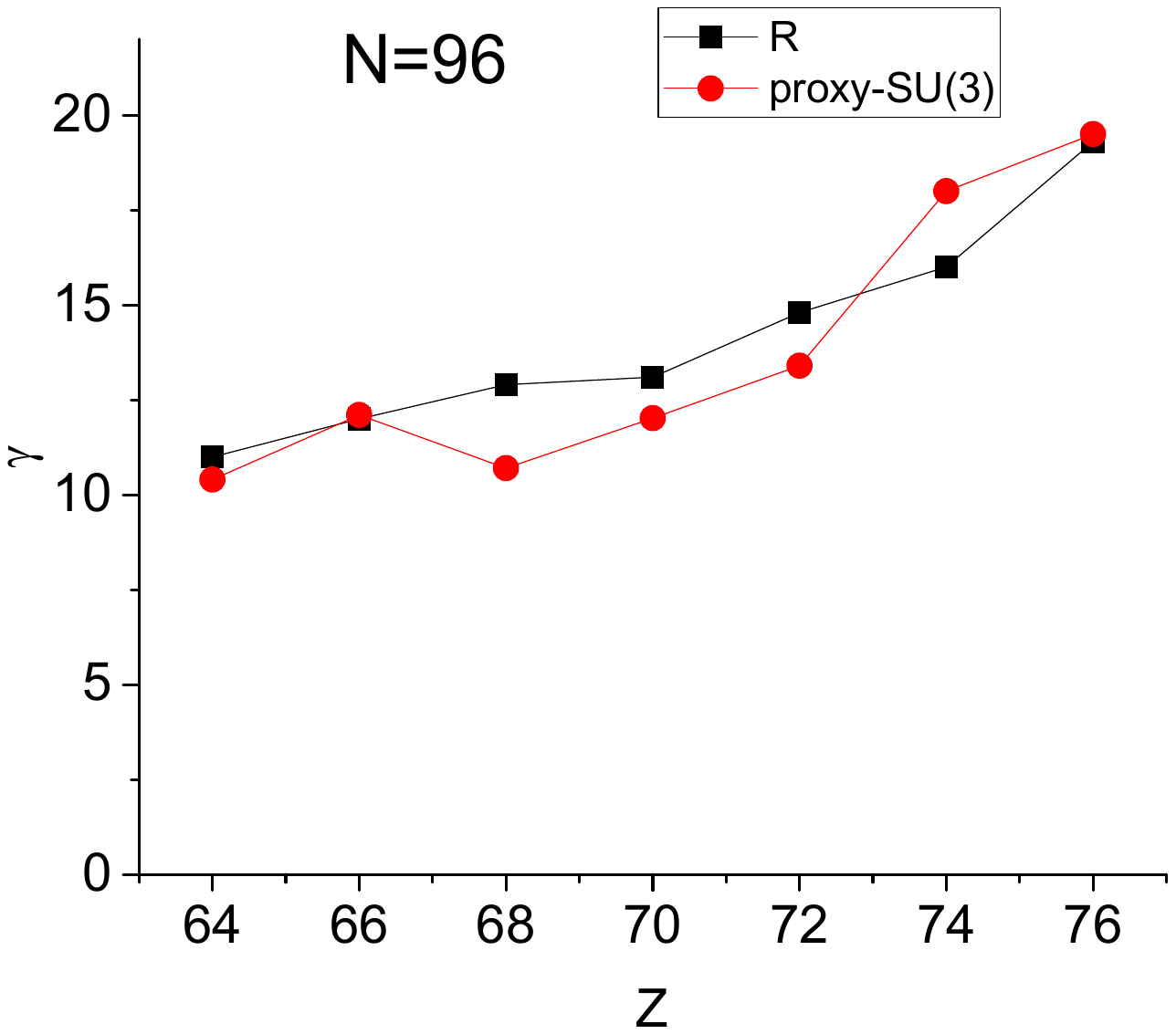} }
    {\includegraphics[width=75mm]{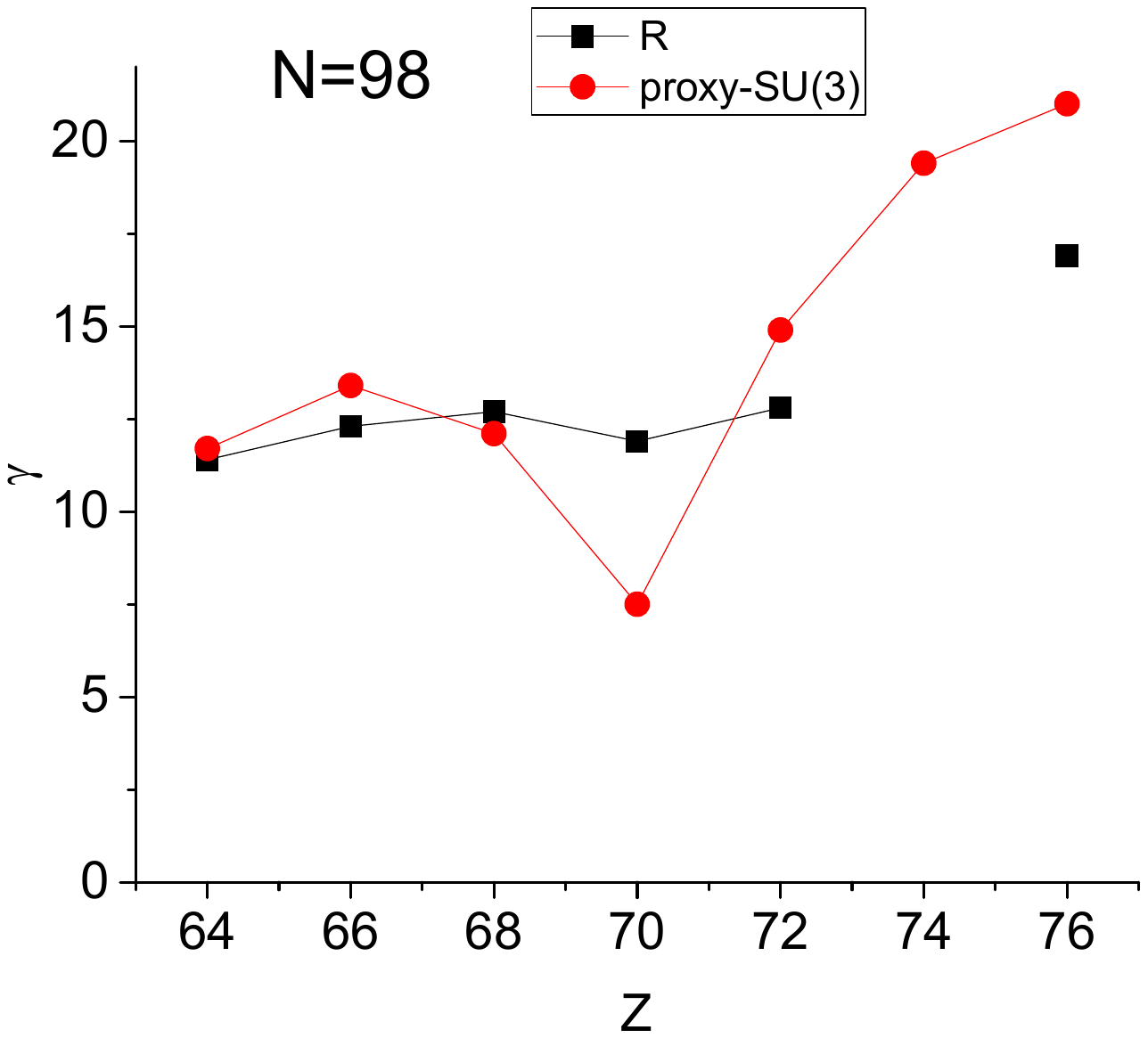}  \hspace{5mm}   \includegraphics[width=75mm]{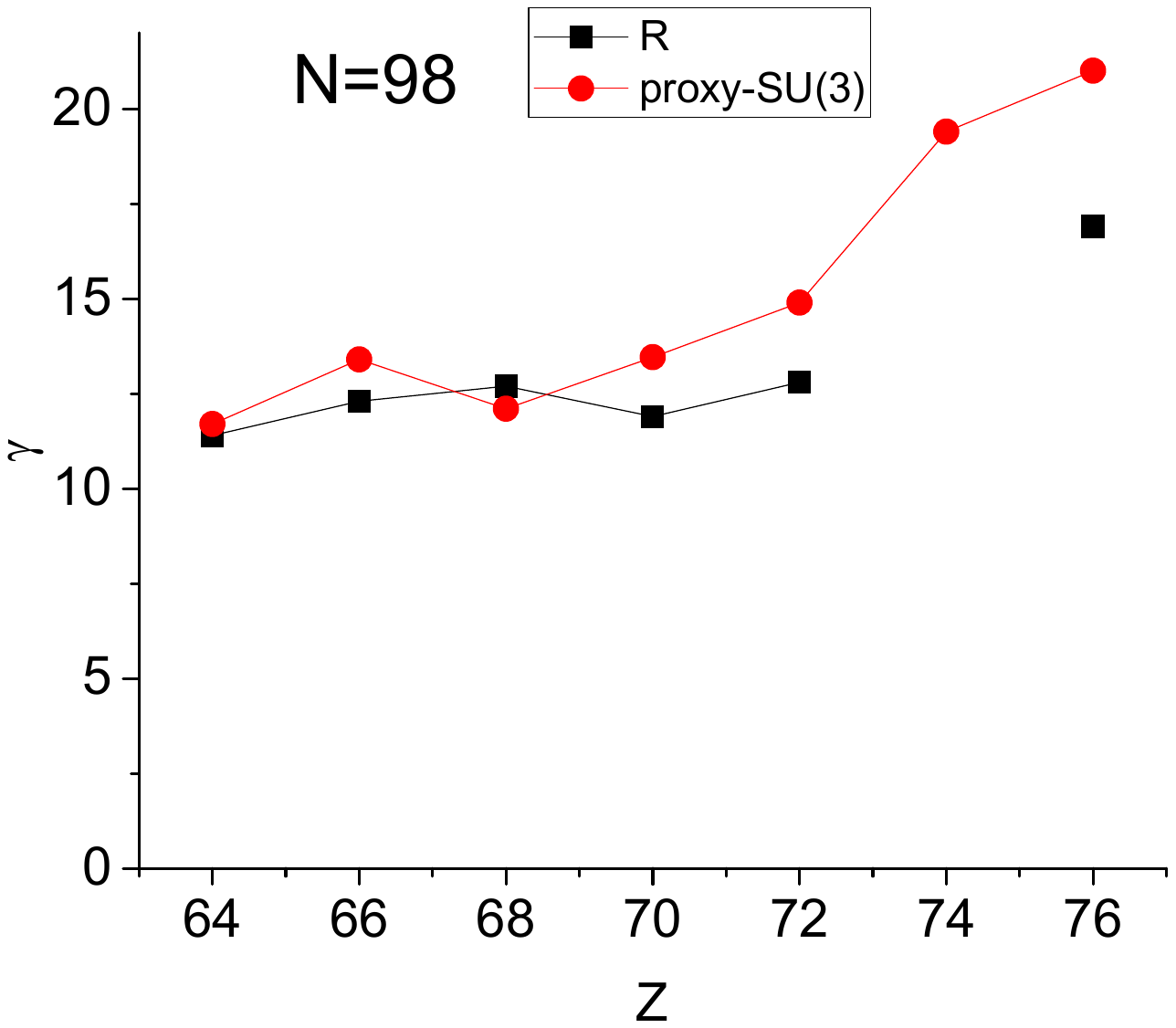} }
     
    \caption{The empirical values of the collective variable $\gamma$, obtained from Eq. (\ref{R}) and labeled by $R$, are compared a) (left column) to the parameter-independent predictions  provided by the proxy-SU(3) scheme using the hw irreps, and b) (right column) to the parameter-independent predictions  provided by the proxy-SU(3) scheme when taking into account also the nhw irreps in cases in which fully symmetric irreps (with $\mu =0$) occur. The proxy-SU(3) predictions are taken from Tables II and III.  See Secs. \ref{bg} and \ref{mixing} for further discussion.} 
    
\end{figure*}


\begin{figure*} [htb]

    {\includegraphics[width=75mm]{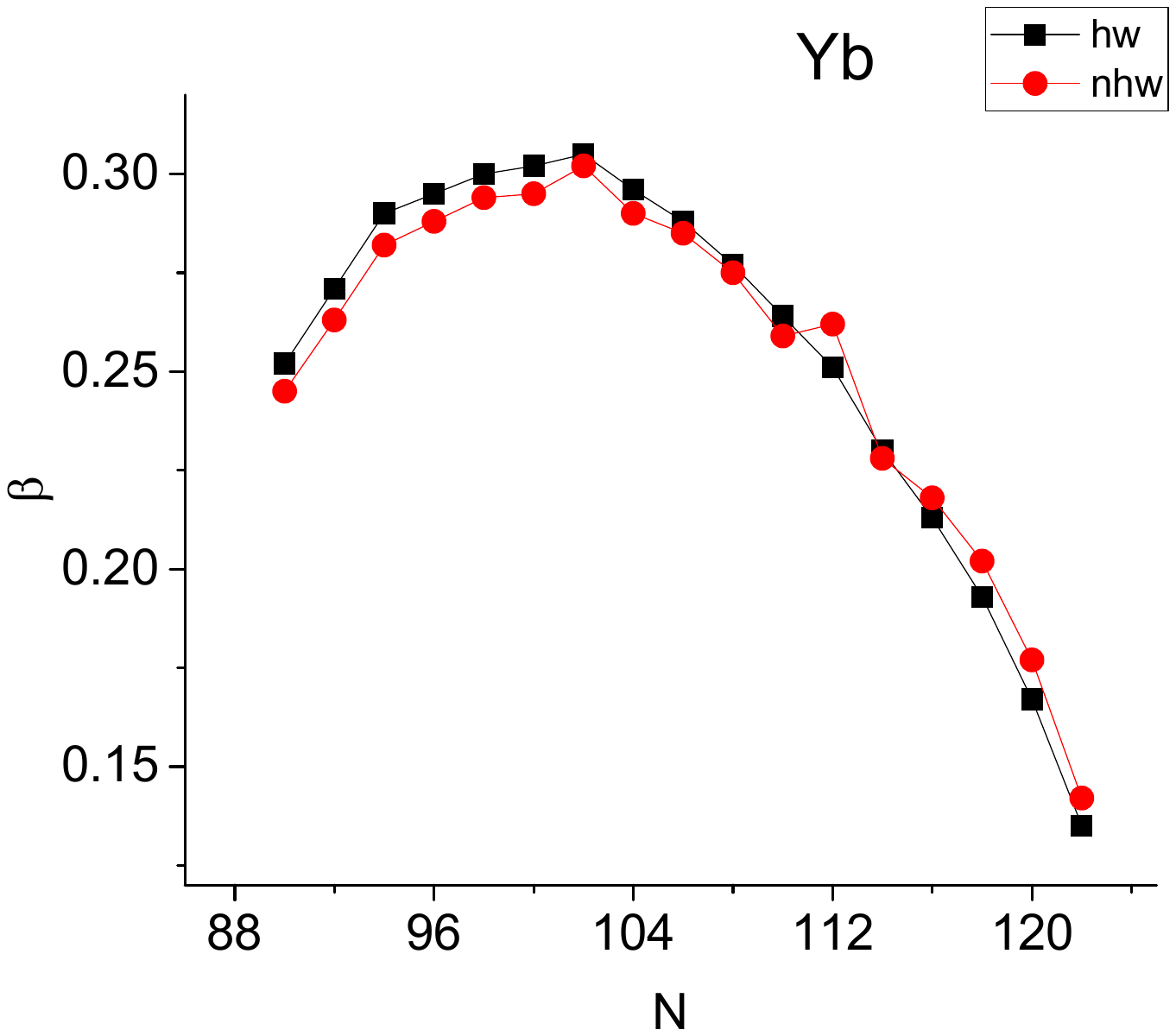} \hspace{5mm}   \includegraphics[width=75mm]{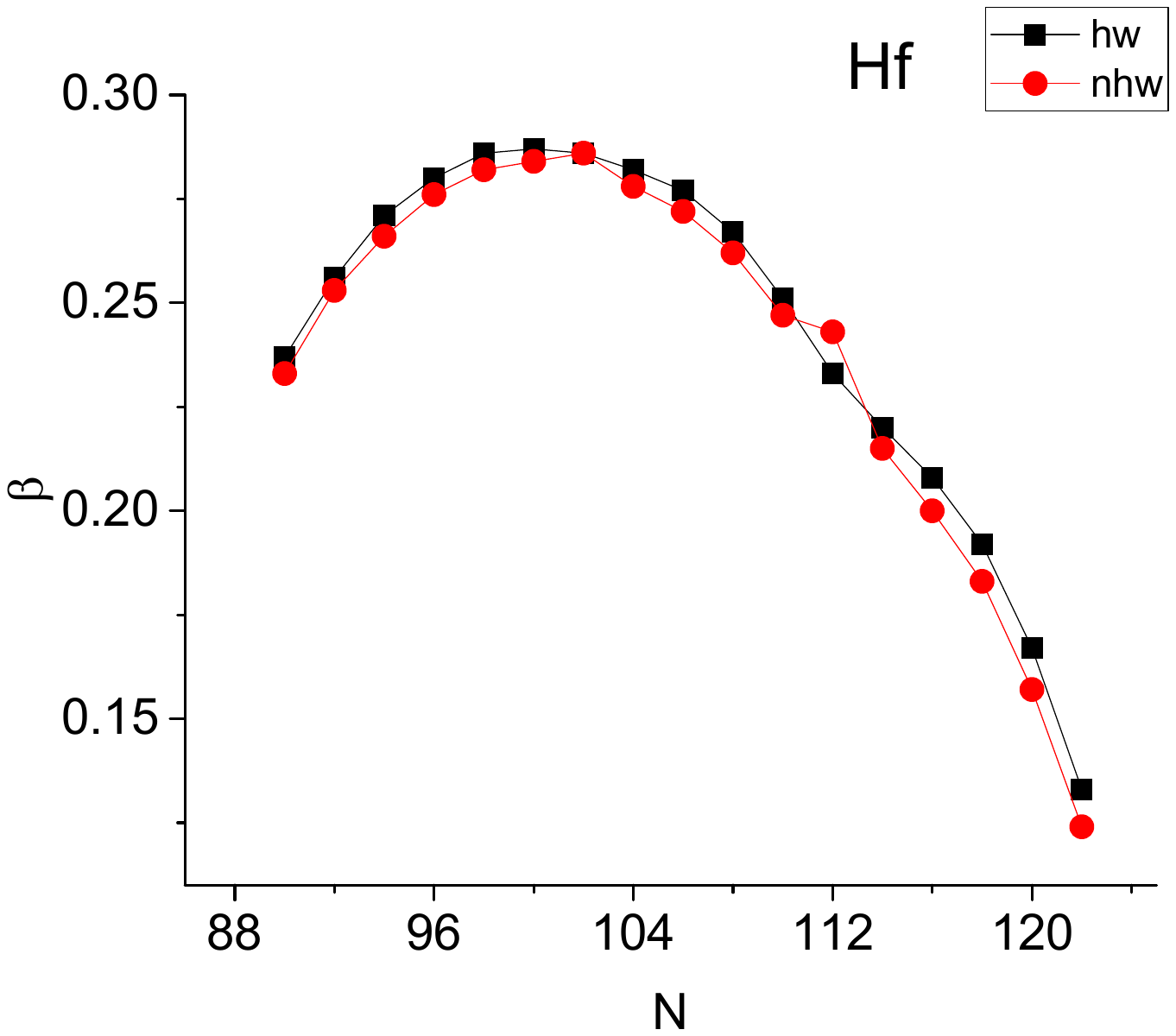} }
    {\includegraphics[width=75mm]{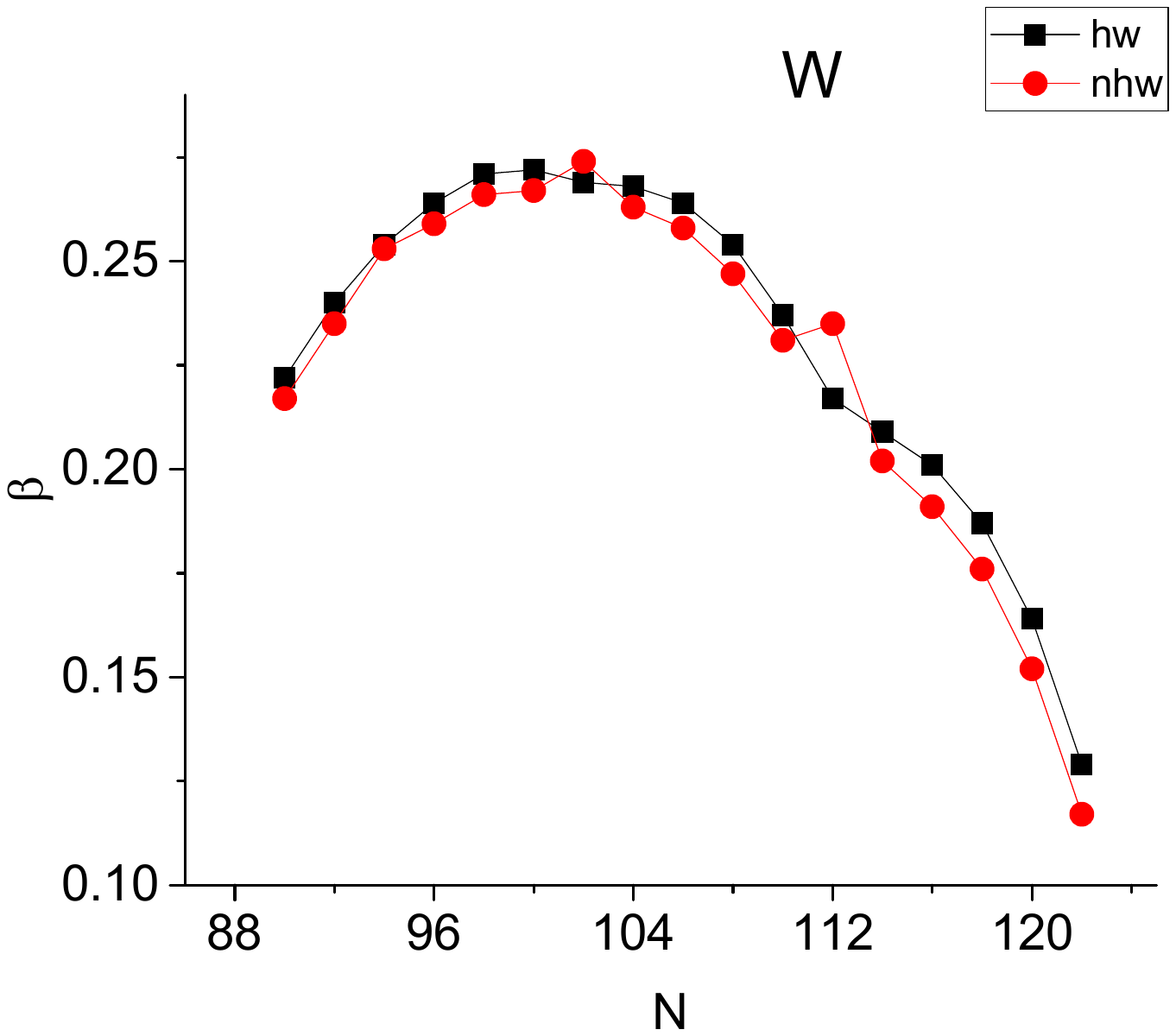} \hspace{5mm}    \includegraphics[width=75mm]{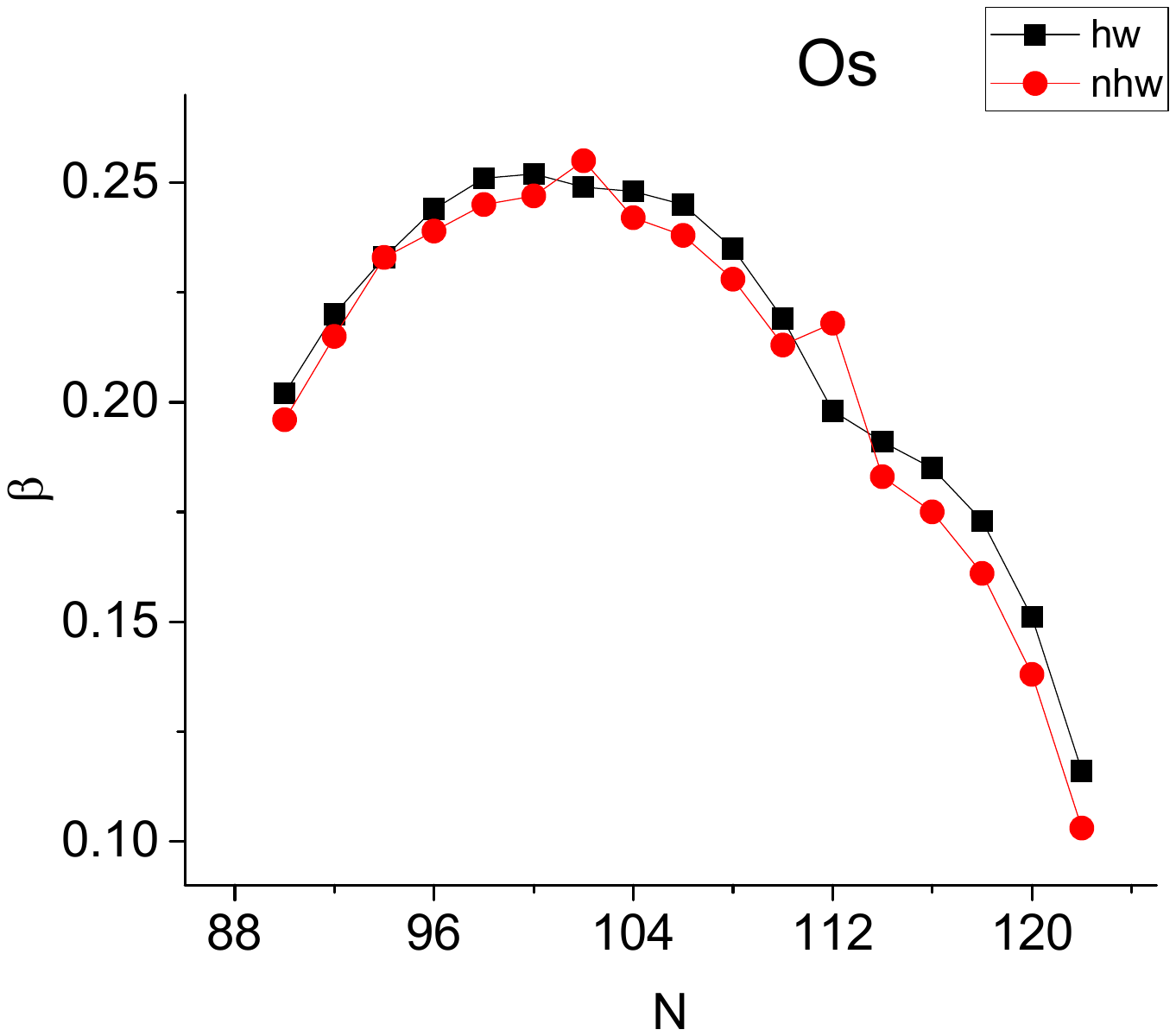} }
    {\includegraphics[width=75mm]{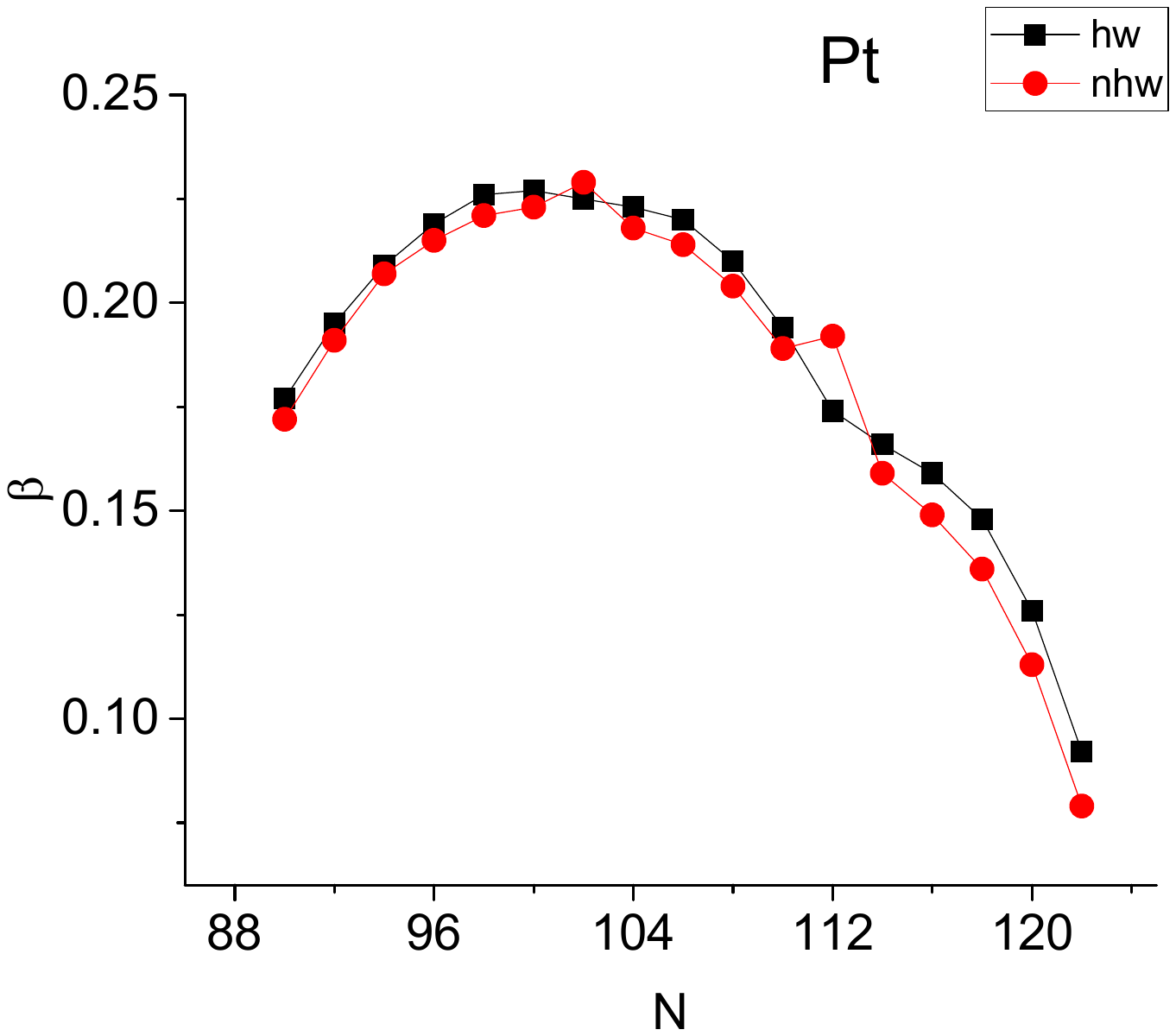}  \hspace{5mm}   \includegraphics[width=75mm]{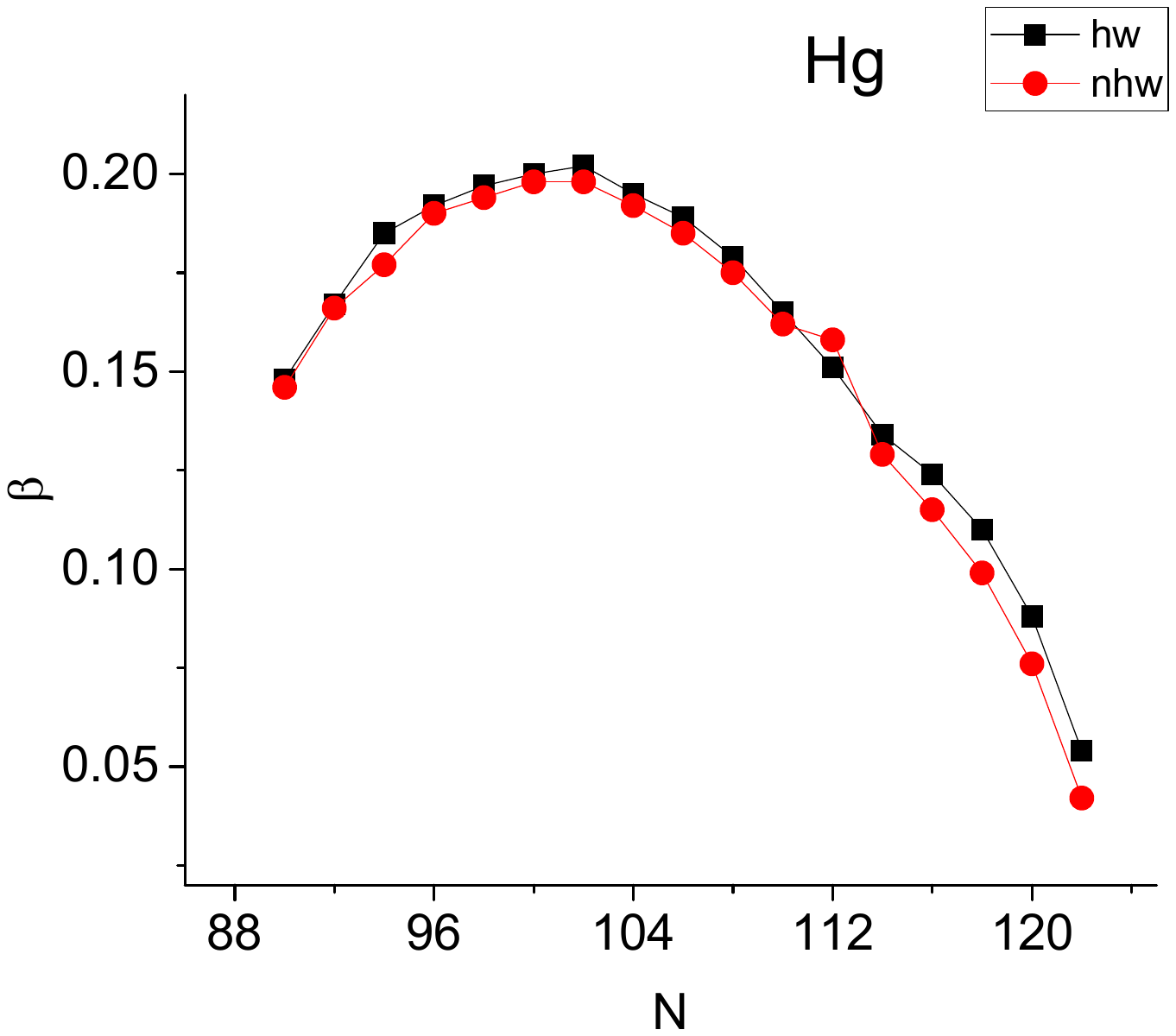} }
     
    \caption{Parameter-independent predictions for the collective variable $\beta$ provided within the proxy-SU(3) symmetry by the highest weight irreducible representation (hw irrep) of SU(3) and by the next highest weight (nhw) irrep of SU(3) in the $Z=70$-80, $N=90$-122 region, taken from Table II, are plotted vs. $N$.  See Sec. \ref{nhw} for further discussion.} 
    
\end{figure*}


\begin{figure*} [htb]

    {\includegraphics[width=75mm]{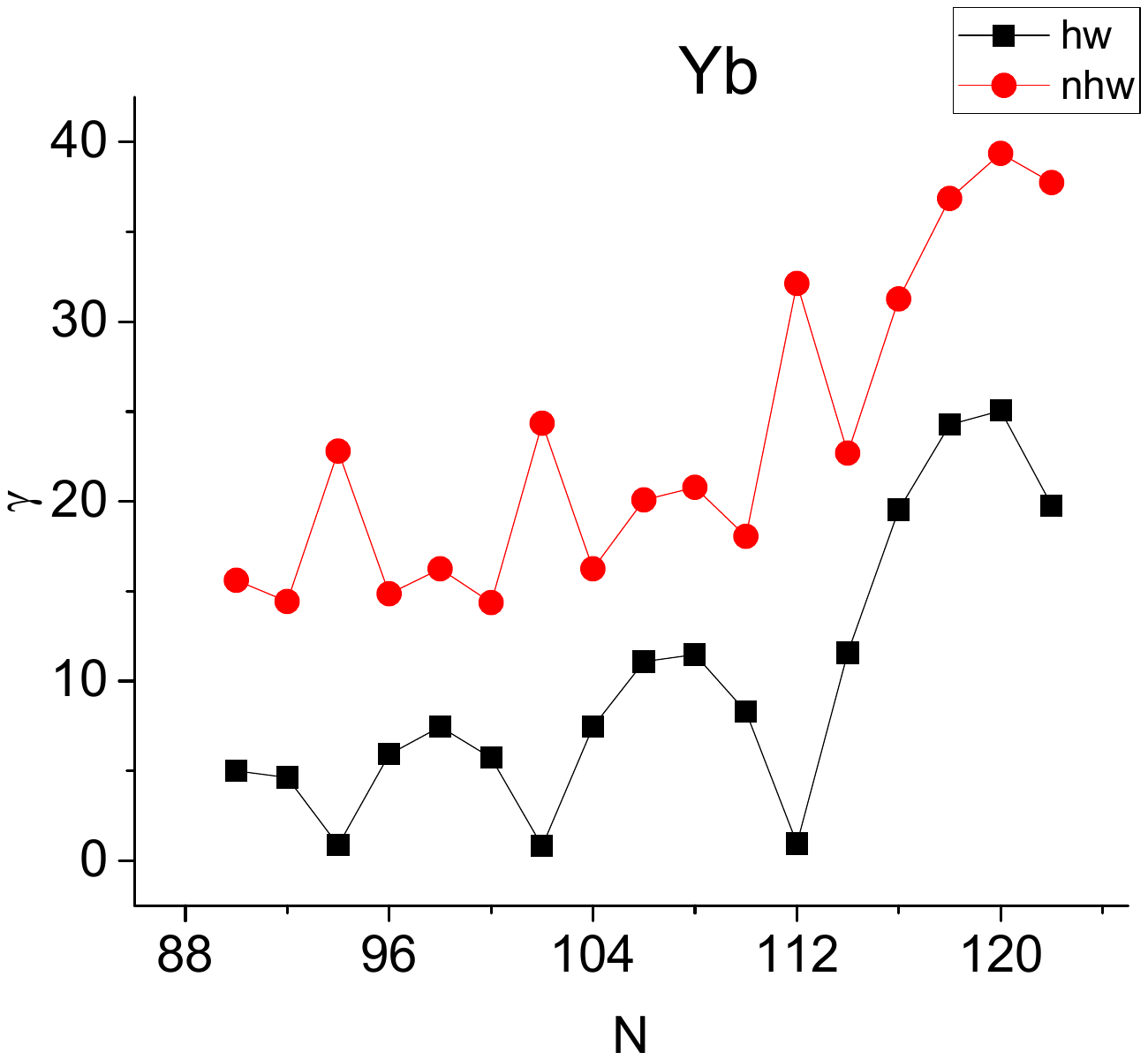} \hspace{5mm}   \includegraphics[width=75mm]{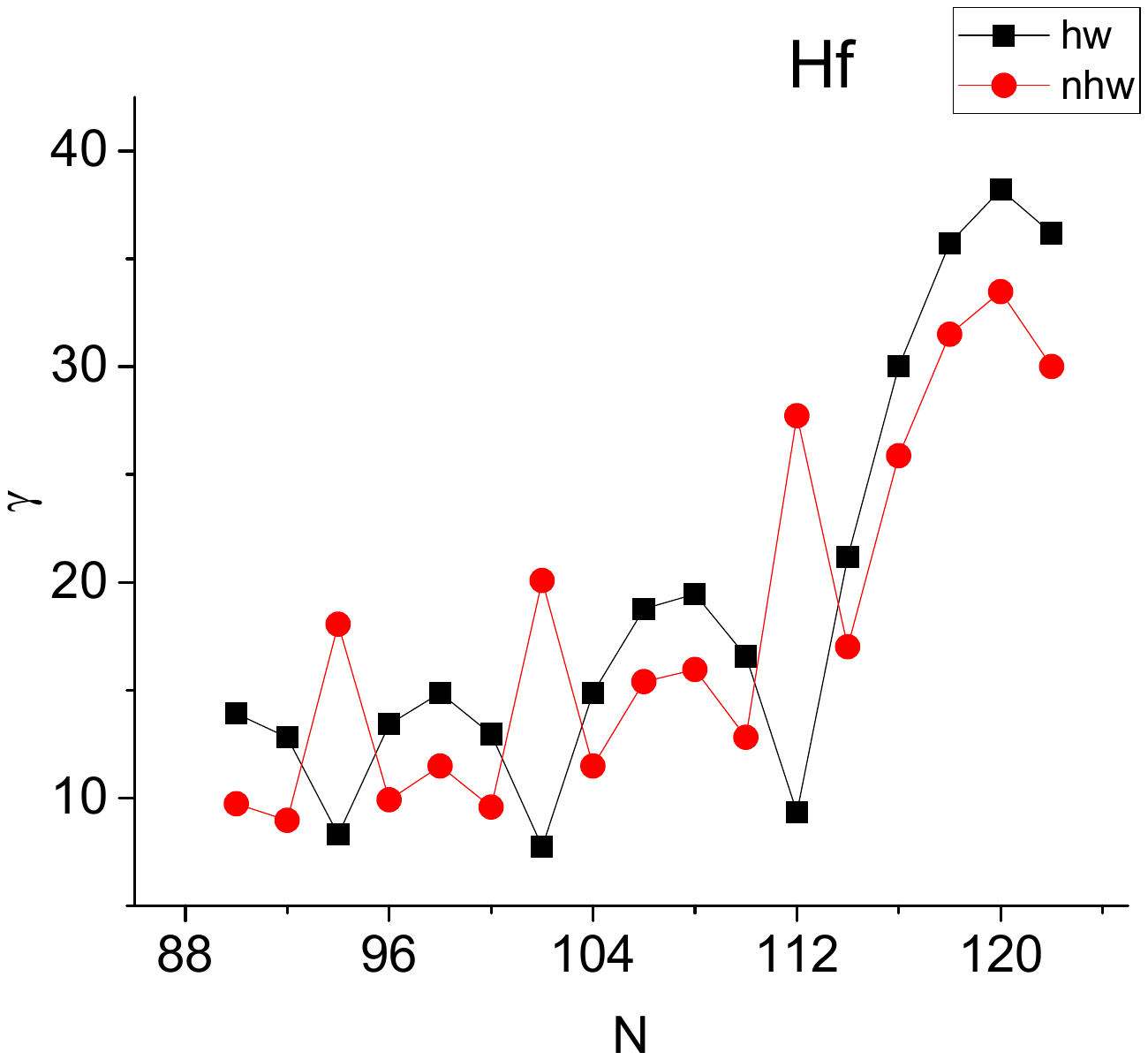} }
    {\includegraphics[width=75mm]{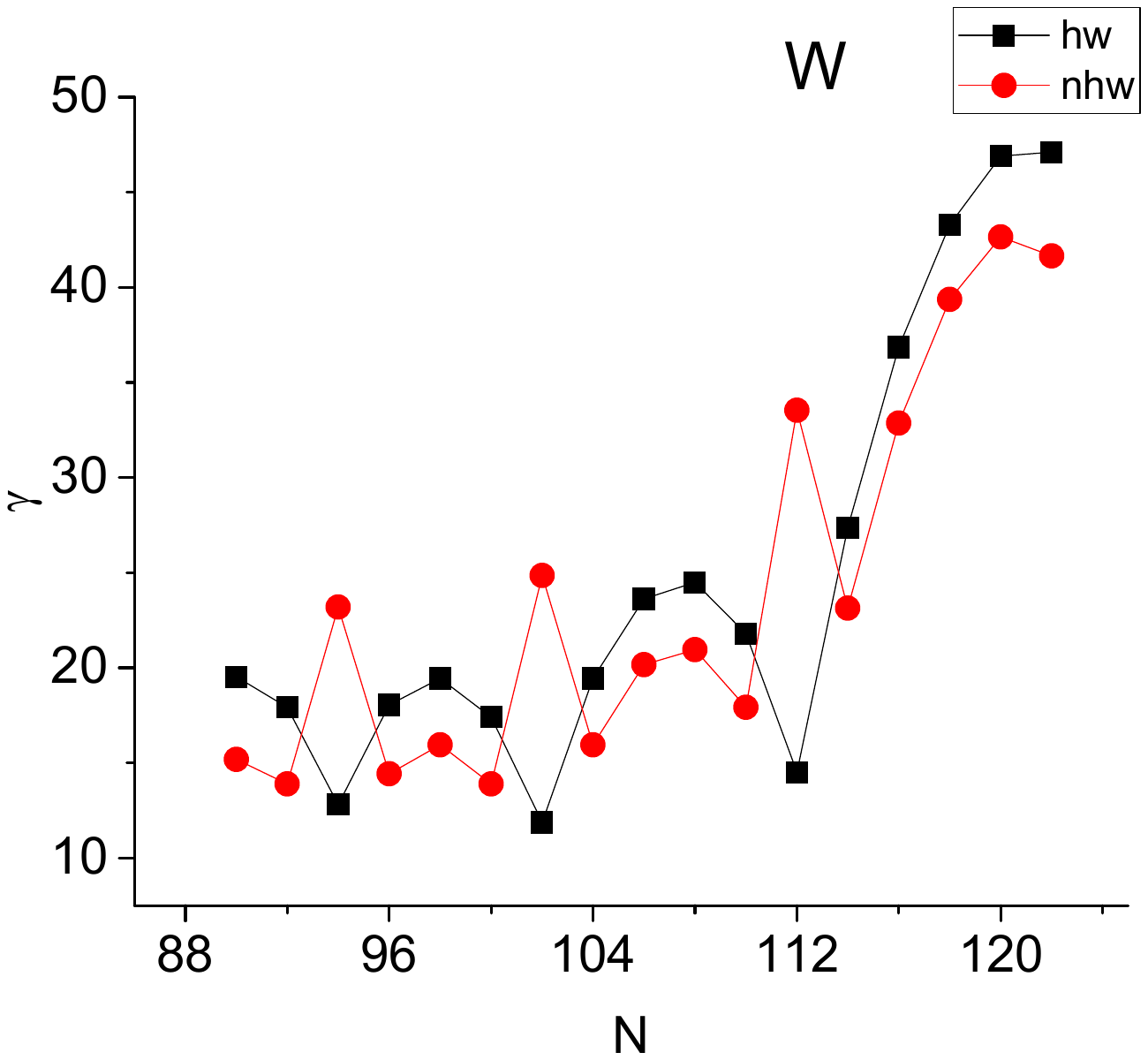} \hspace{5mm}    \includegraphics[width=75mm]{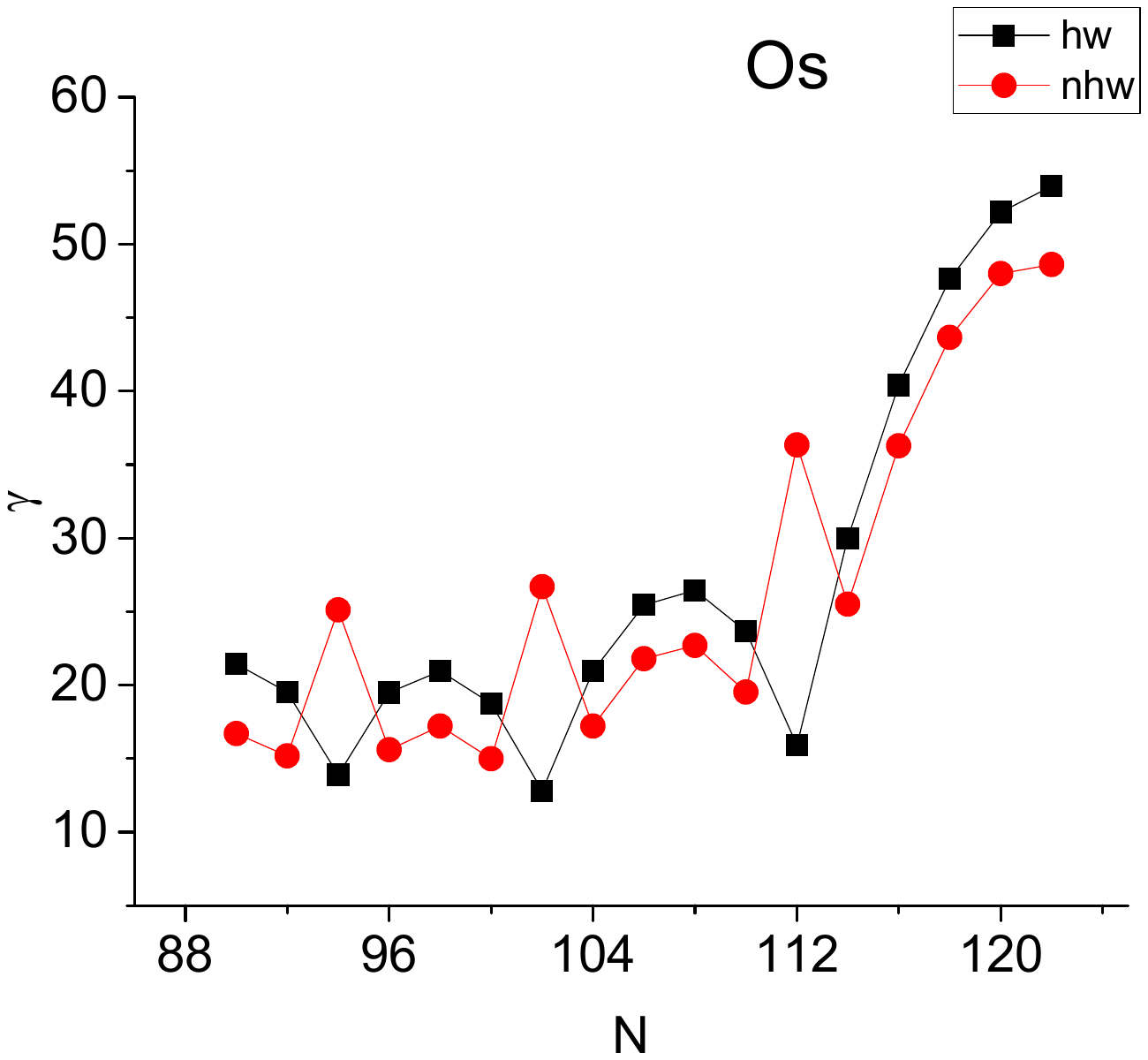} }
    {\includegraphics[width=75mm]{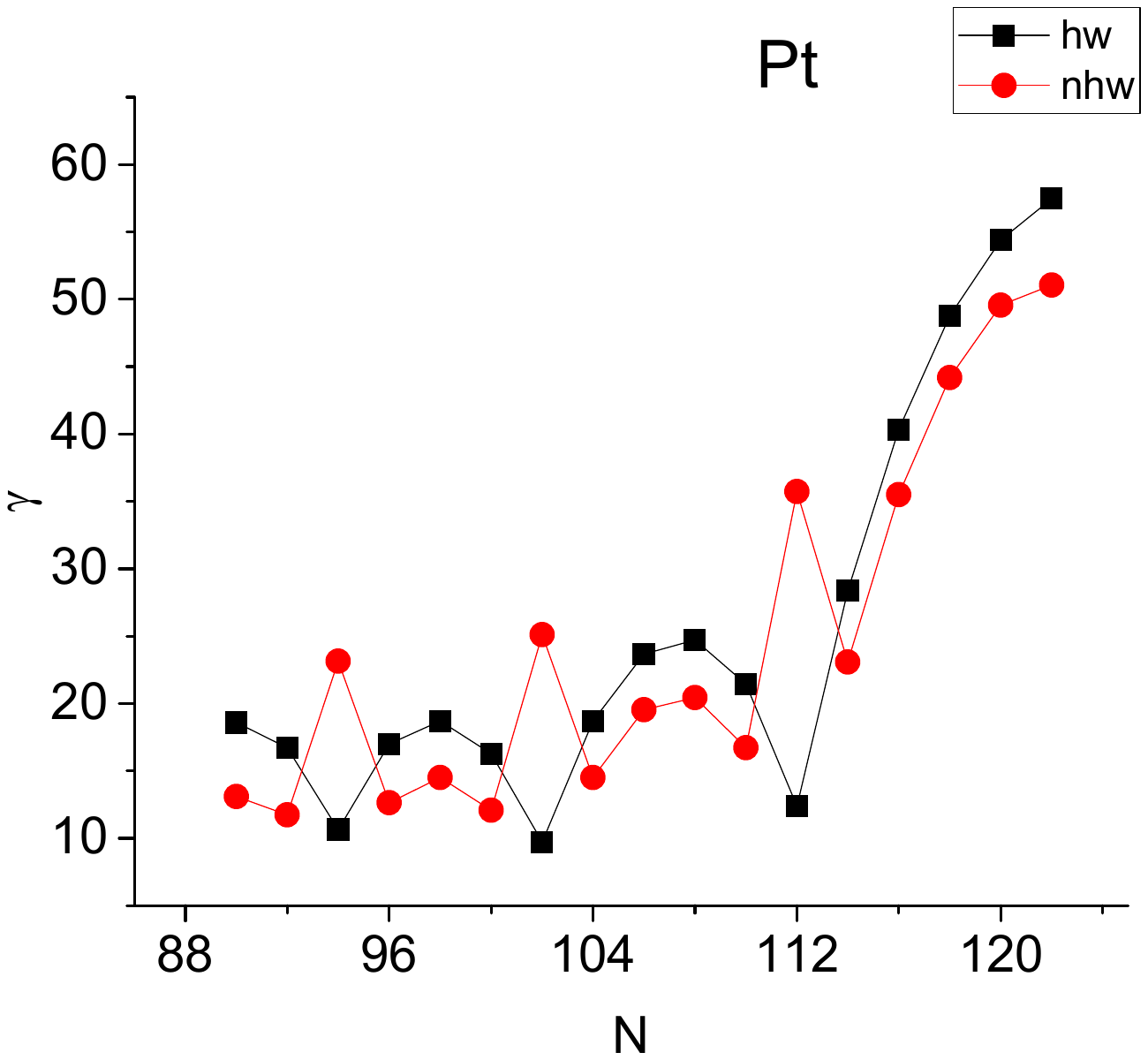}  \hspace{5mm}   \includegraphics[width=75mm]{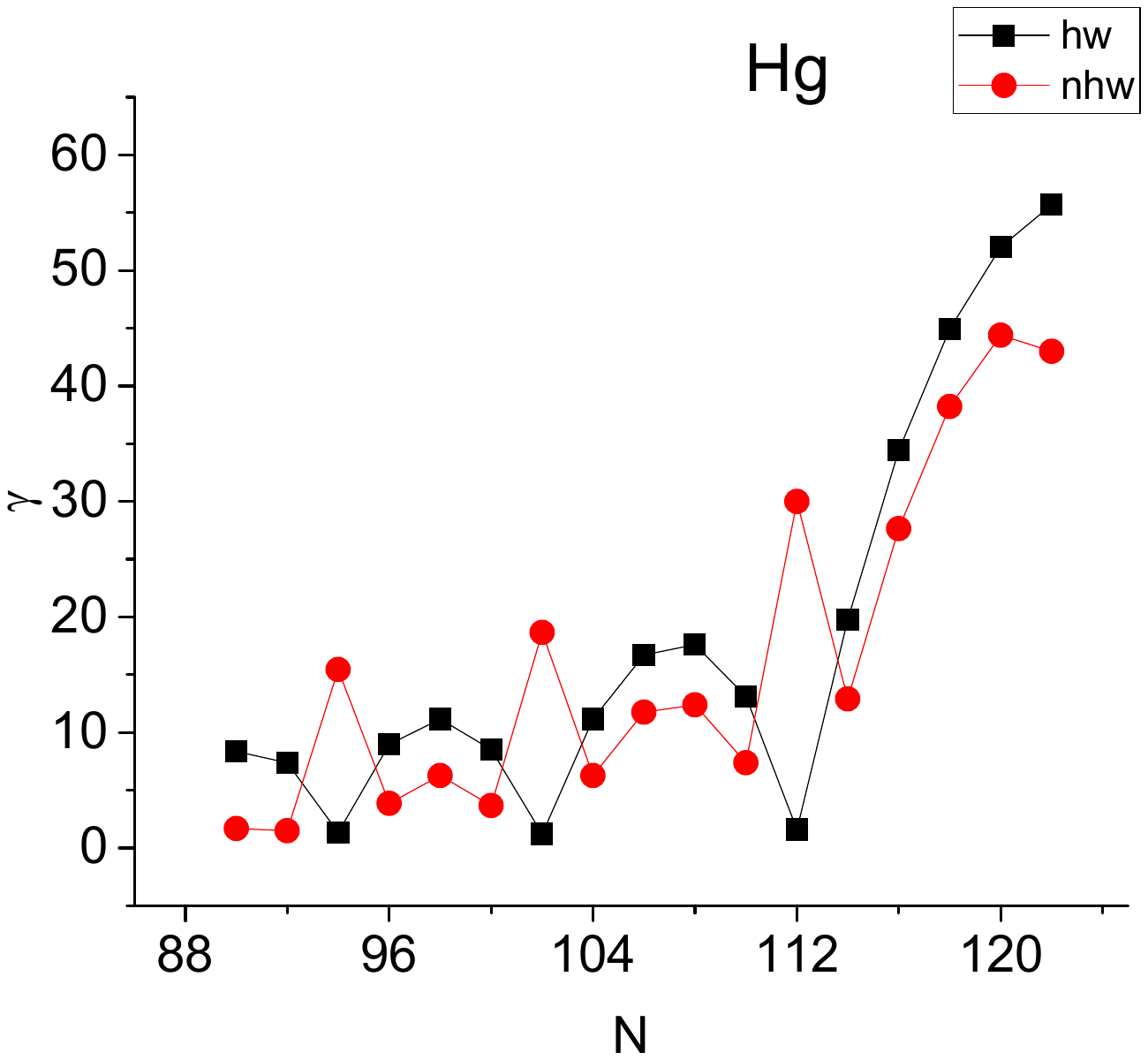} }
     
    \caption{Parameter-independent predictions for the collective variable $\gamma$ provided within the proxy-SU(3) symmetry by the highest weight irreducible representation (hw irrep) of SU(3) and by the next highest weight (nhw) irrep of SU(3) in the $Z=70$-80, $N=90$-122 region, taken from Table II, are plotted vs. $N$.  See Sec. \ref{nhw} for further discussion.}

\end{figure*}


\begin{figure*} [htb]

    {\includegraphics[width=75mm]{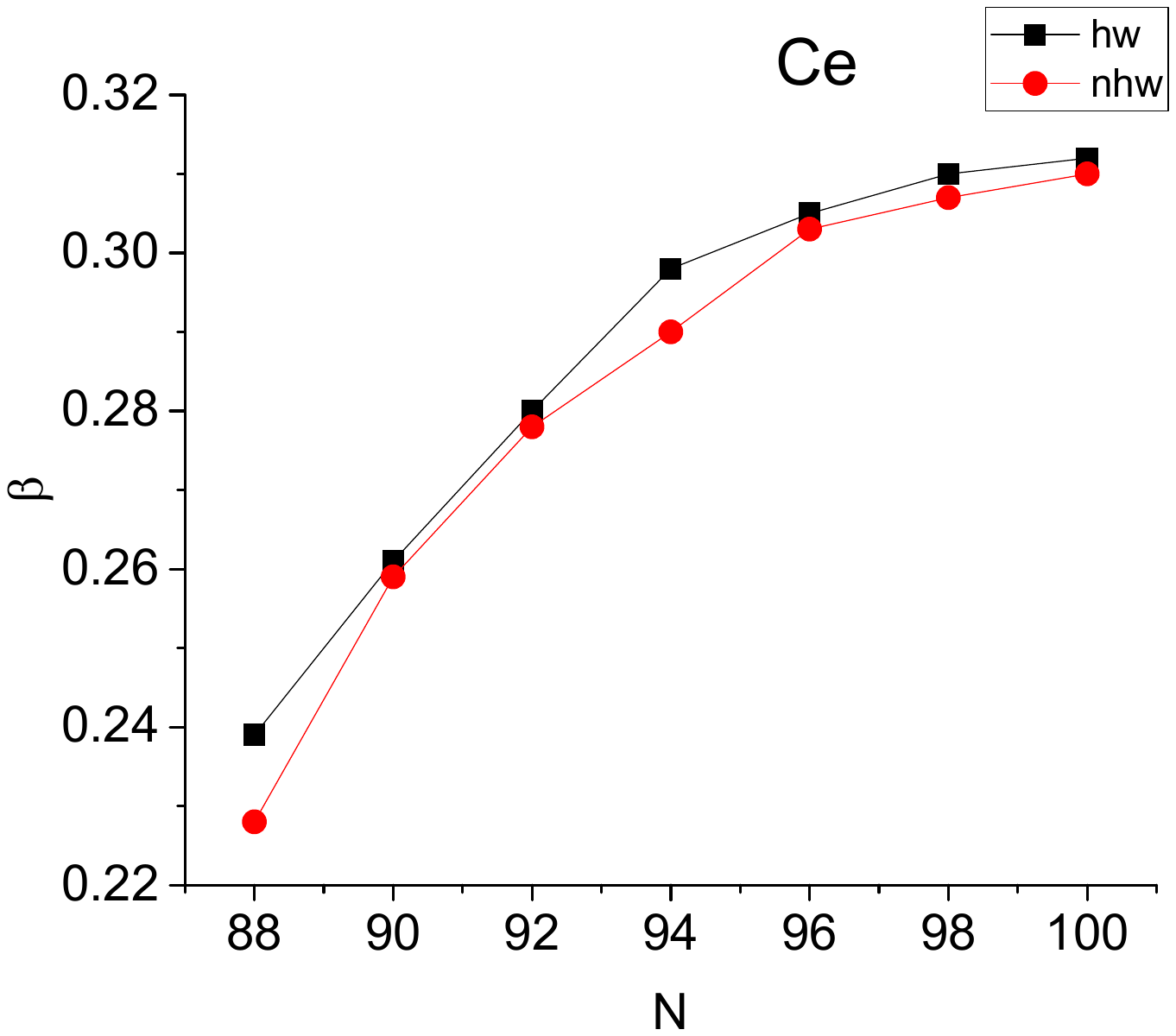} \hspace{5mm}   \includegraphics[width=75mm]{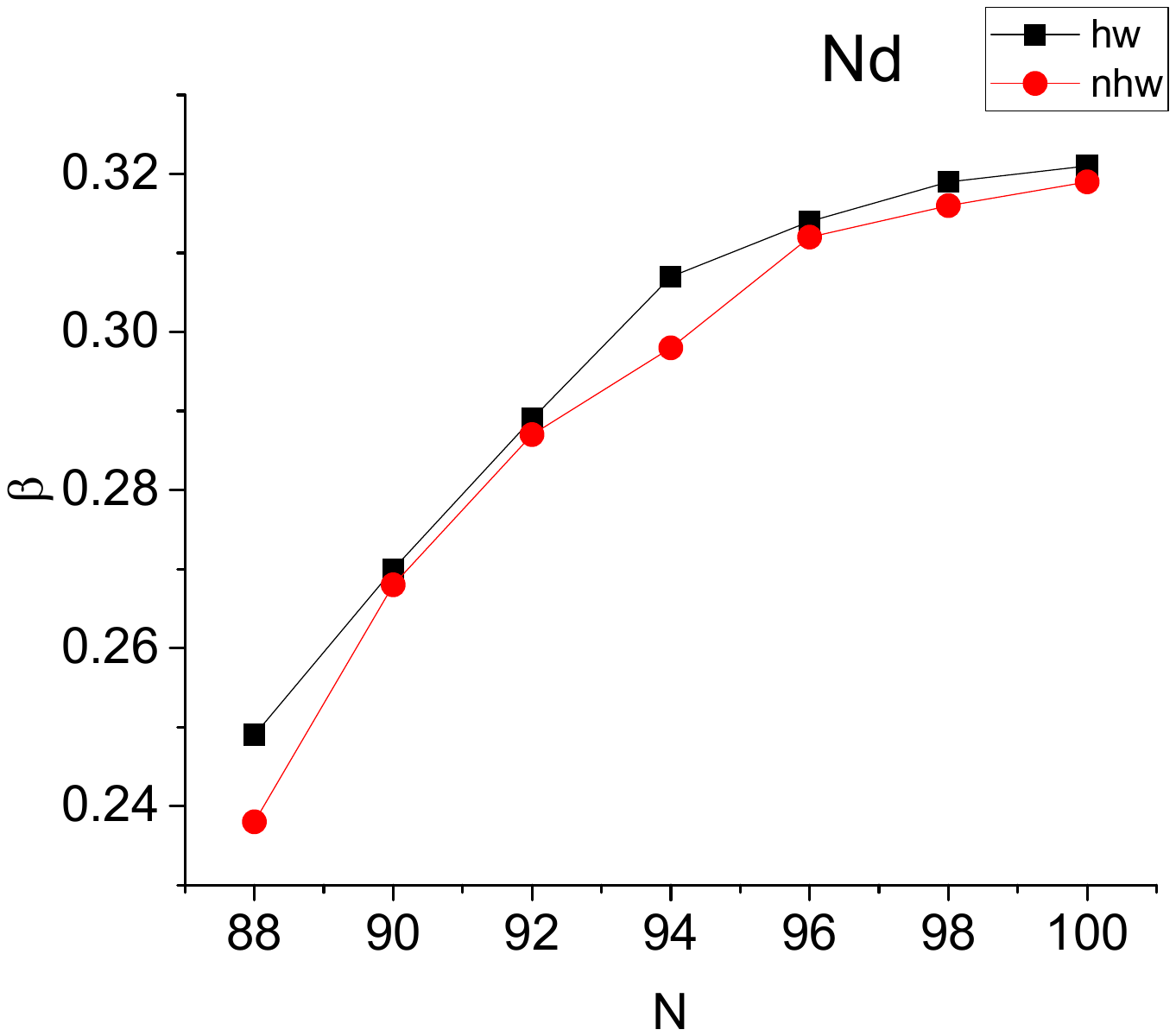} }
    {\includegraphics[width=75mm]{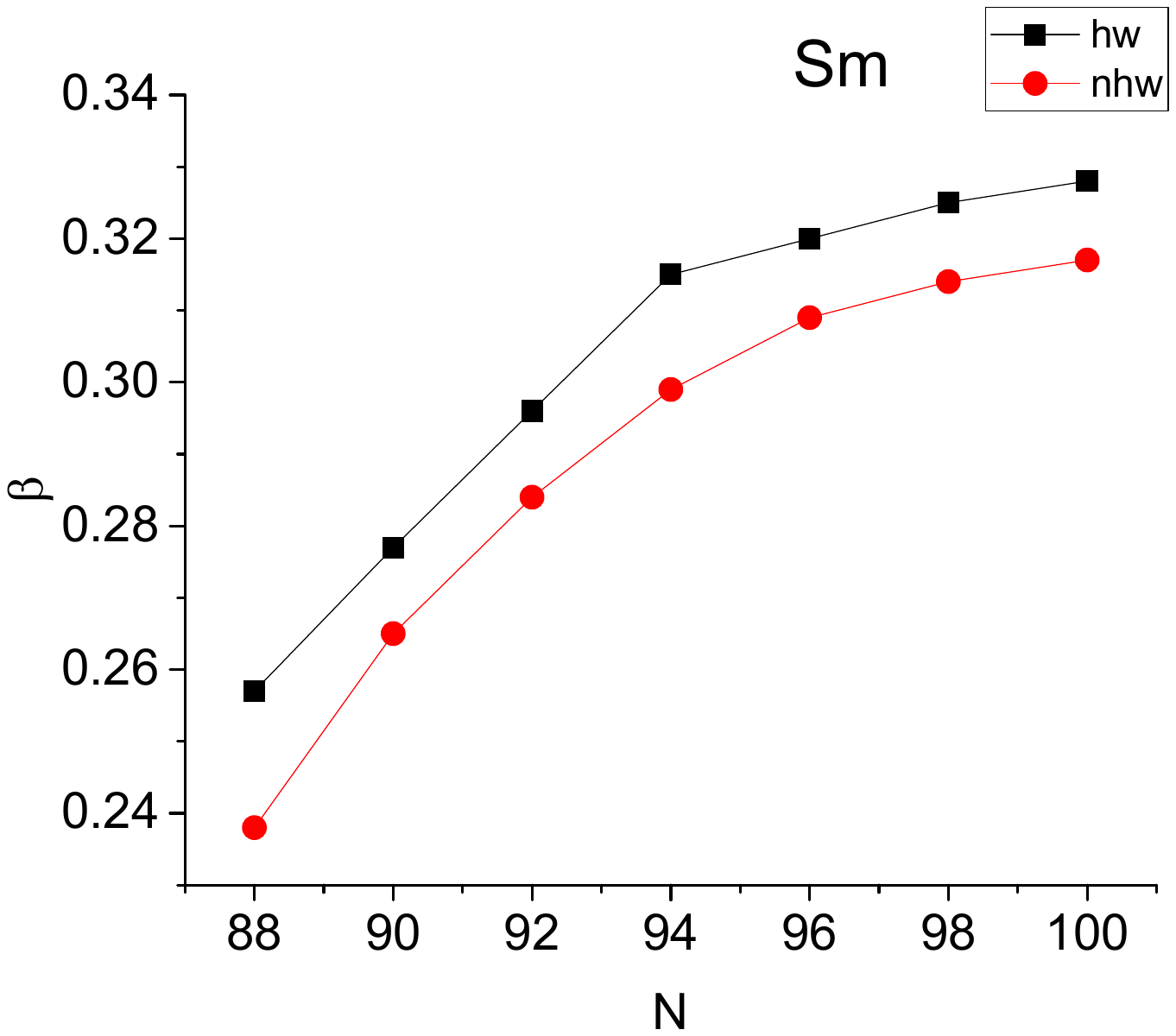} \hspace{5mm}    \includegraphics[width=75mm]{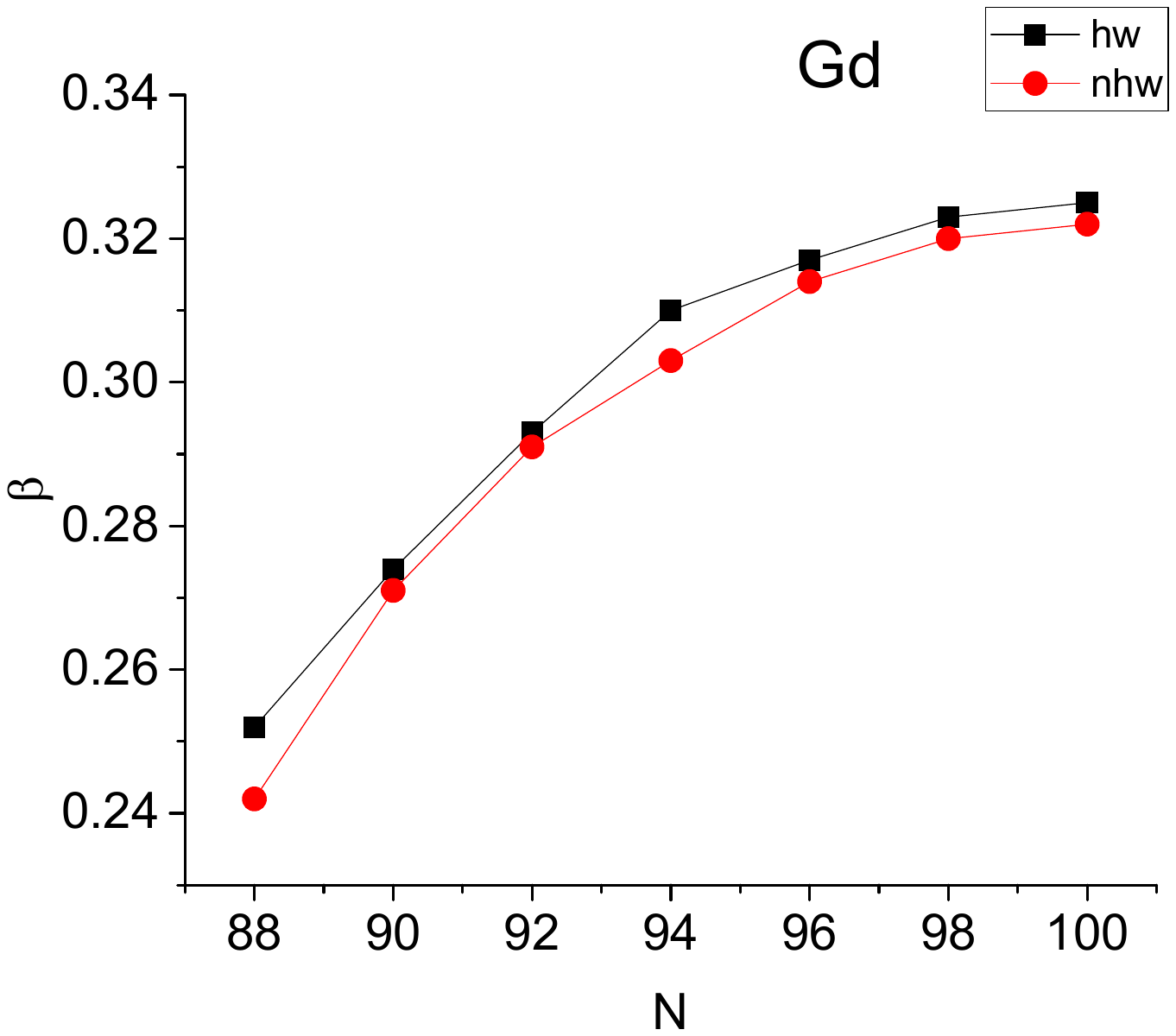} }
    {\includegraphics[width=75mm]{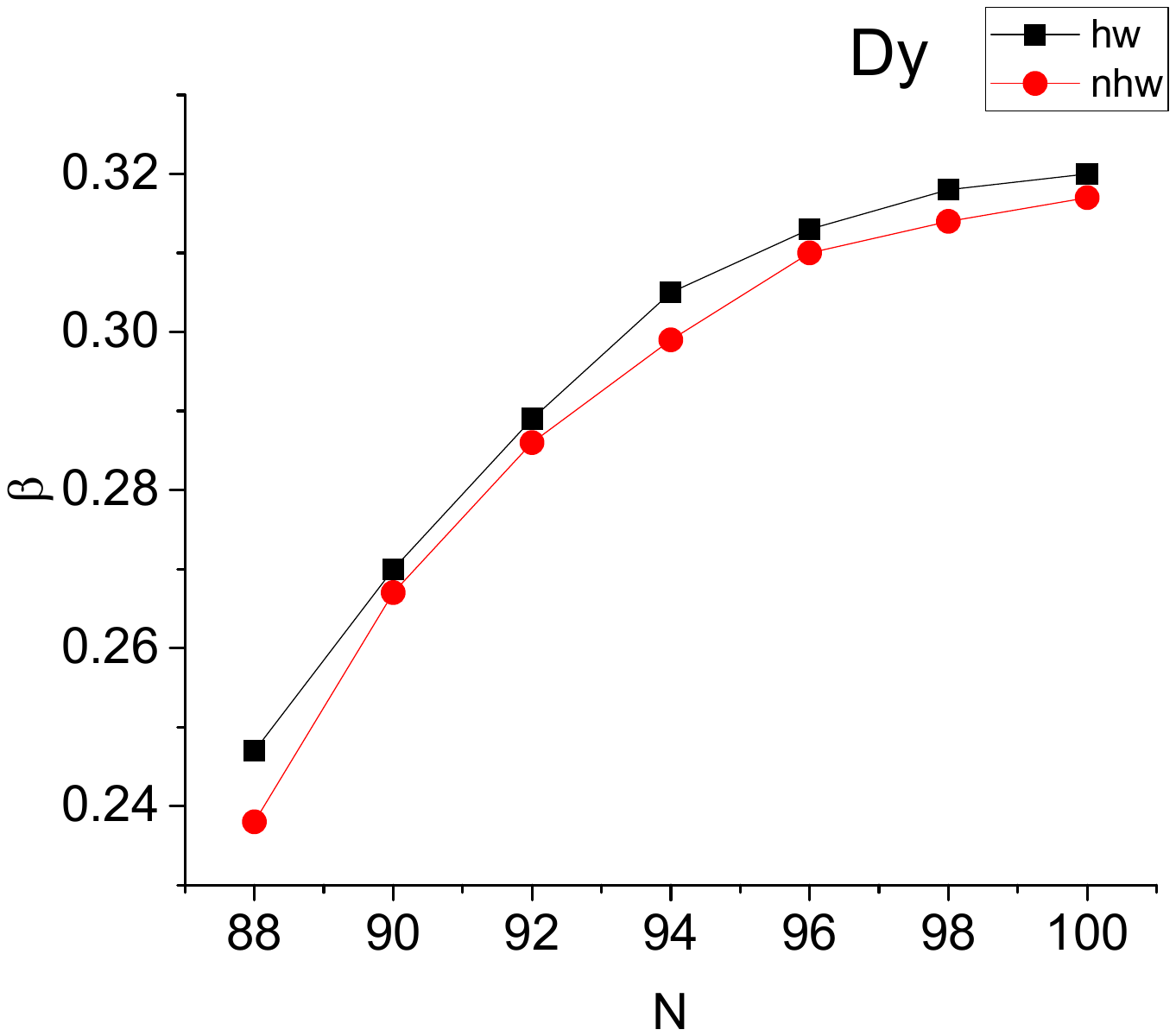}  \hspace{5mm}   \includegraphics[width=75mm]{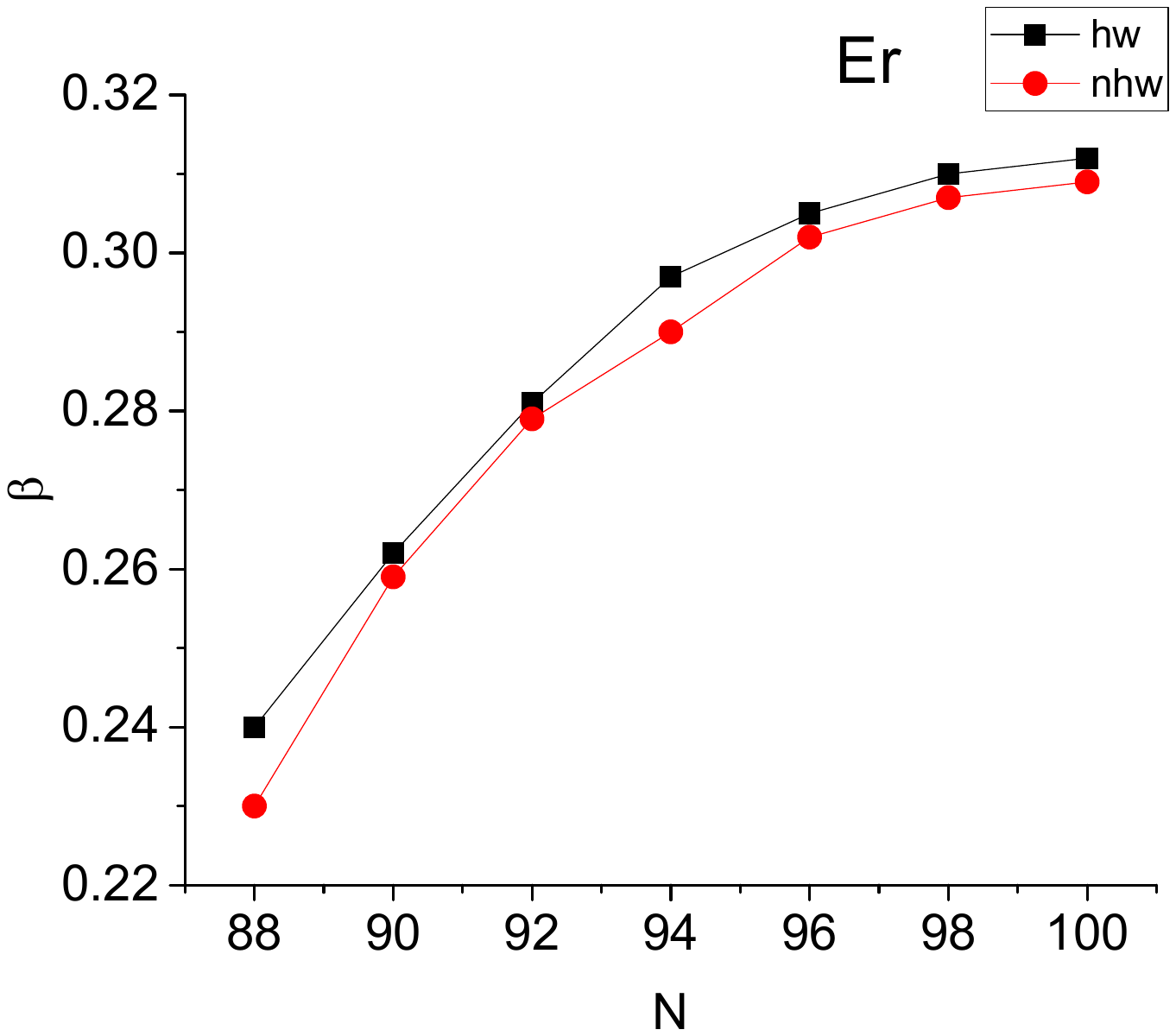} }
     
    \caption{Same as Fig. 3, but for the $Z=58$-68, $N=88$-100 region, with predictions taken from Table III.  See Sec. \ref{nhw} for further discussion.} 
    
\end{figure*}


\begin{figure*} [htb]

    {\includegraphics[width=75mm]{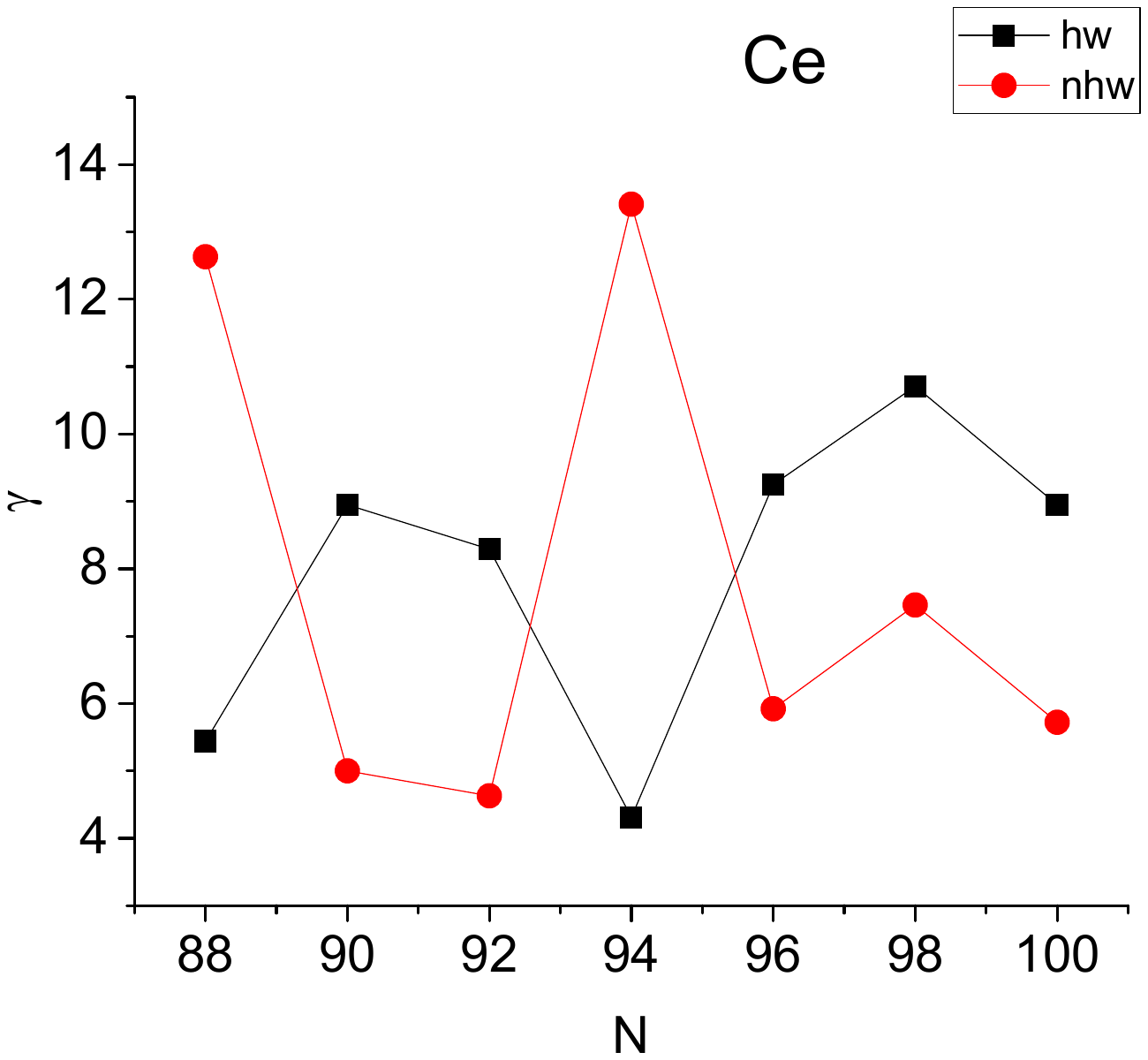} \hspace{5mm}   \includegraphics[width=75mm]{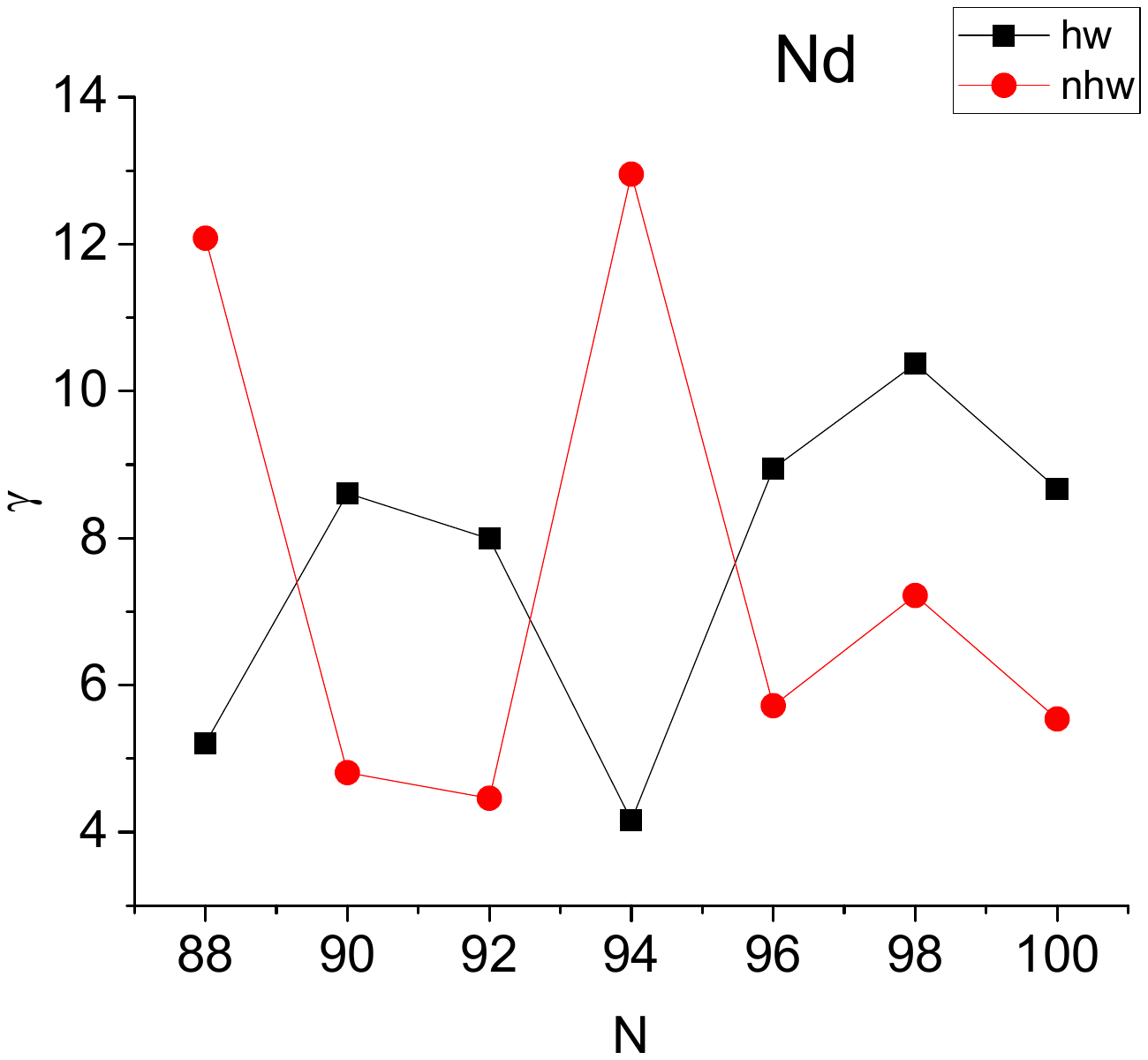} }
    {\includegraphics[width=75mm]{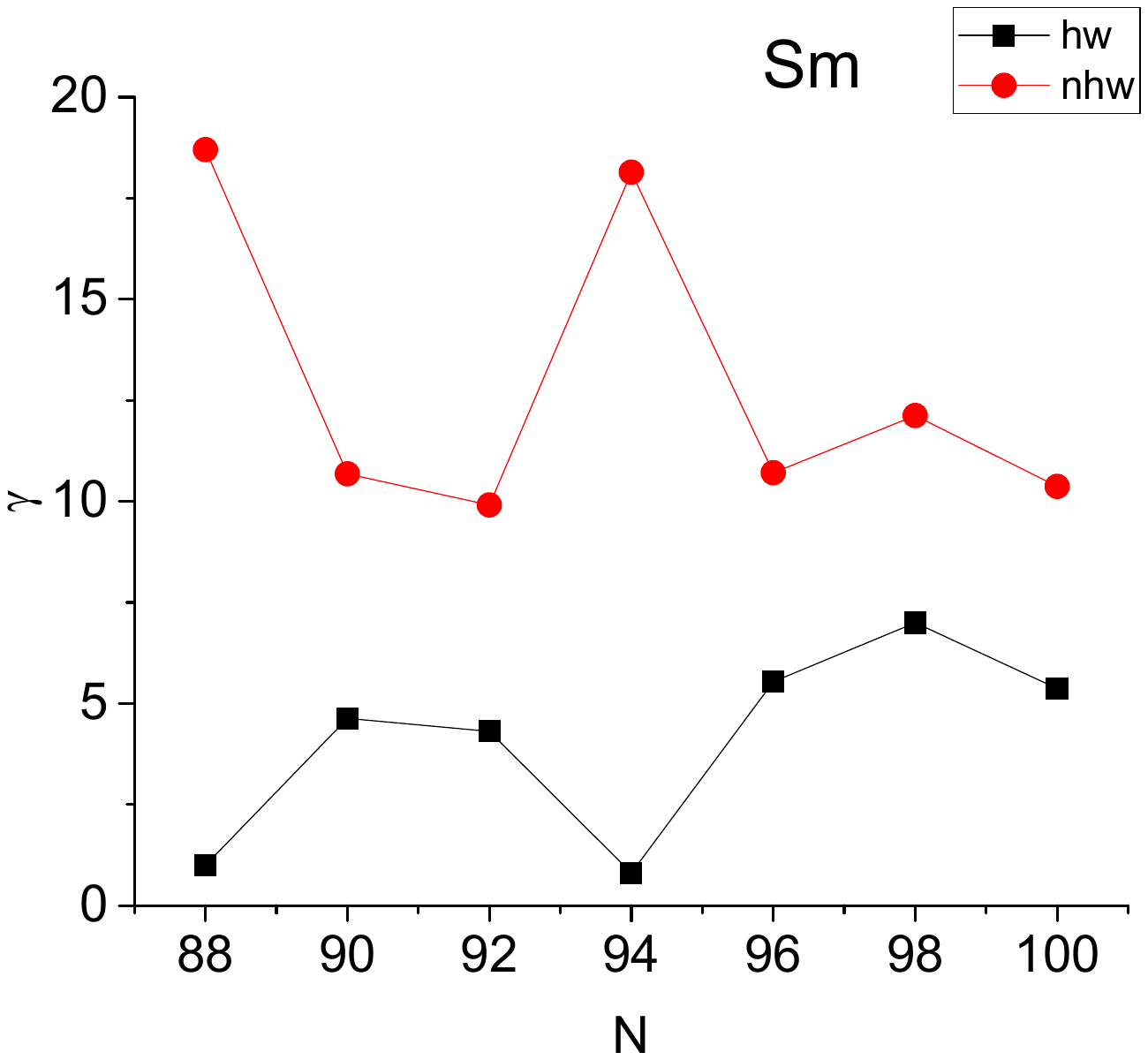} \hspace{5mm}    \includegraphics[width=75mm]{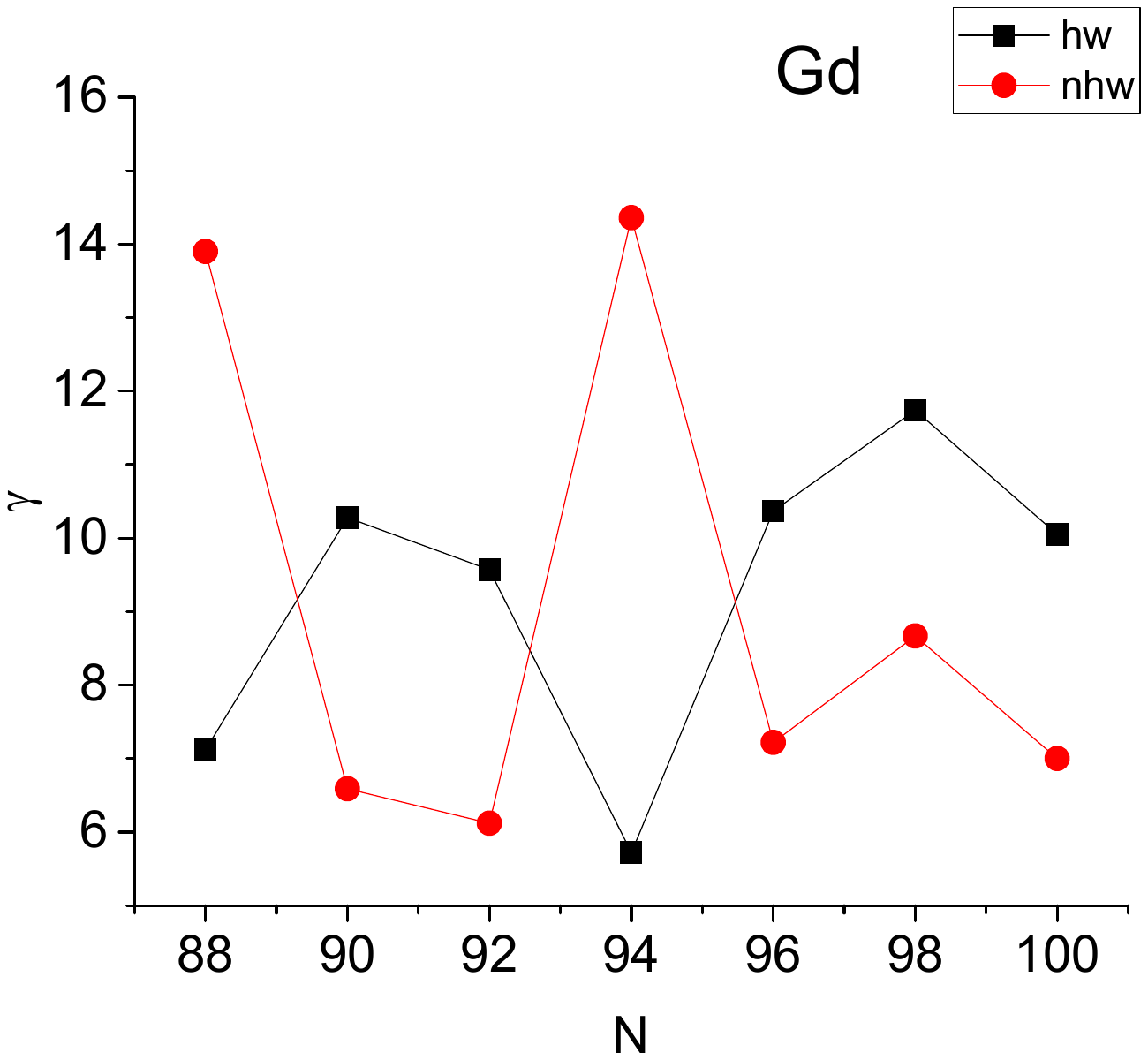} }
    {\includegraphics[width=75mm]{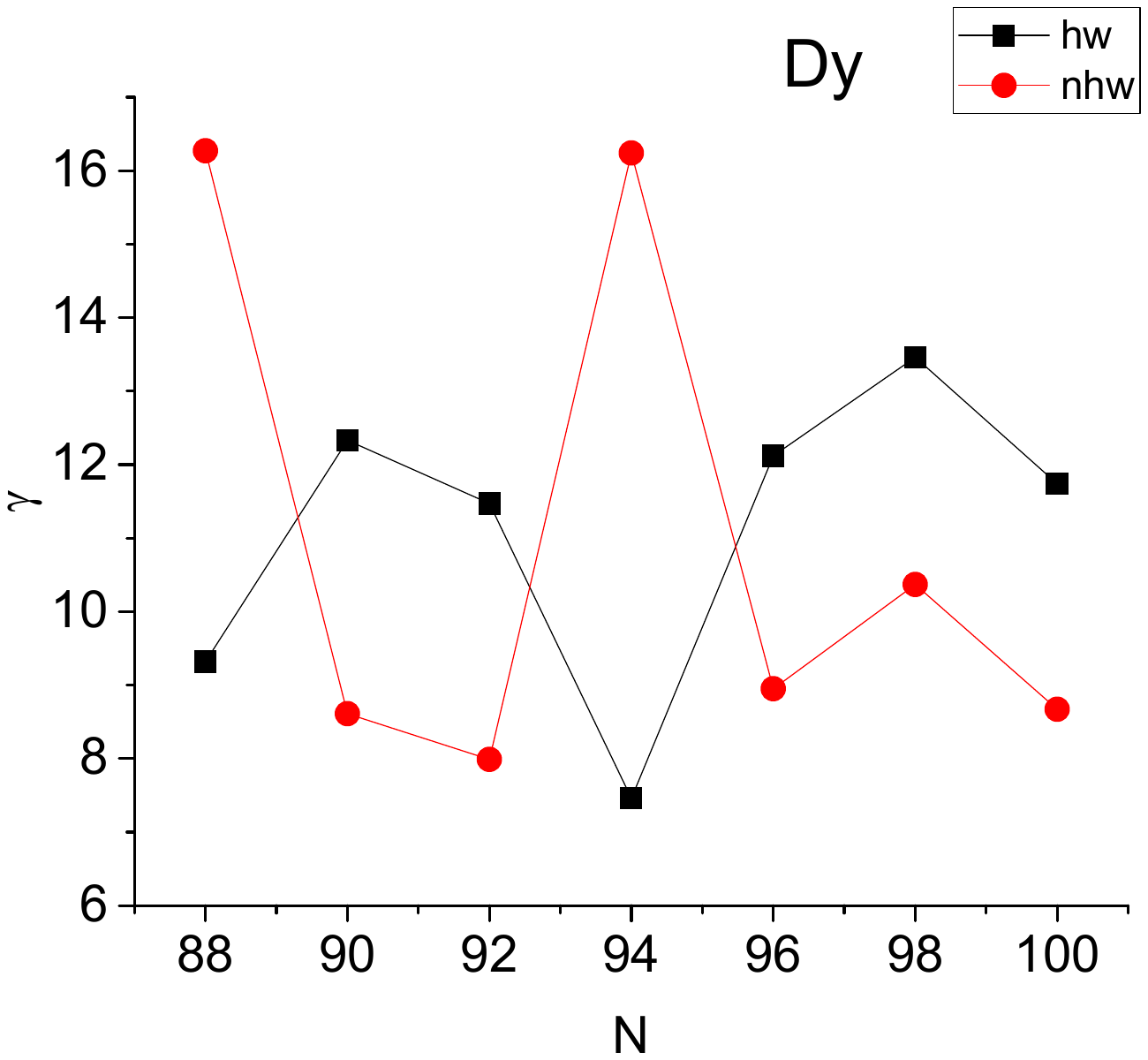}  \hspace{5mm}   \includegraphics[width=75mm]{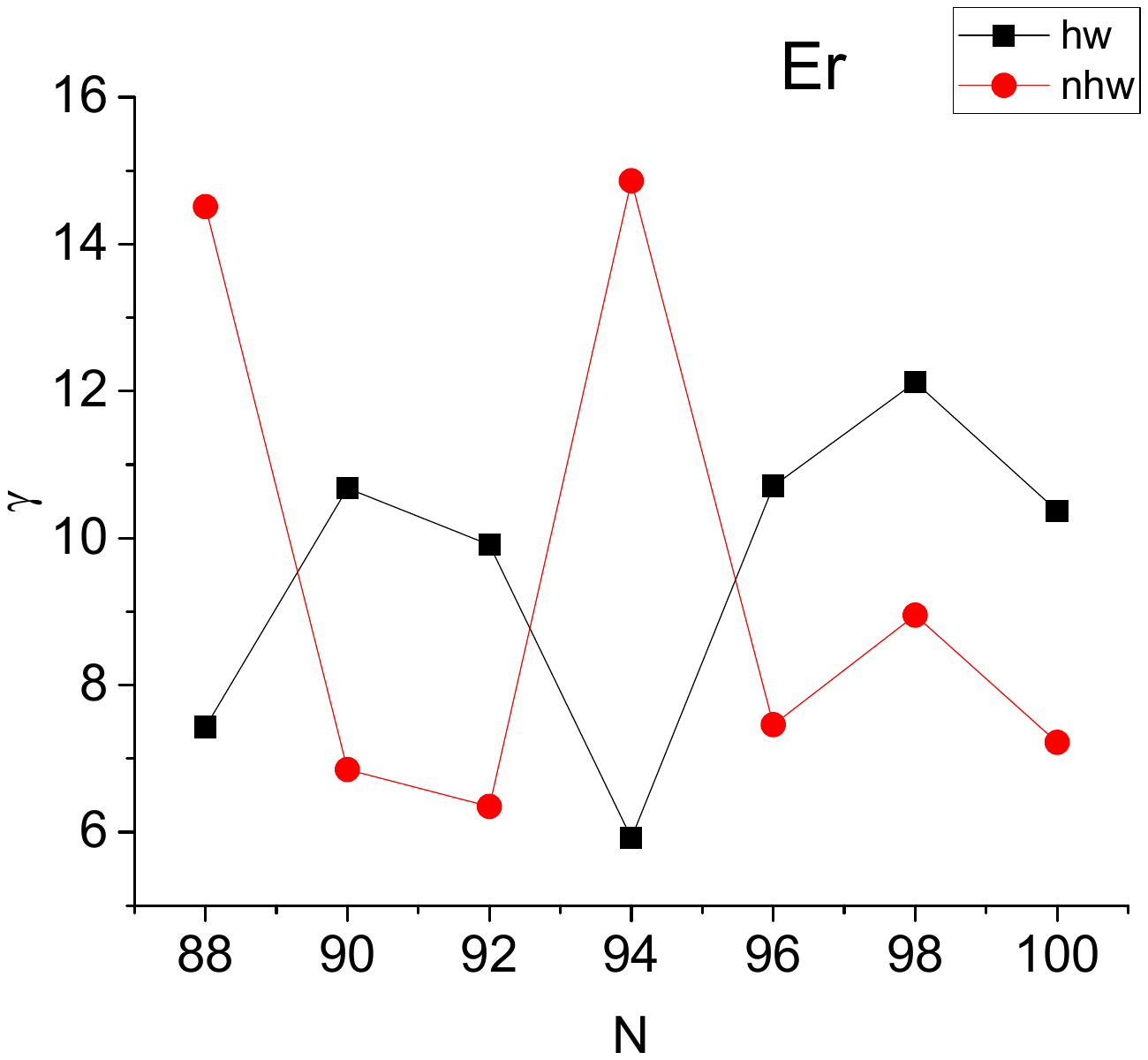} }
     
    \caption{Same as Fig. 4, but for the $Z=58$-68, $N=88$-100 region, with predictions taken from Table III. See Sec. \ref{nhw} for further discussion.} 
    
\end{figure*}


\begin{figure*} [htb]

    {\includegraphics[width=75mm]{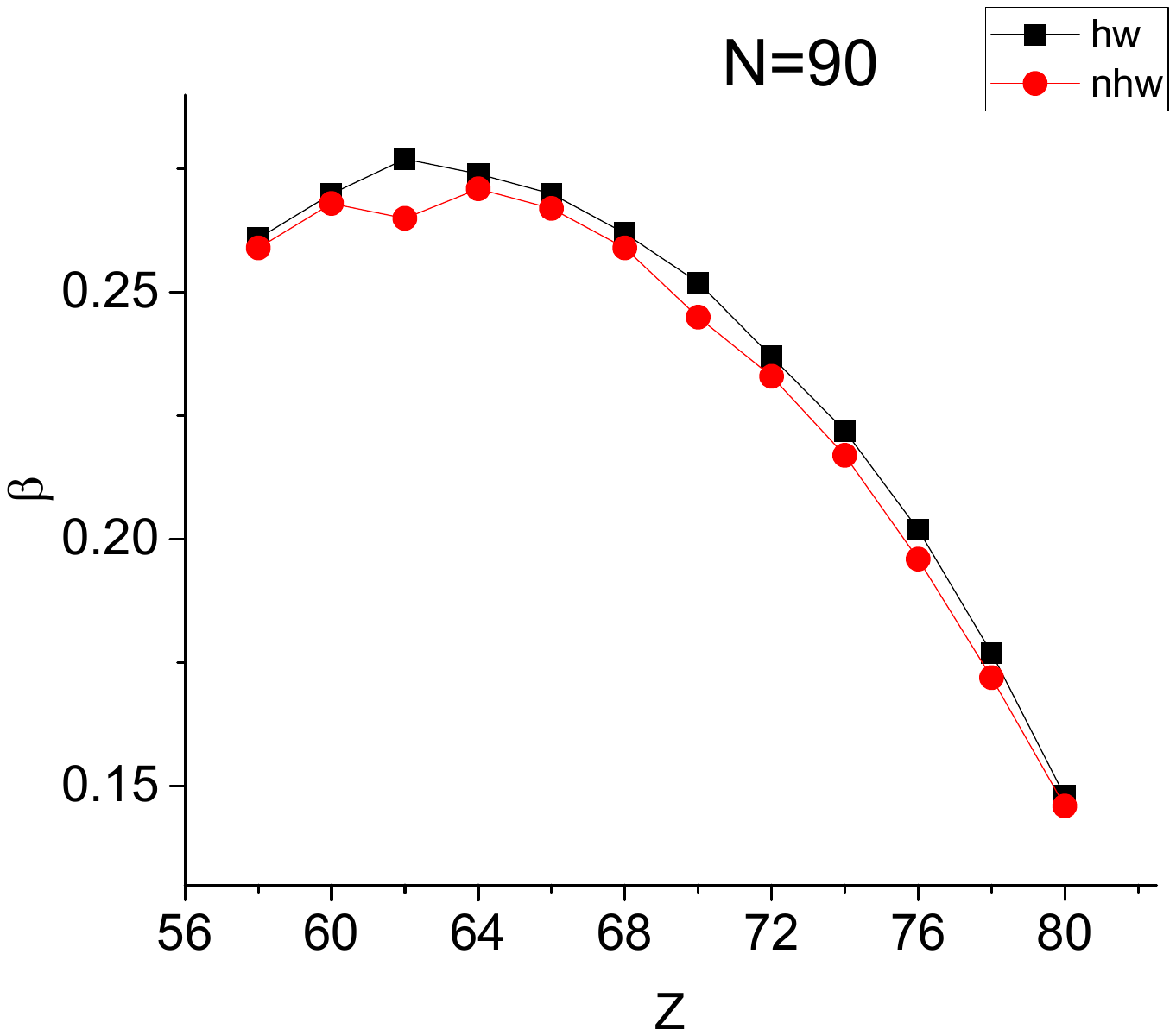} \hspace{5mm}   \includegraphics[width=75mm]{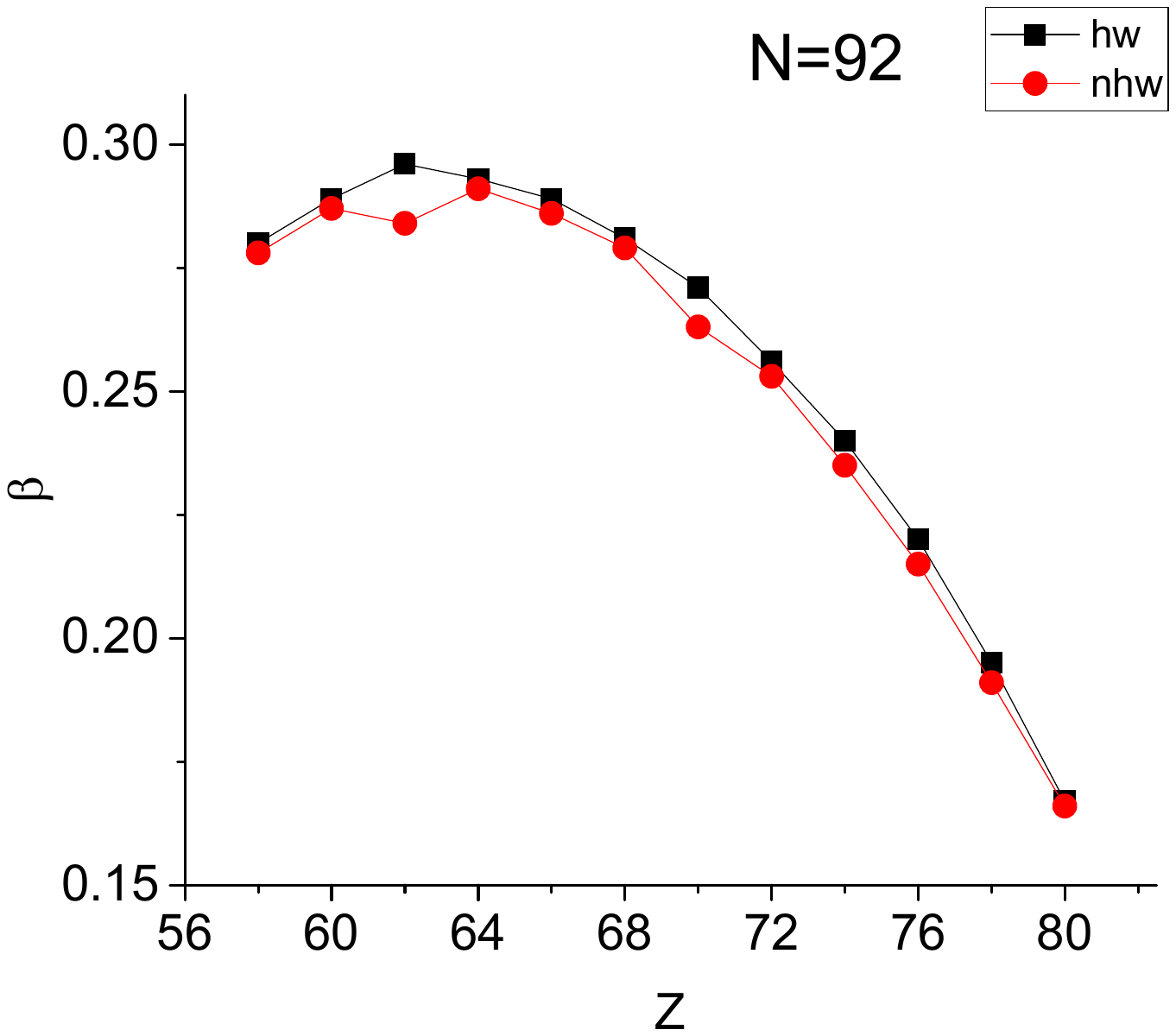} }
    {\includegraphics[width=75mm]{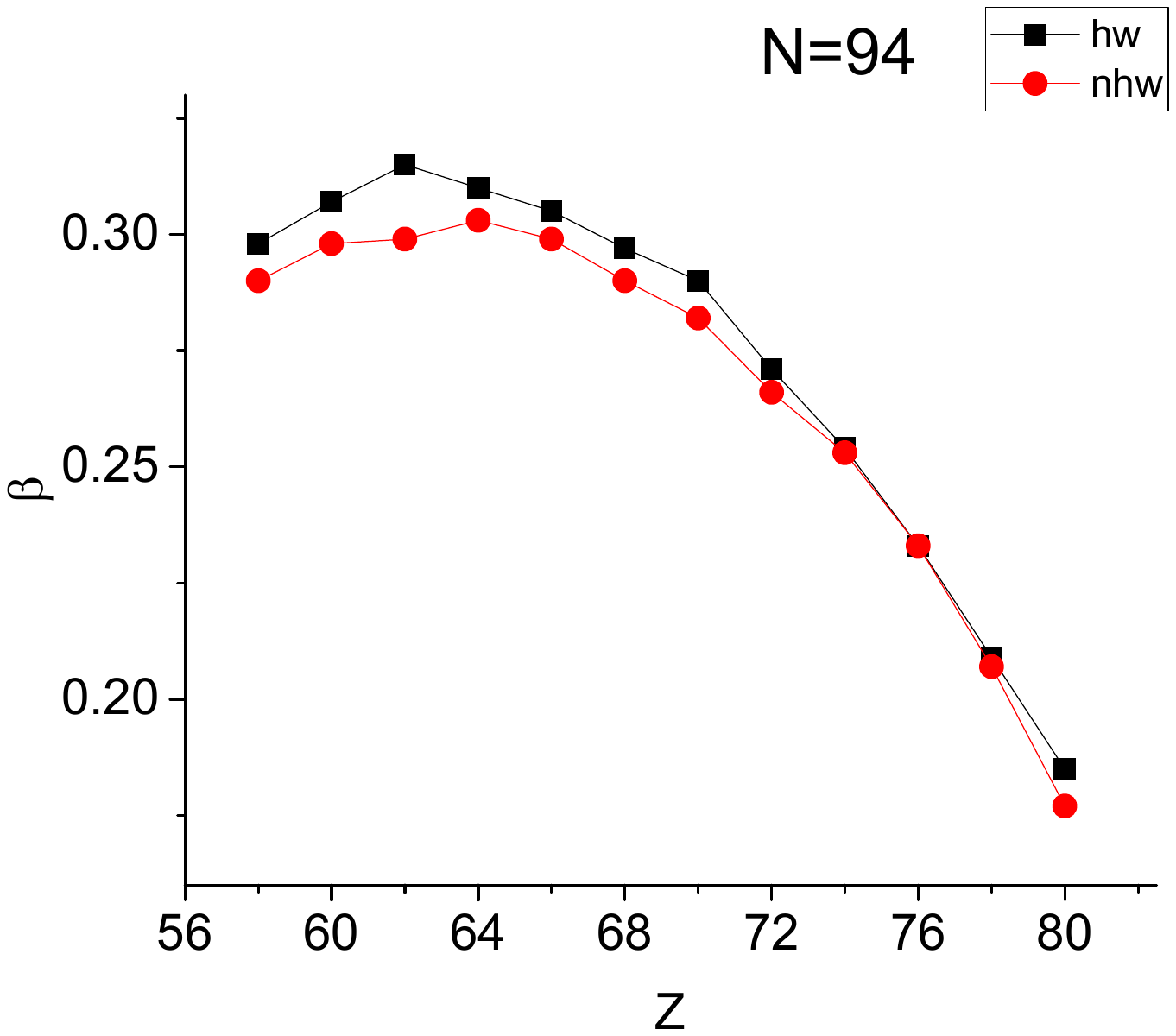} \hspace{5mm}    \includegraphics[width=75mm]{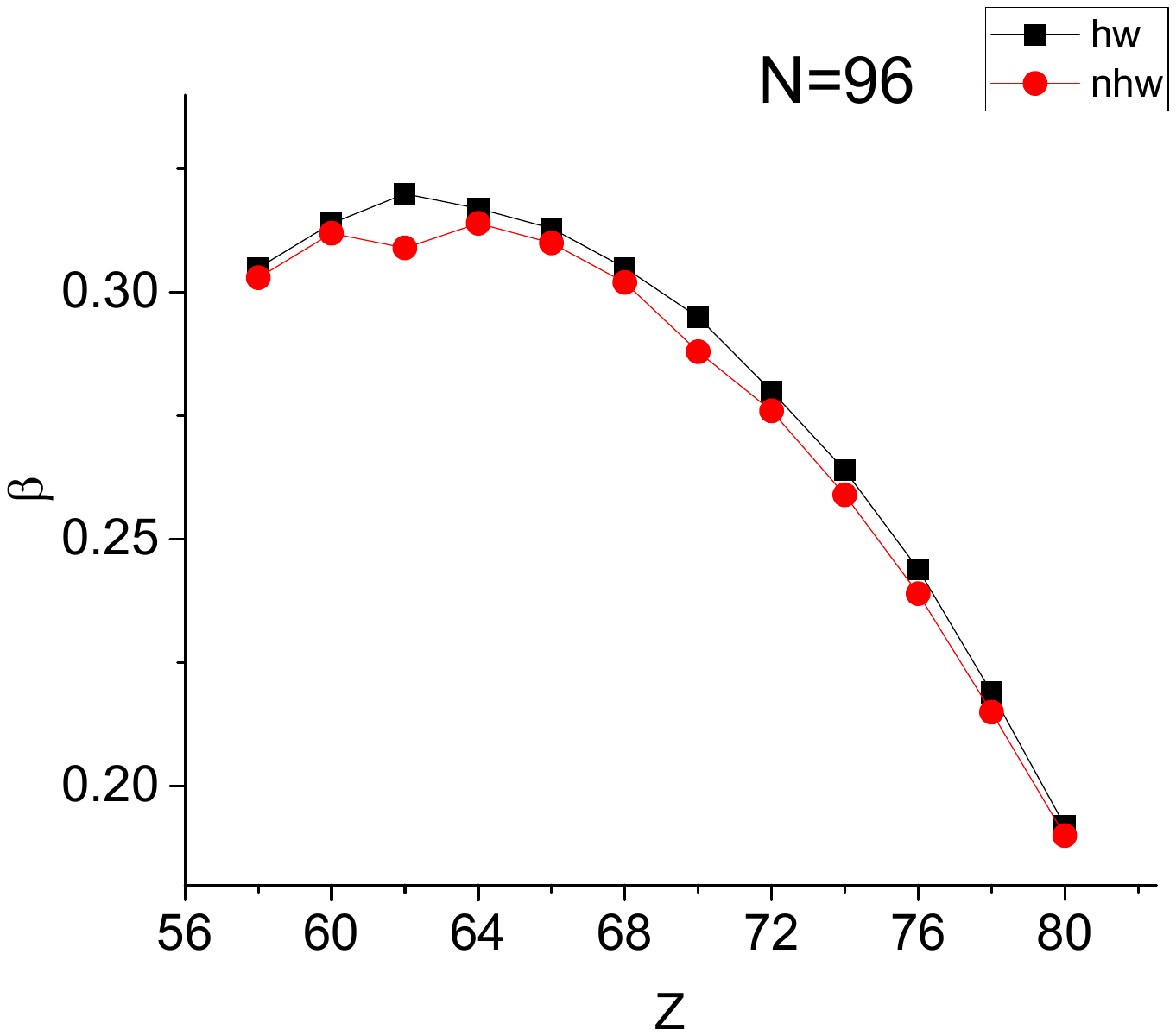} }
    {\includegraphics[width=75mm]{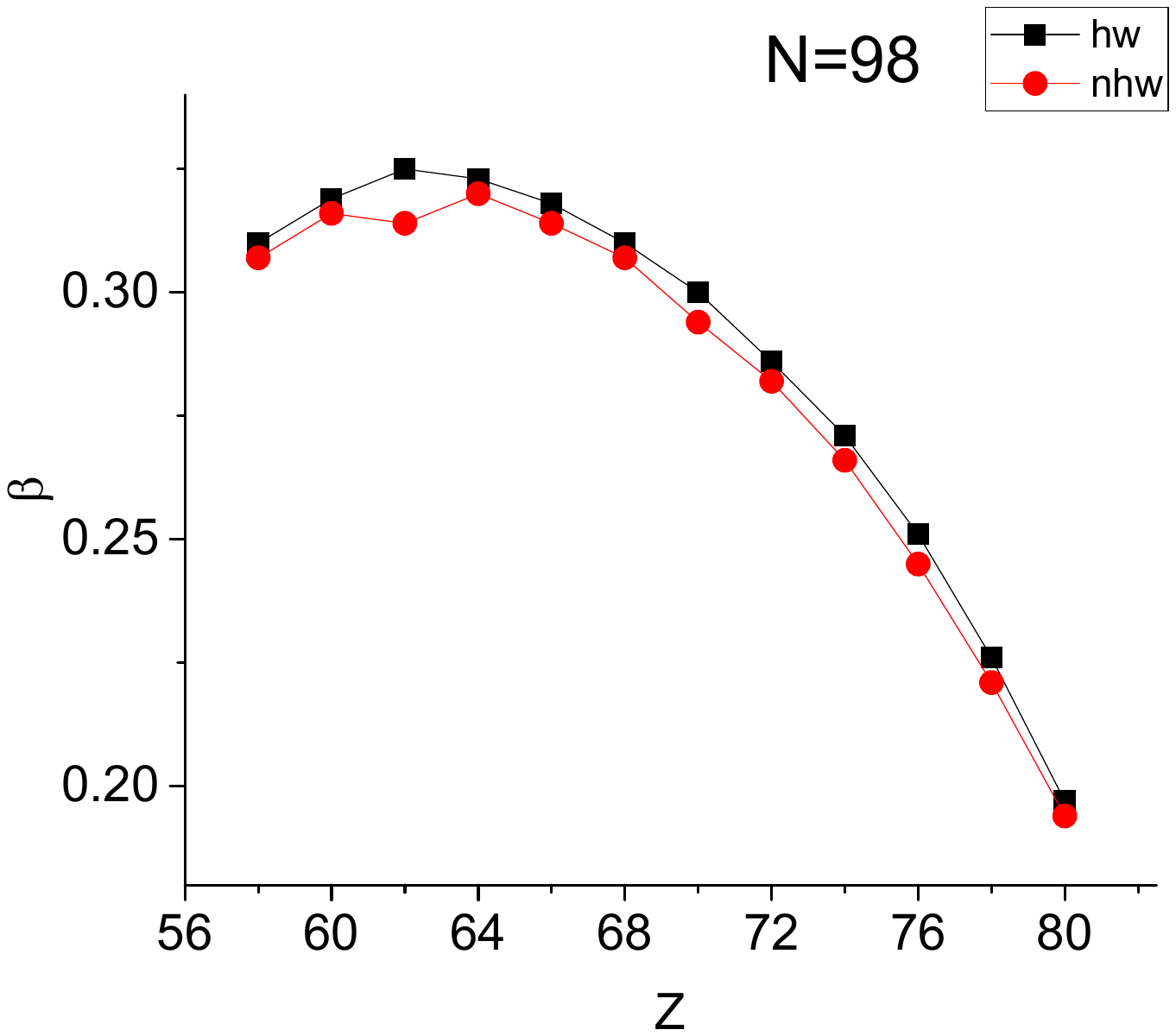}  \hspace{5mm}   \includegraphics[width=75mm]{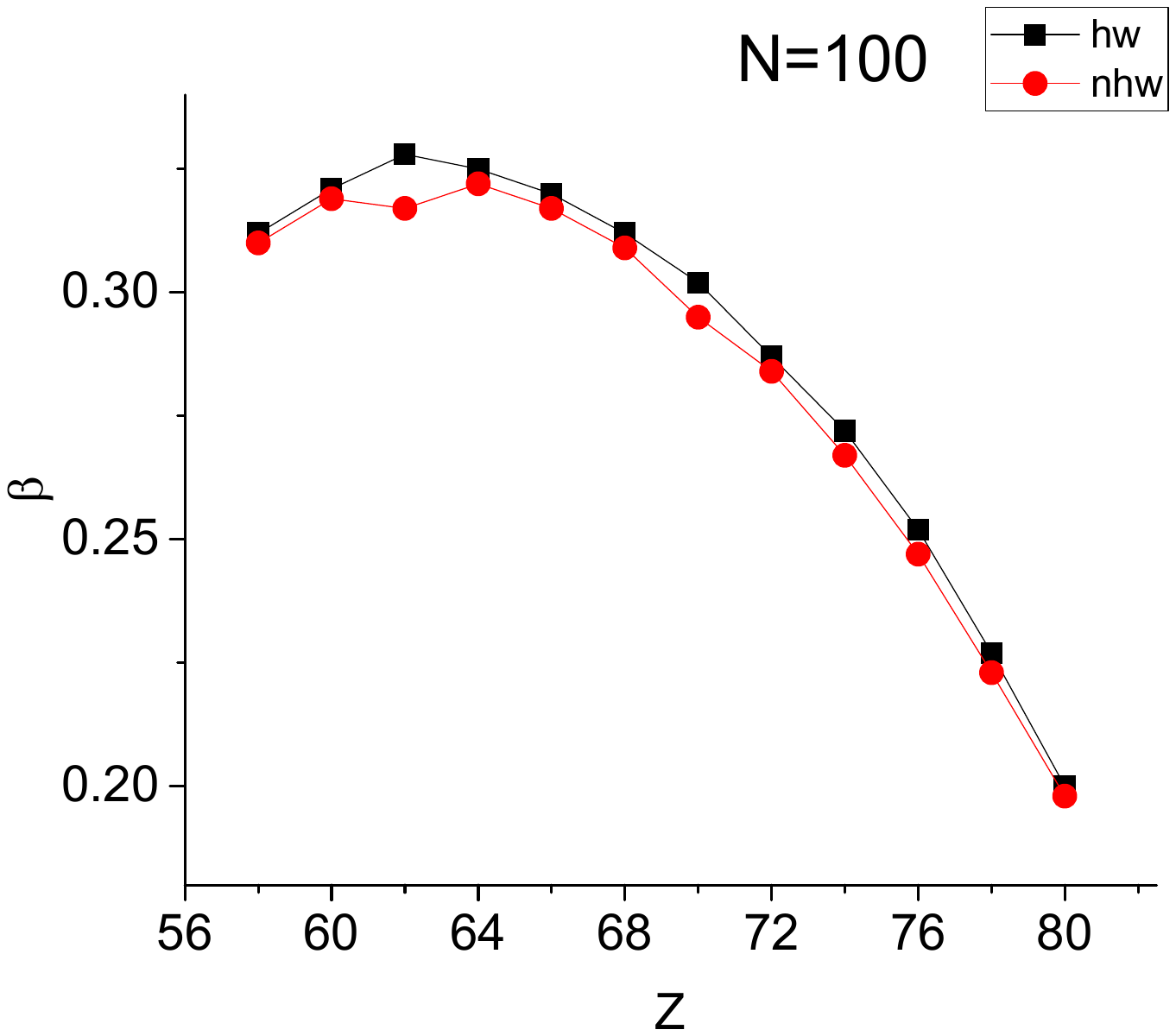} }
     
    \caption{Parameter-independent predictions for the collective variable $\beta$ provided within the proxy-SU(3) symmetry by the highest weight irreducible representation (hw irrep) of SU(3) and by the next highest weight (nhw) irrep of SU(3) in the $Z=58$-80, $N=90$-100 region, taken from Tables II and III,  are plotted vs. $Z$. See Sec. \ref{nhw} for further discussion.} 
    
\end{figure*}


\begin{figure*} [htb]

    {\includegraphics[width=75mm]{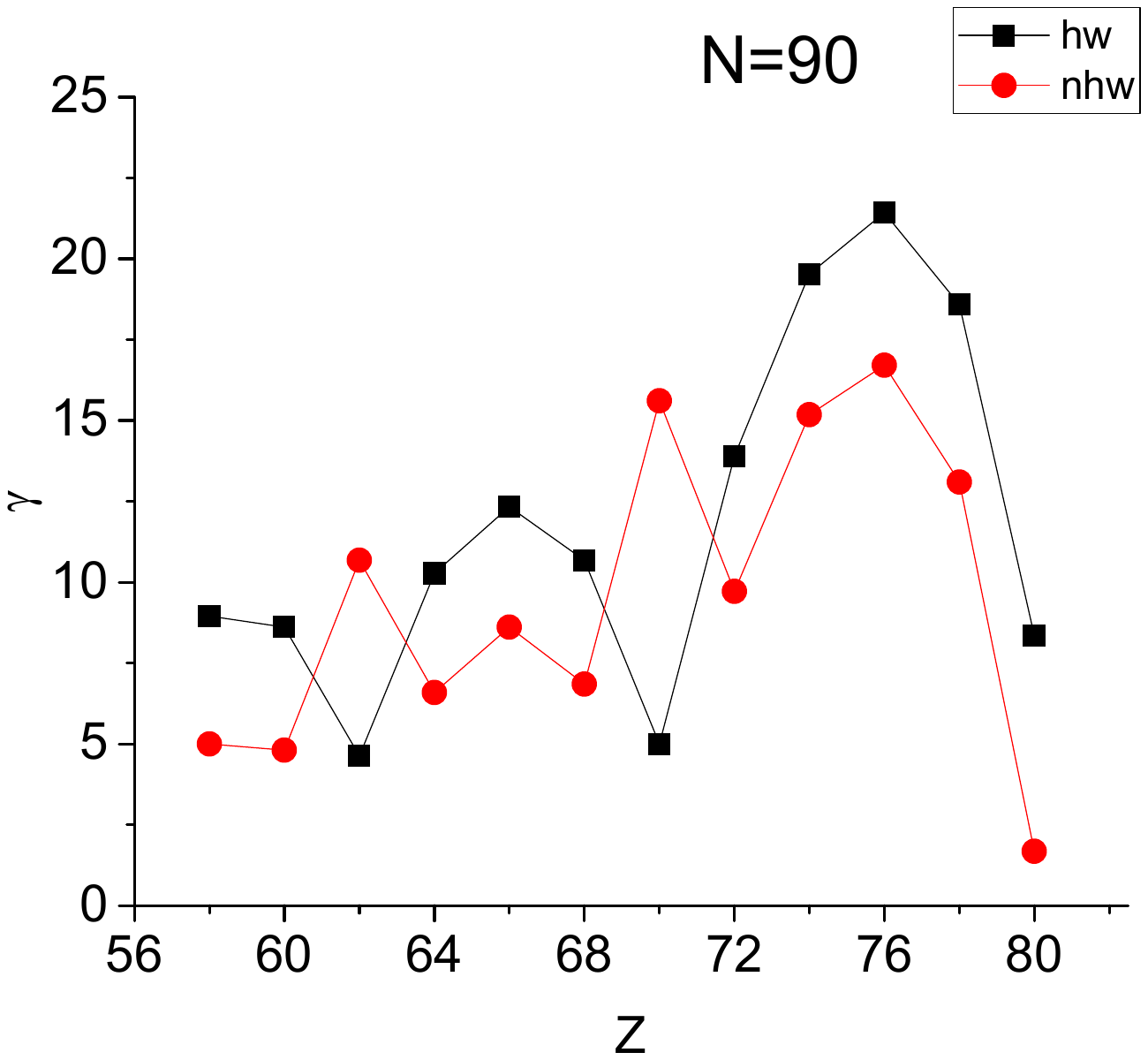} \hspace{5mm}   \includegraphics[width=75mm]{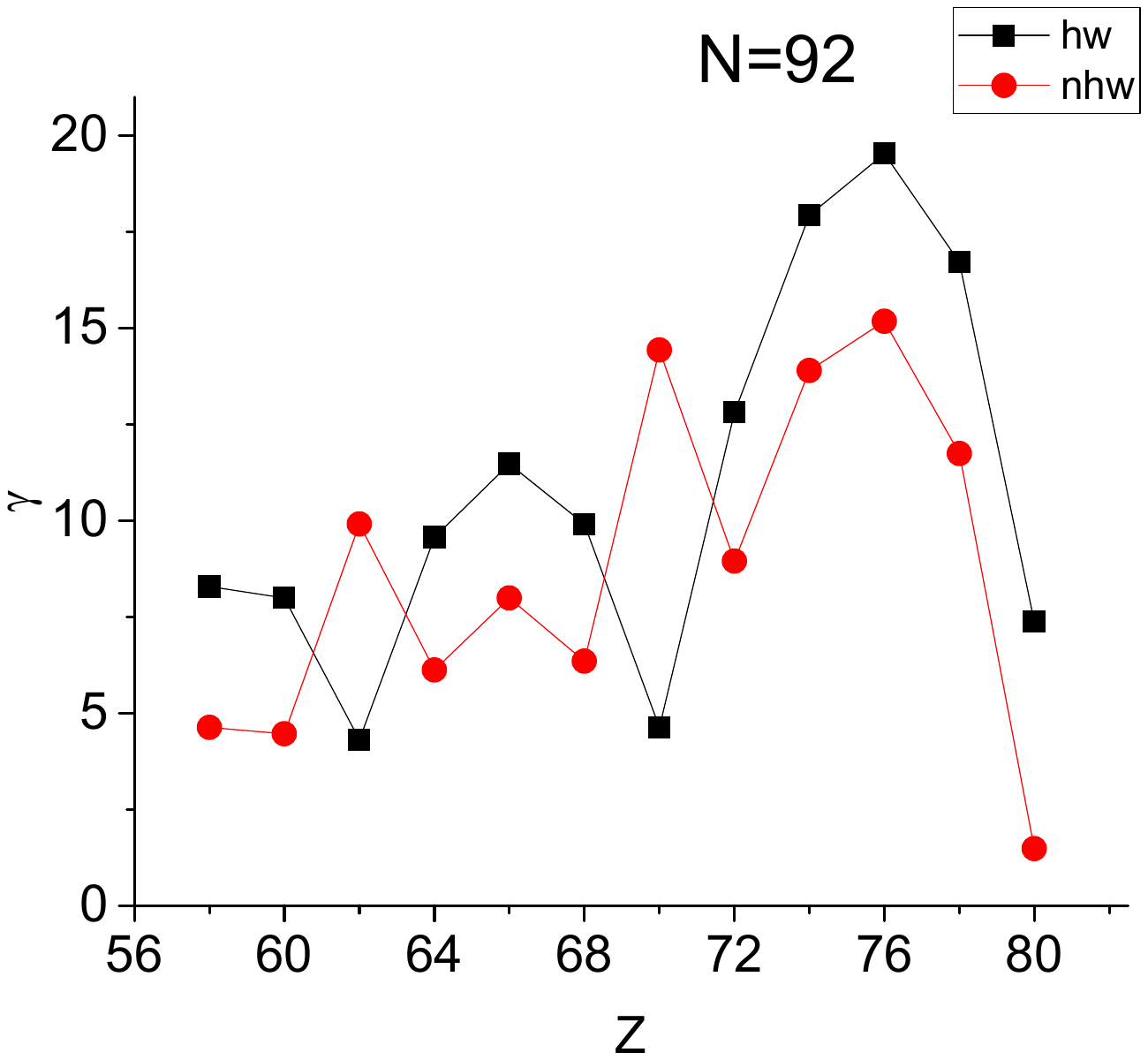} }
    {\includegraphics[width=75mm]{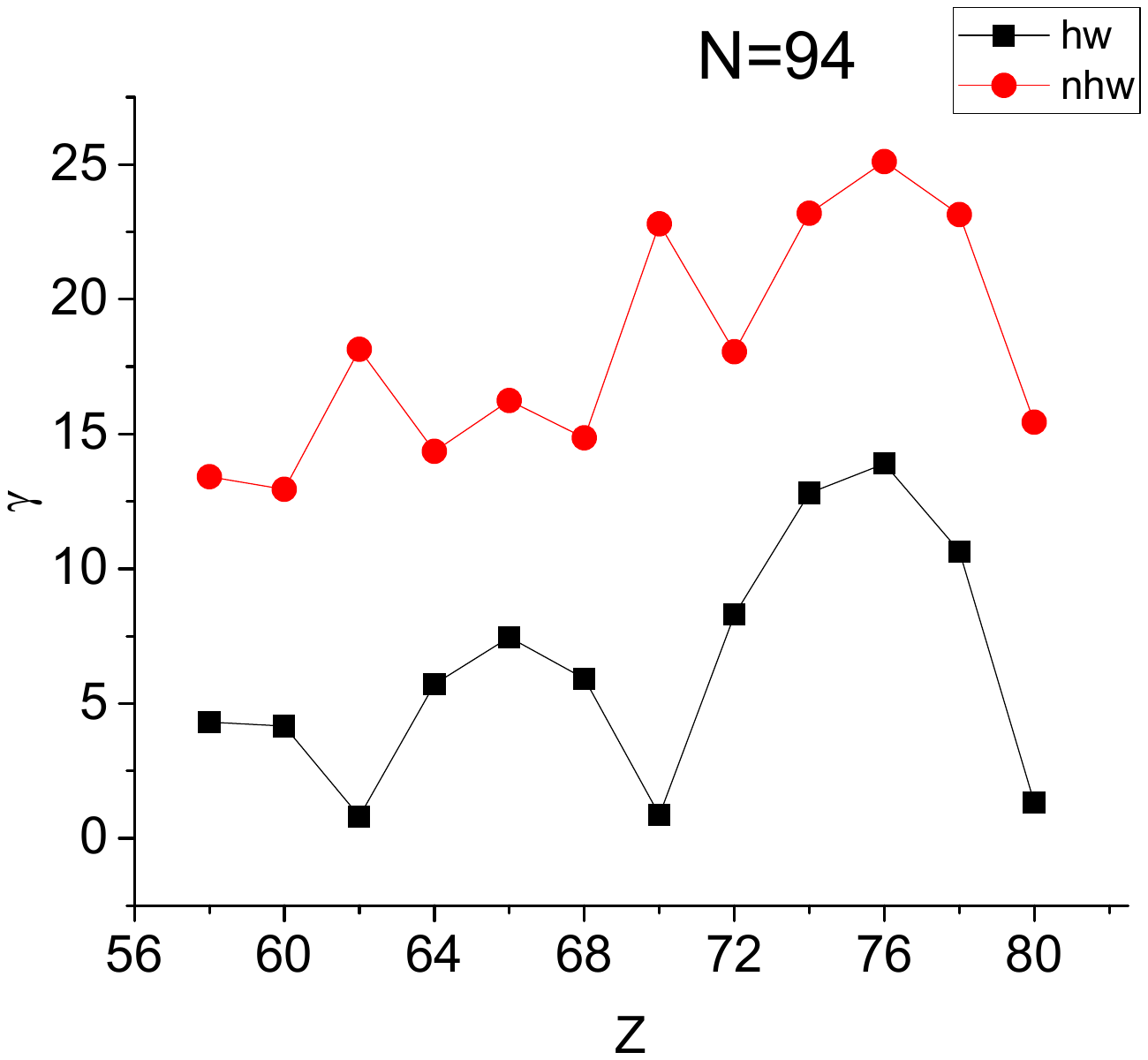} \hspace{5mm}    \includegraphics[width=75mm]{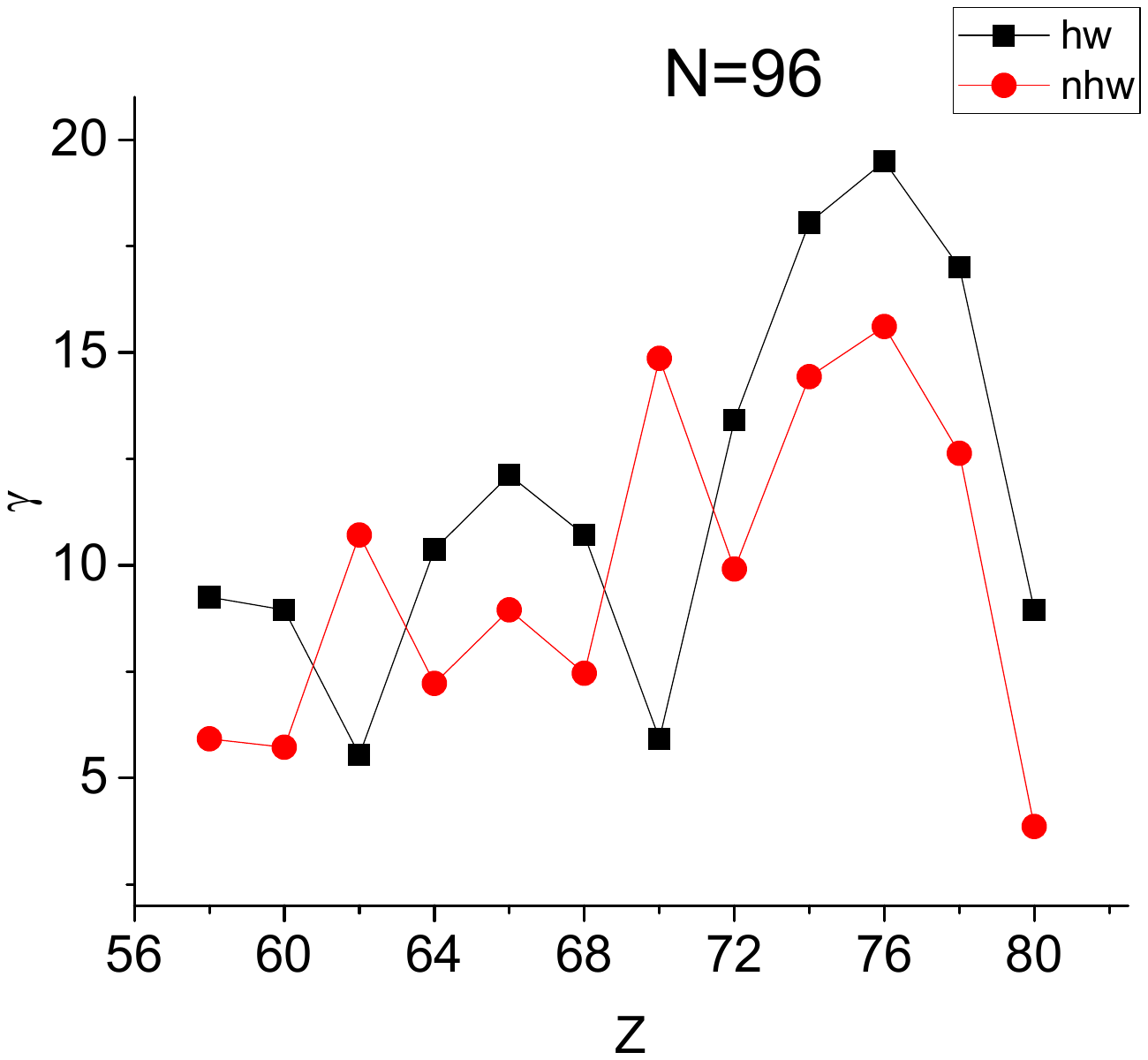} }
    {\includegraphics[width=75mm]{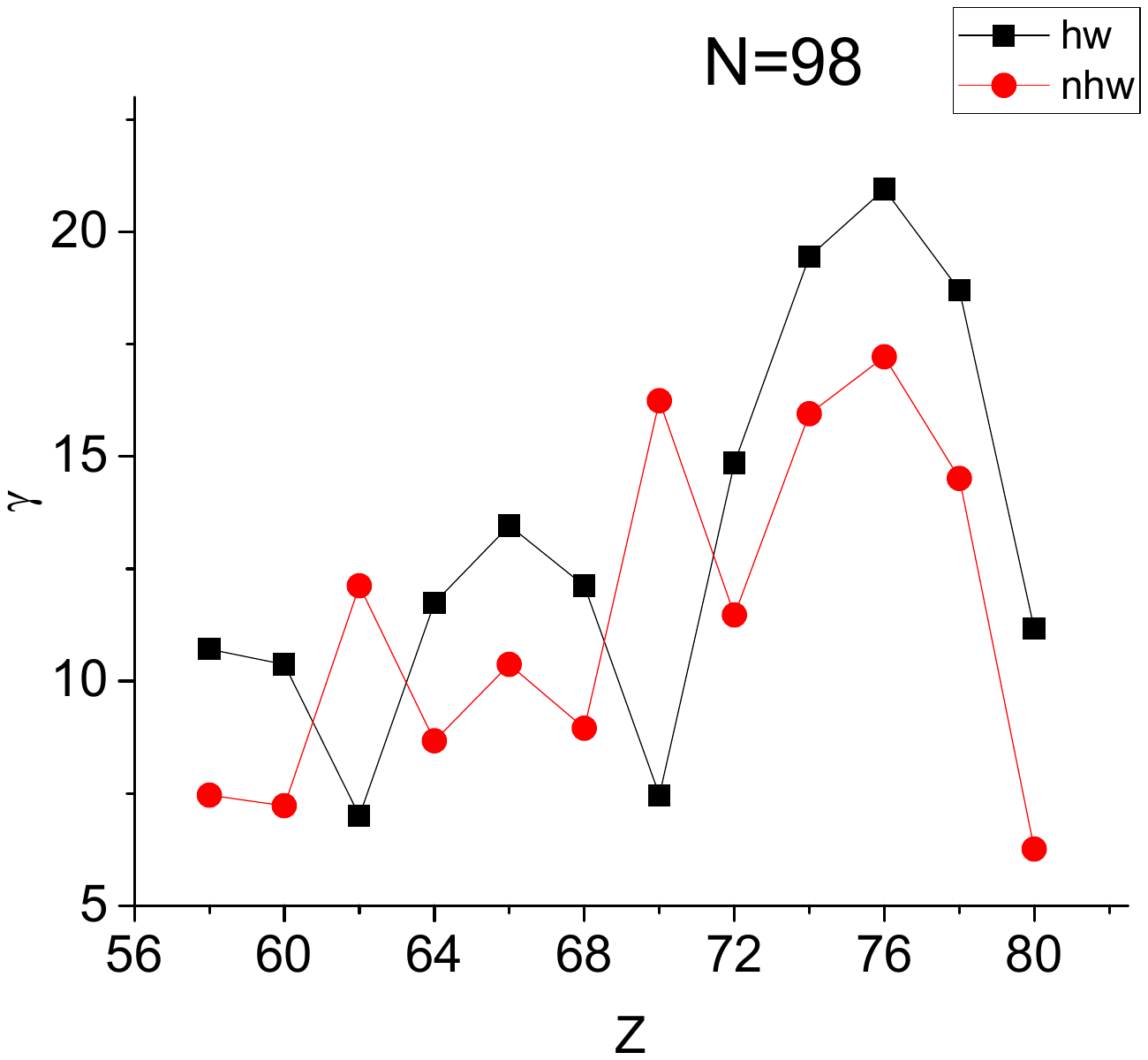}  \hspace{5mm}   \includegraphics[width=75mm]{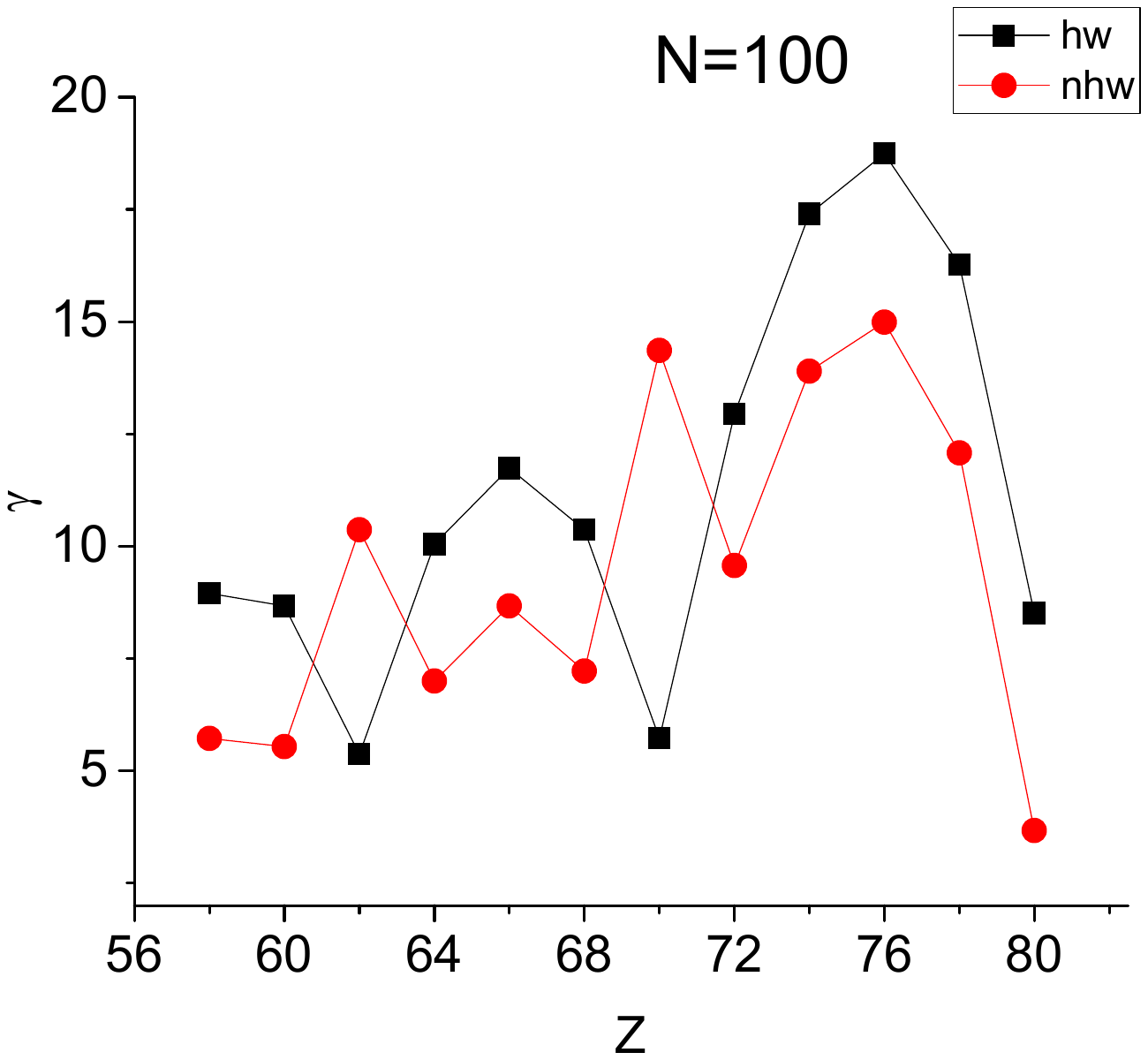} }
    
    \caption{Parameter-independent predictions for the collective variable $\gamma
    $ provided within the proxy-SU(3) symmetry by the highest weight irreducible representation (hw irrep) of SU(3) and by the next highest weight (nhw) irrep of SU(3) in the $Z=58$-80, $N=90$-100 region, taken from Tables II and III,  are plotted vs. $Z$. See Sec. \ref{nhw} for further discussion.}

\end{figure*}


\begin{figure*} [htb]

  {\includegraphics[width=75mm]{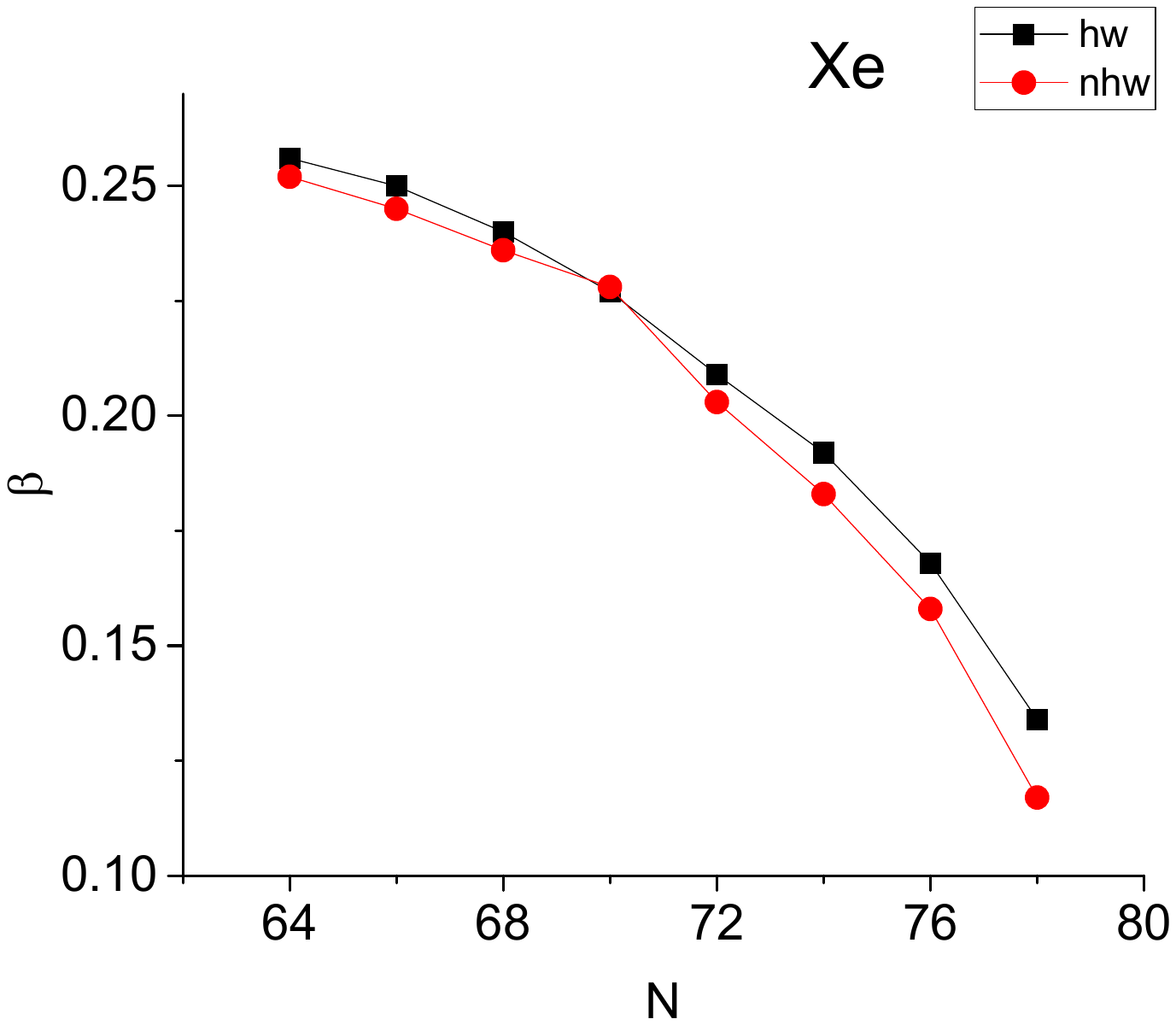}  \hspace{5mm}   \includegraphics[width=75mm]{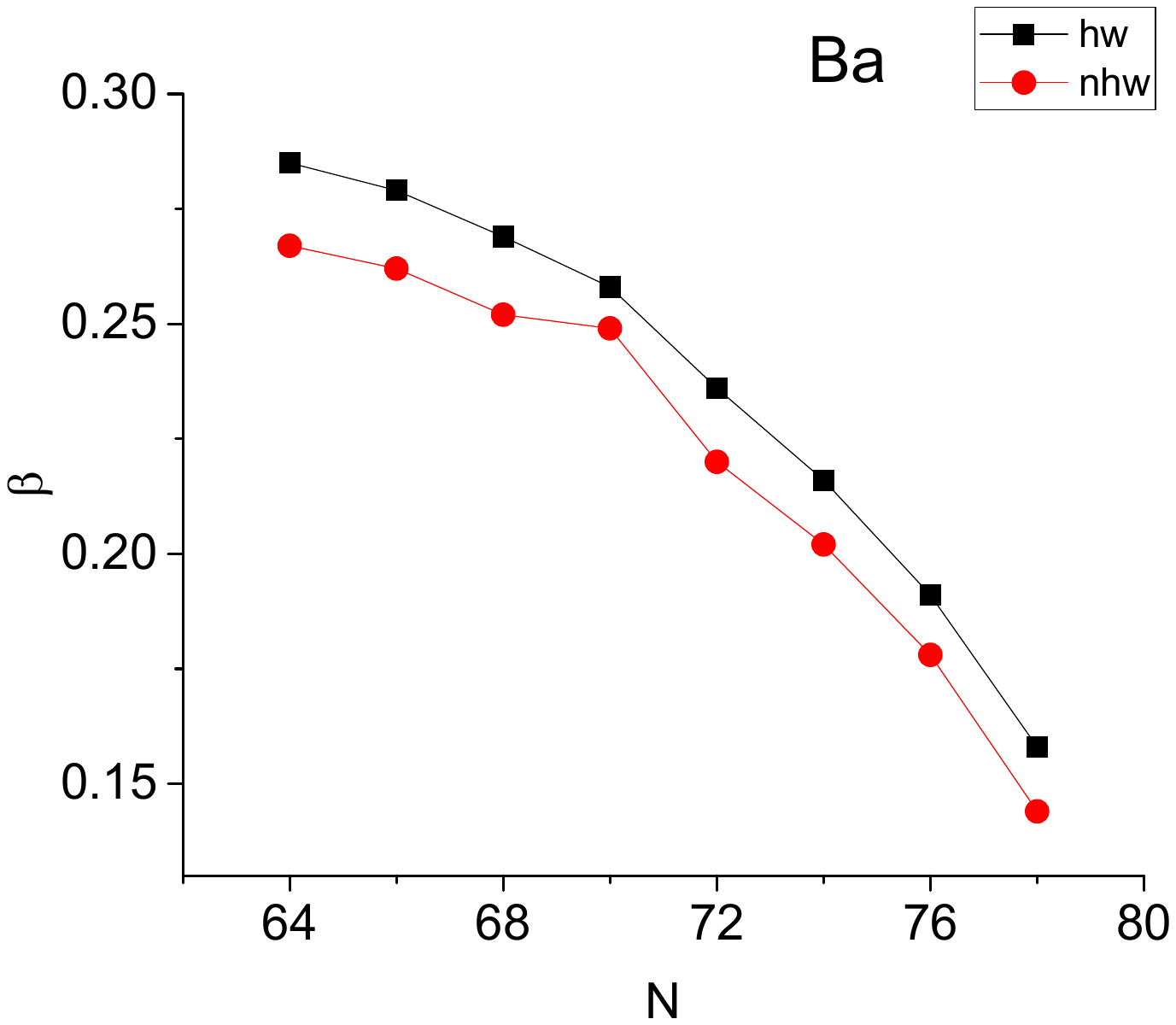} }
    {\includegraphics[width=75mm]{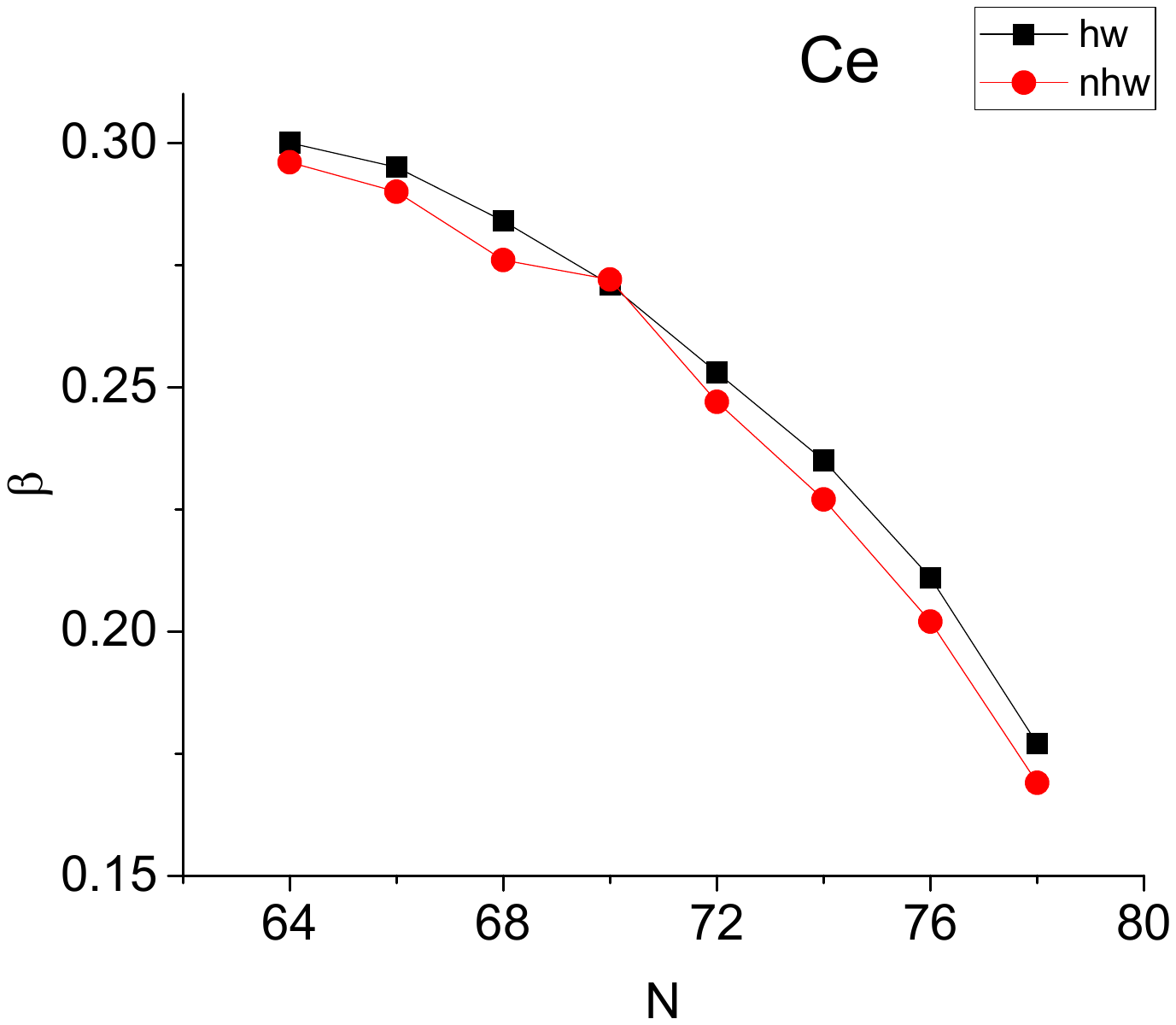} \hspace{5mm}   \includegraphics[width=75mm]{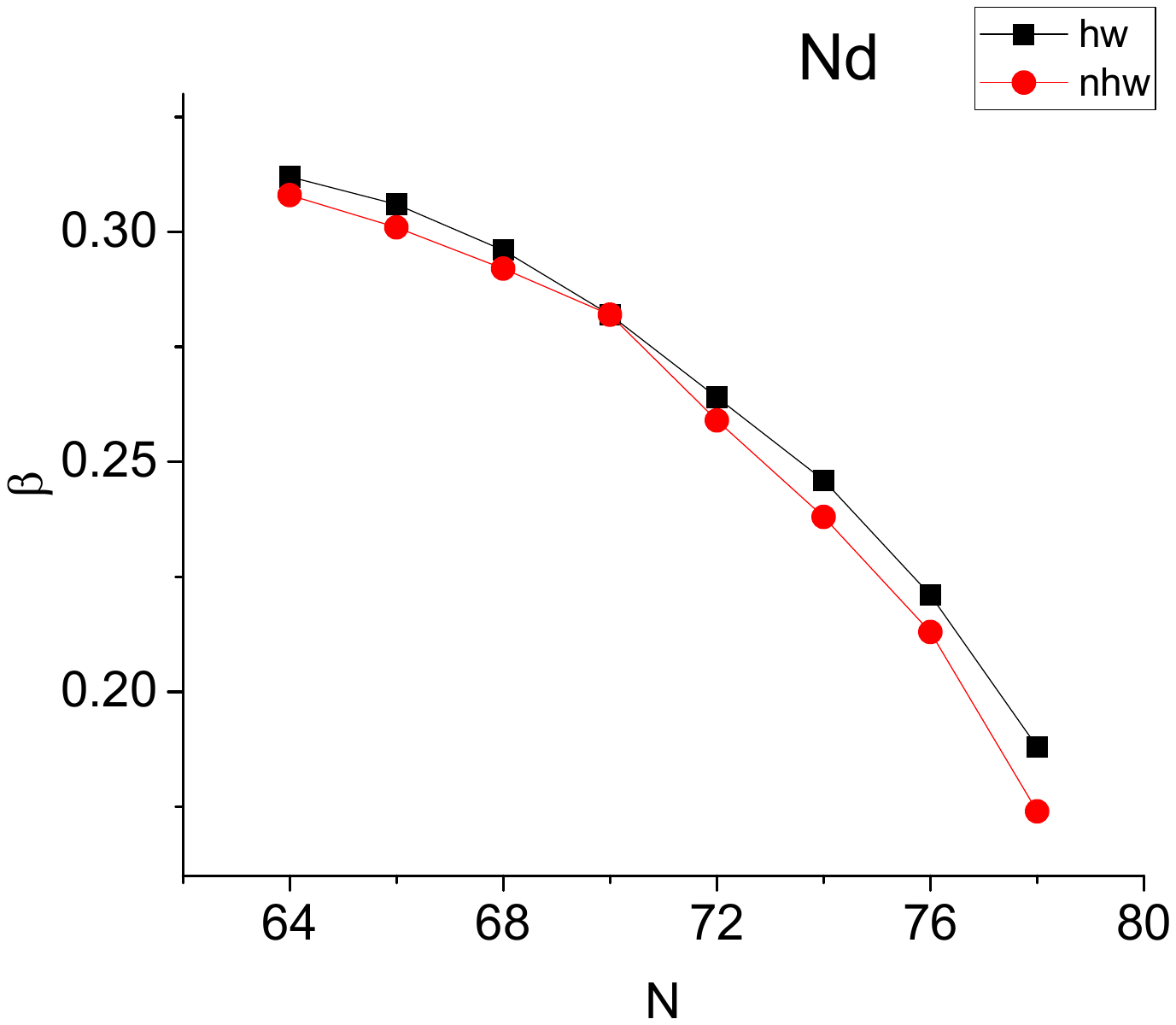} }
    {\includegraphics[width=75mm]{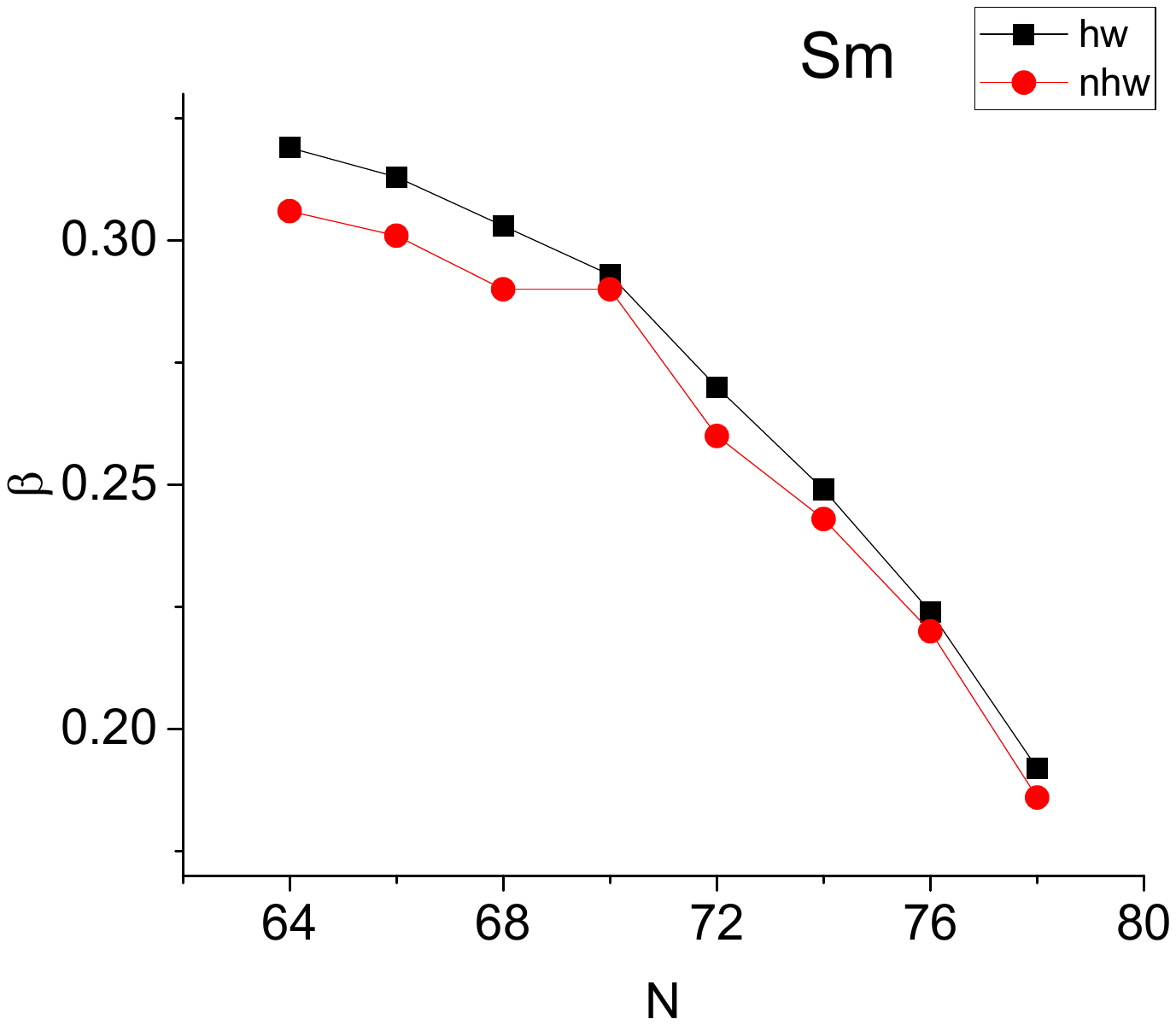} \hspace{5mm}    \includegraphics[width=75mm]{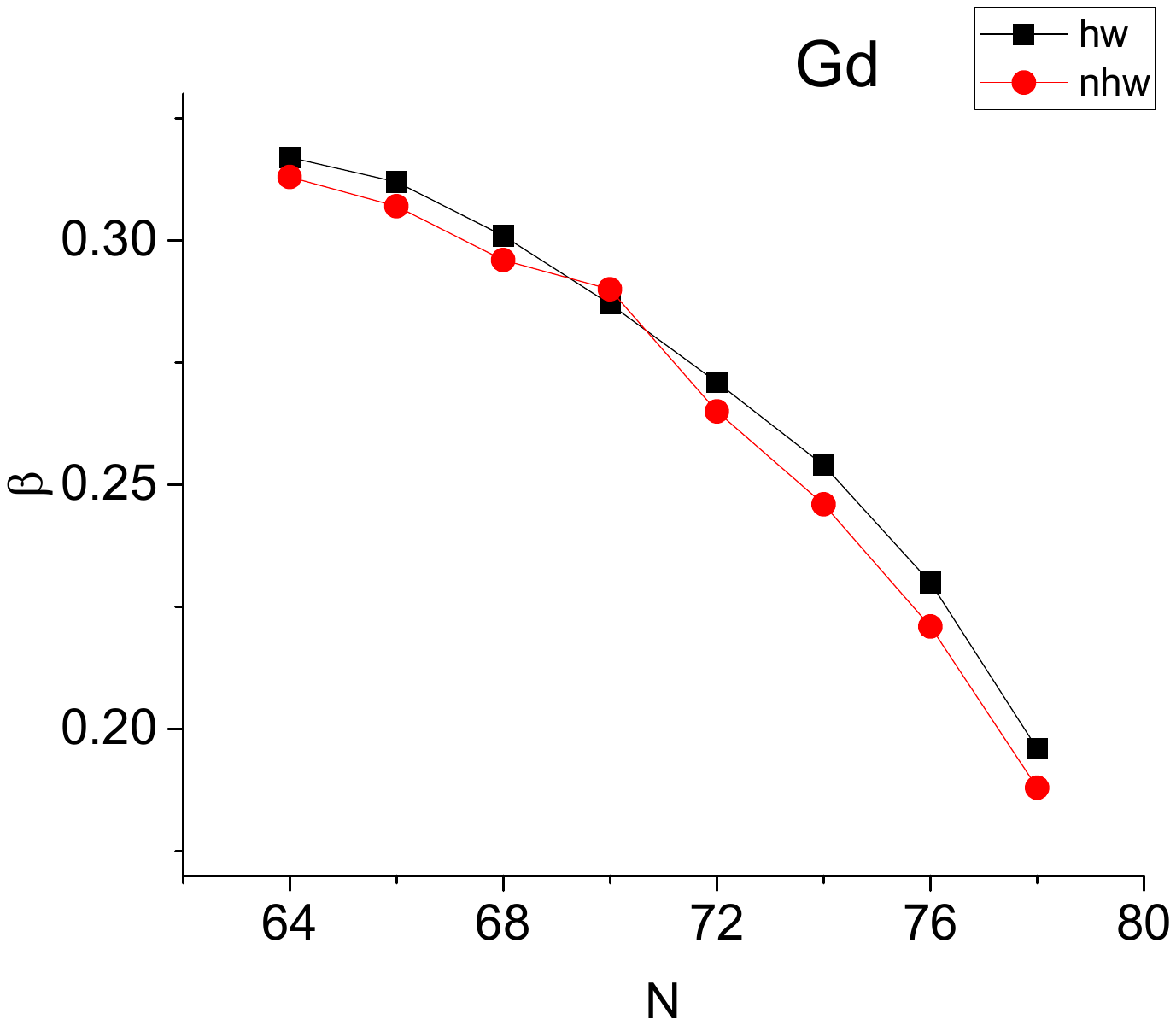} }

    \caption{Same as Fig. 3, but for the $Z=54$-64, $N=64$-78 region, with predictions taken from Table IV. See Sec. \ref{nhw} for further discussion.} 
    
\end{figure*}


\begin{figure*} [htb]

    {\includegraphics[width=75mm]{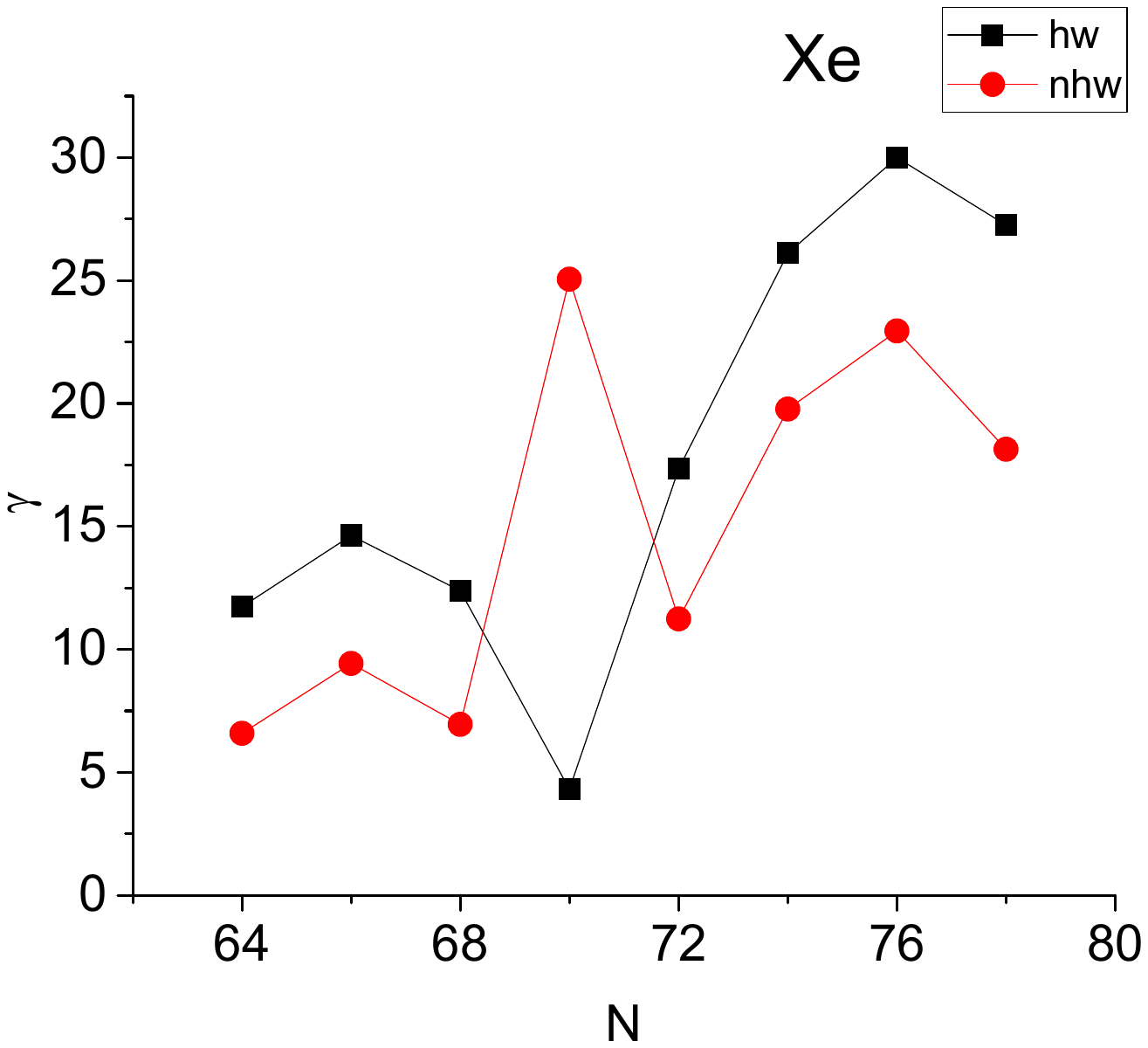}  \hspace{5mm}   \includegraphics[width=75mm]{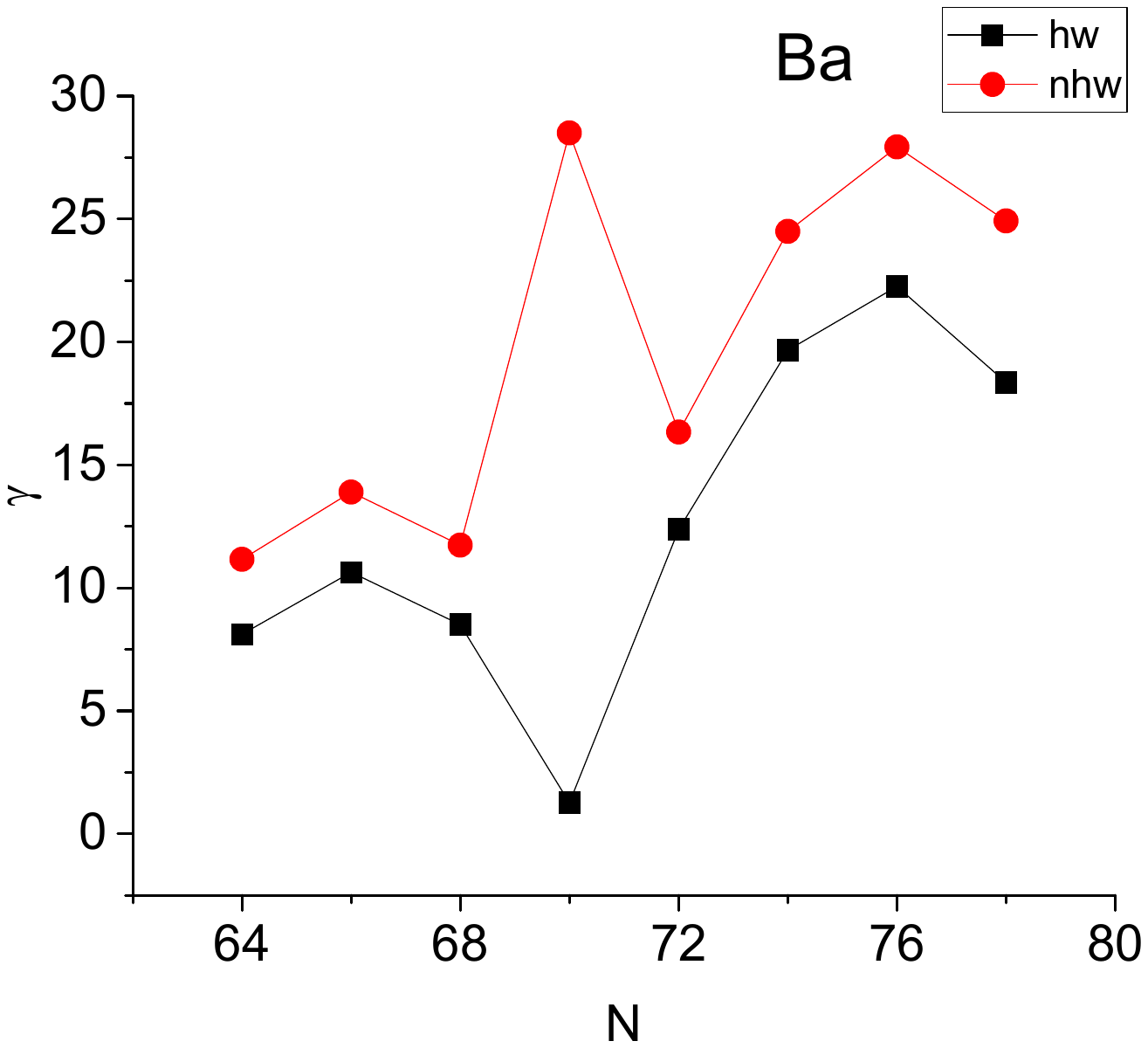} }
    {\includegraphics[width=75mm]{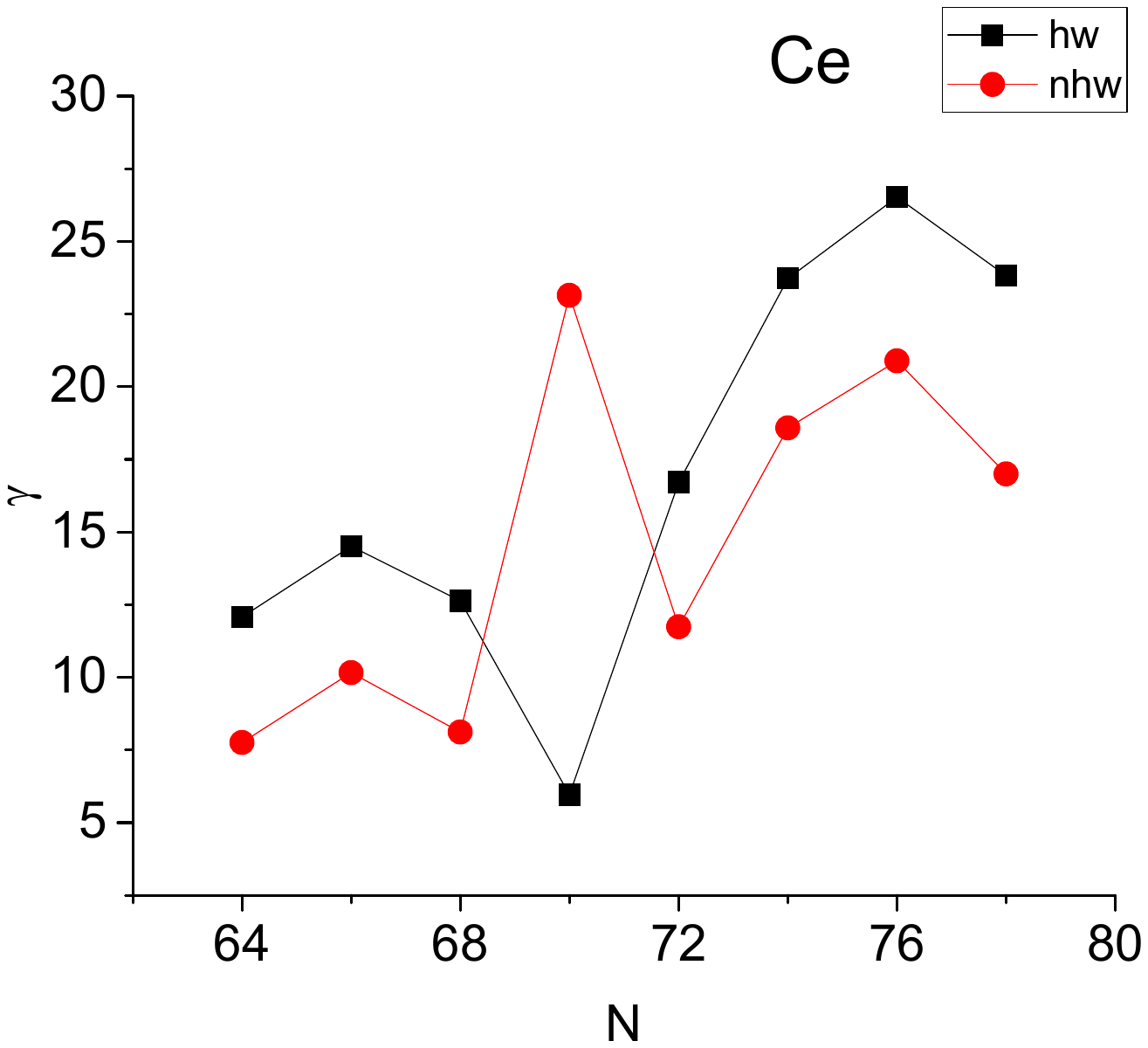} \hspace{5mm}   \includegraphics[width=75mm]{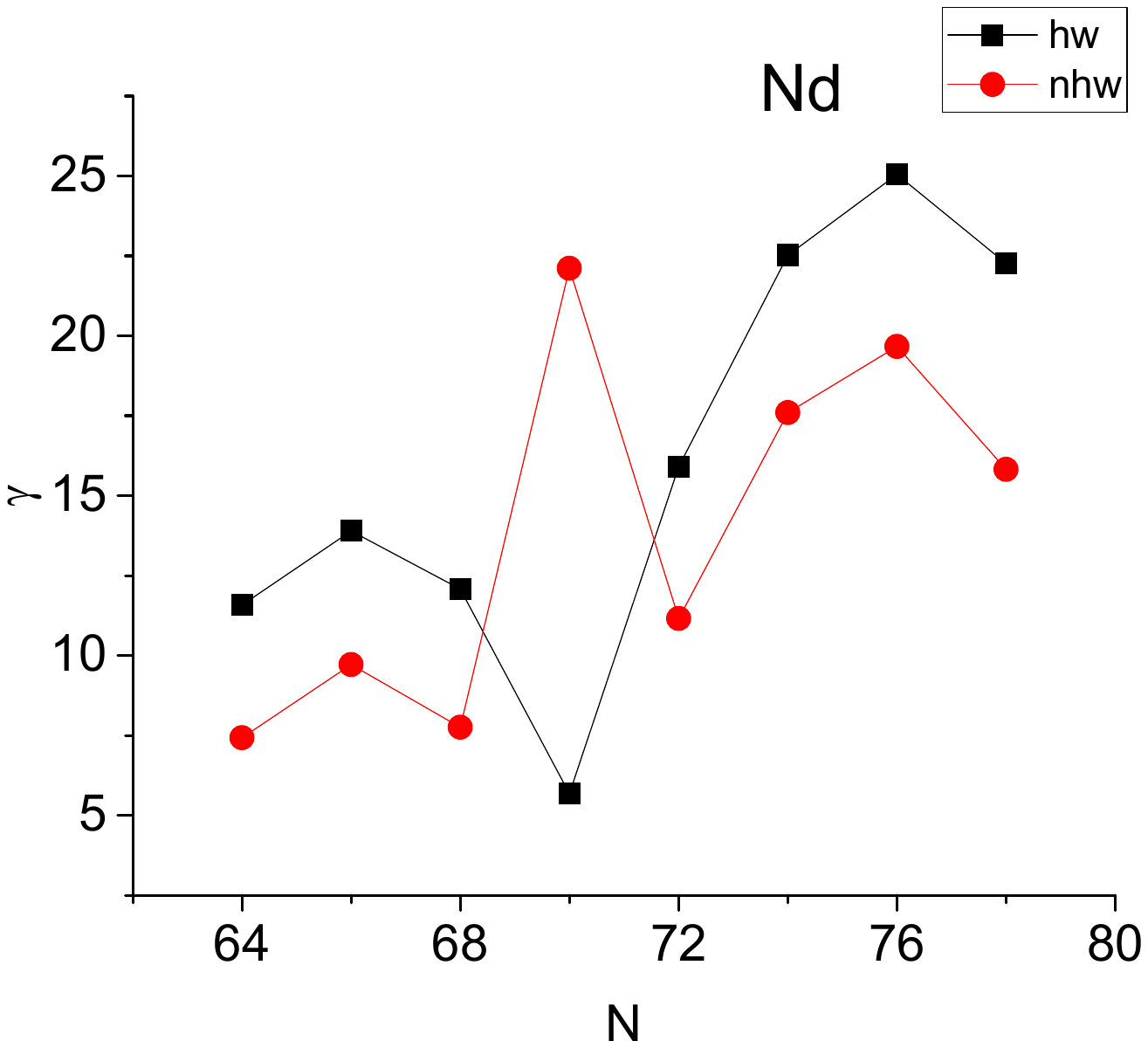} }
    {\includegraphics[width=75mm]{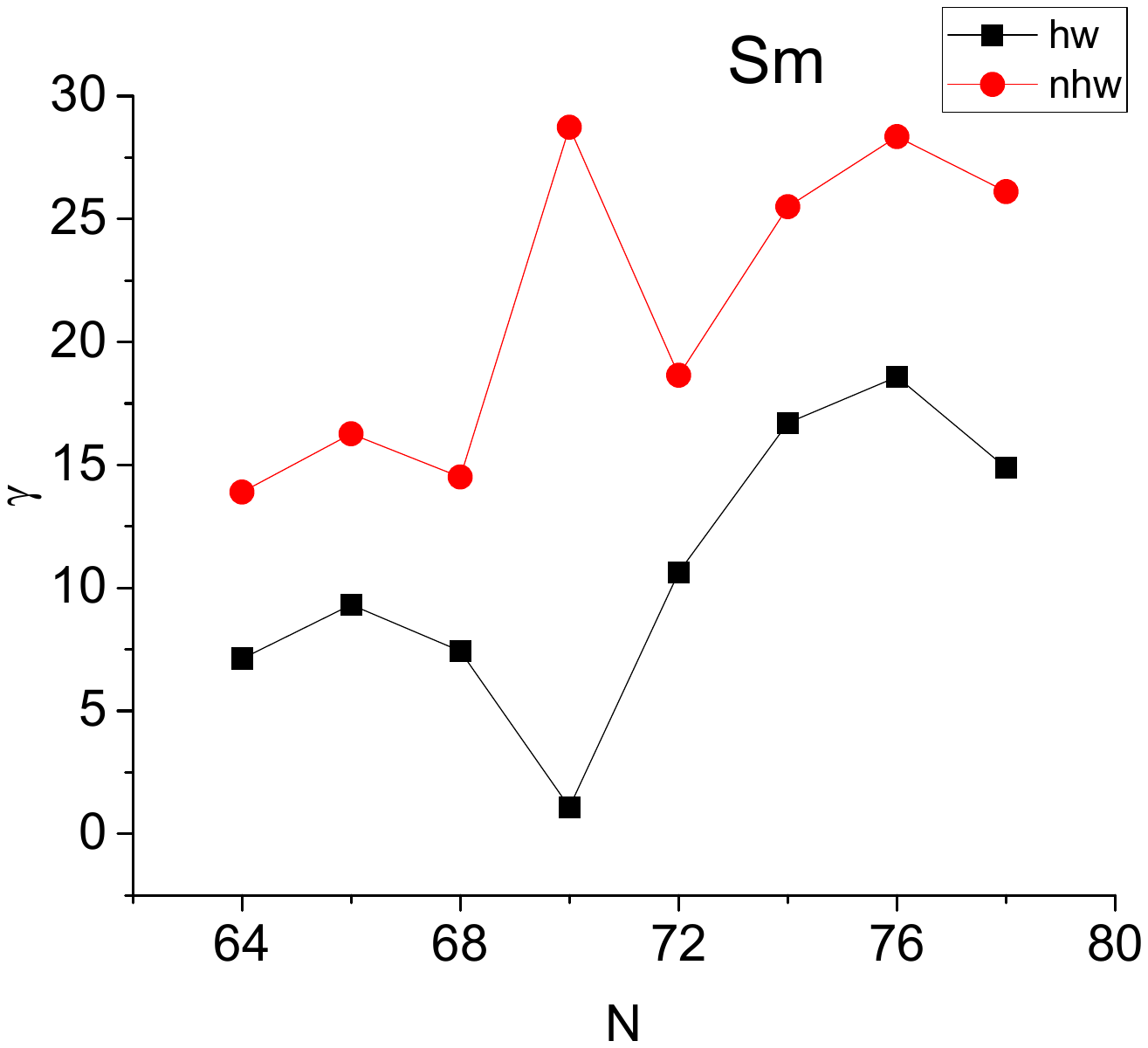} \hspace{5mm}    \includegraphics[width=75mm]{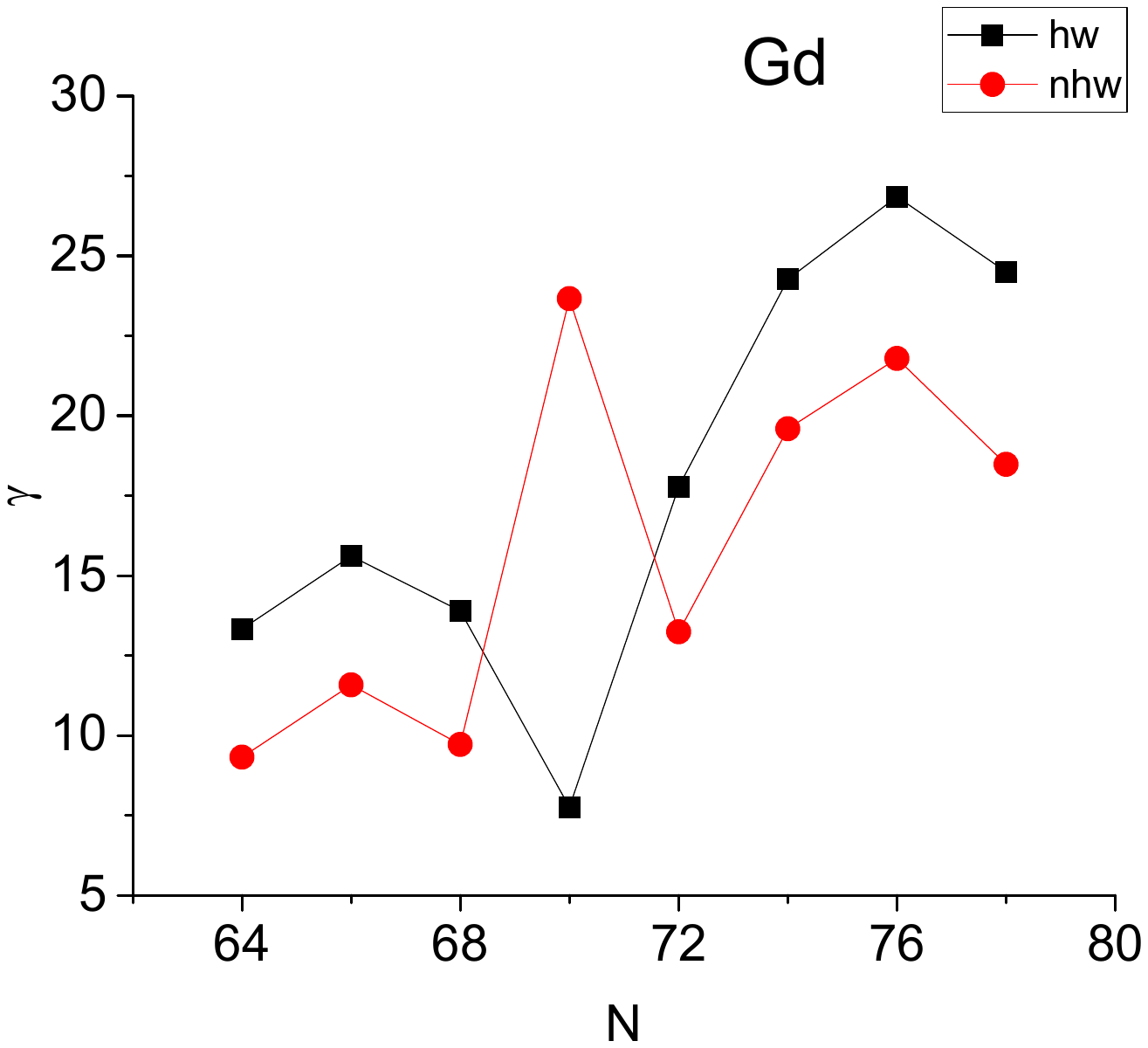} }

    \caption{Same as Fig. 4, but for the $Z=54$-64, $N=64$-78 region, with predictions taken from Table IV. See Sec. \ref{nhw} for further discussion.} 
    
\end{figure*}


\begin{figure*} [htb]

  {\includegraphics[width=75mm]{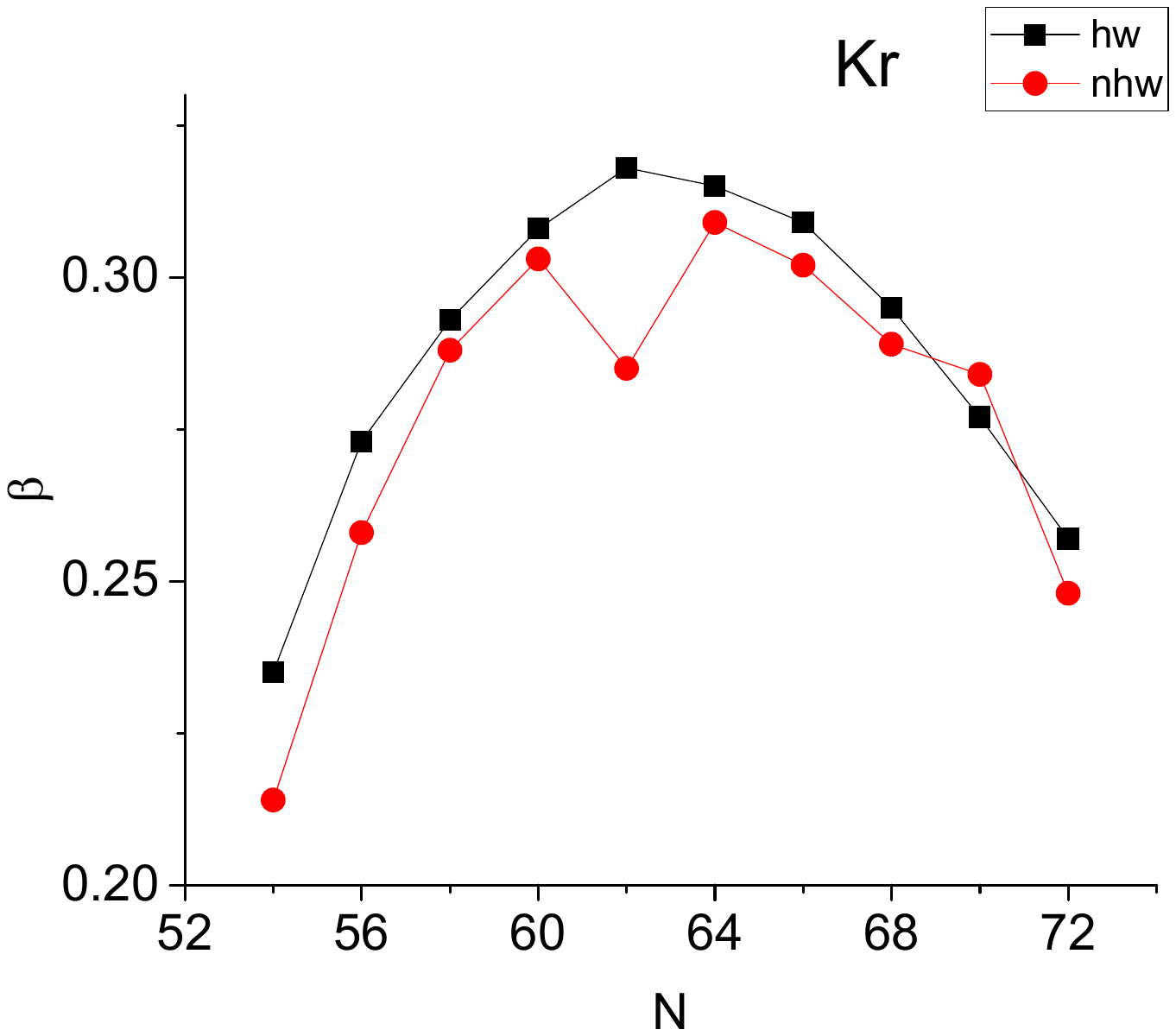}  \hspace{5mm}   \includegraphics[width=75mm]{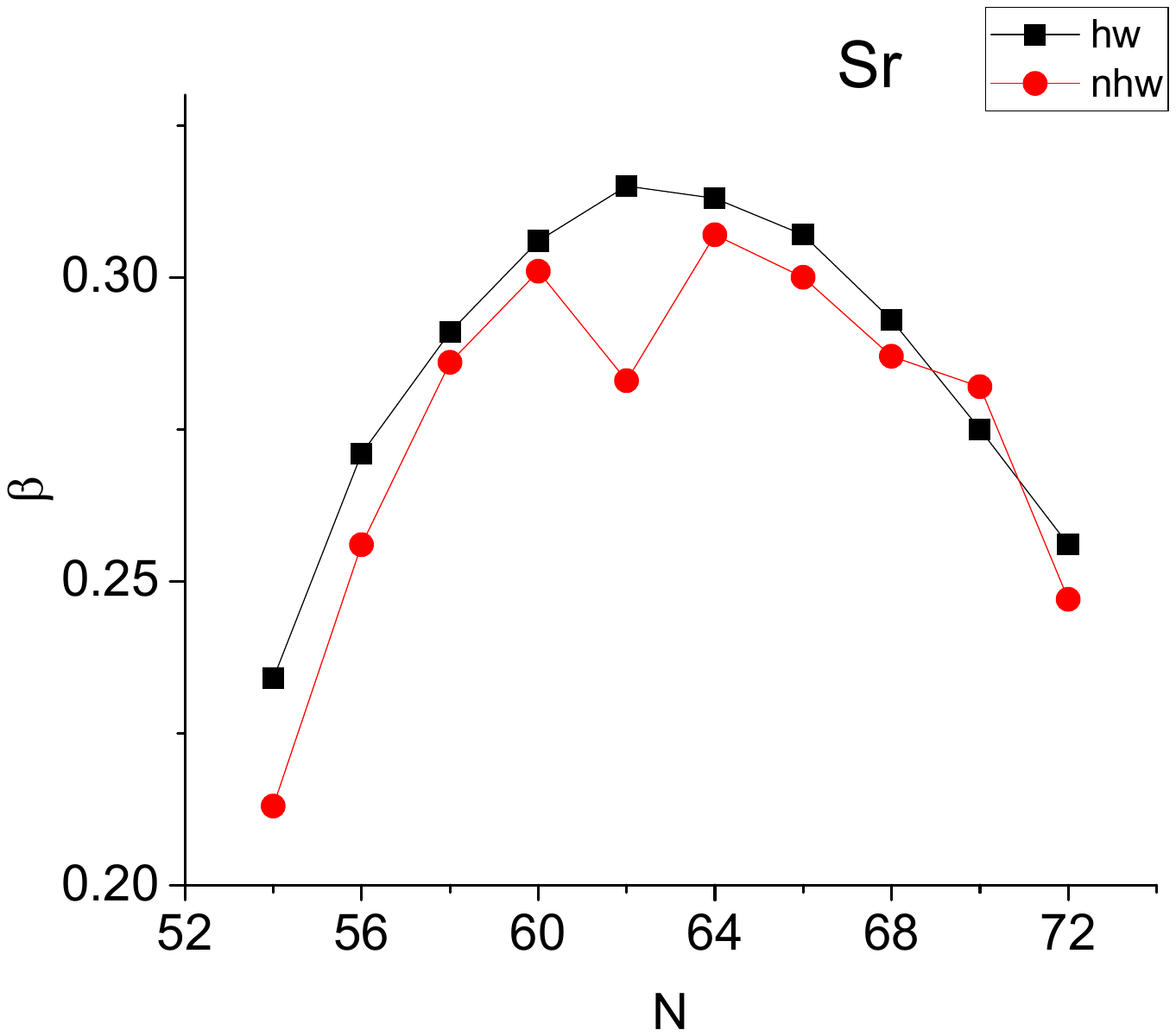} }
    {\includegraphics[width=75mm]{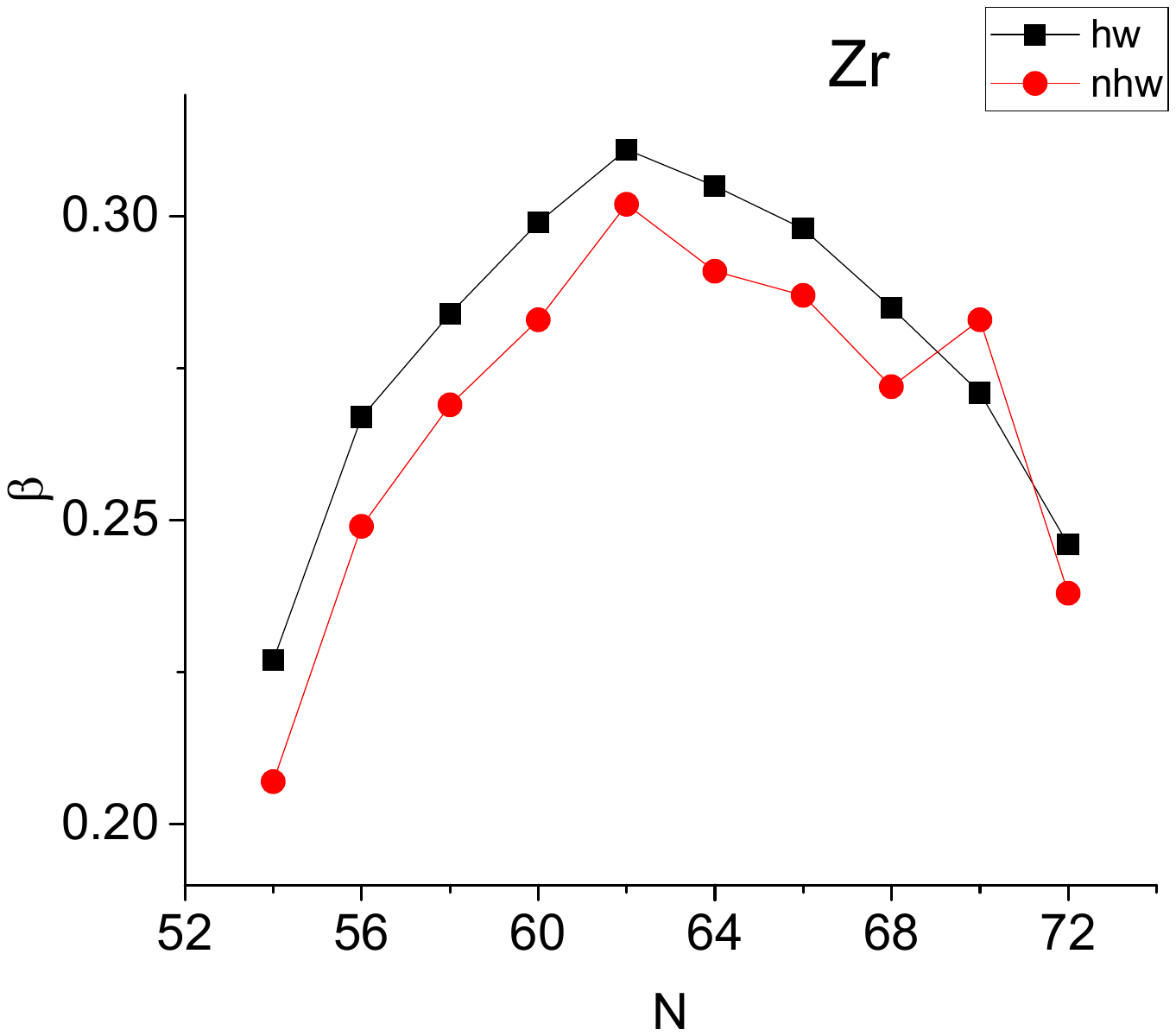} \hspace{5mm}   \includegraphics[width=75mm]{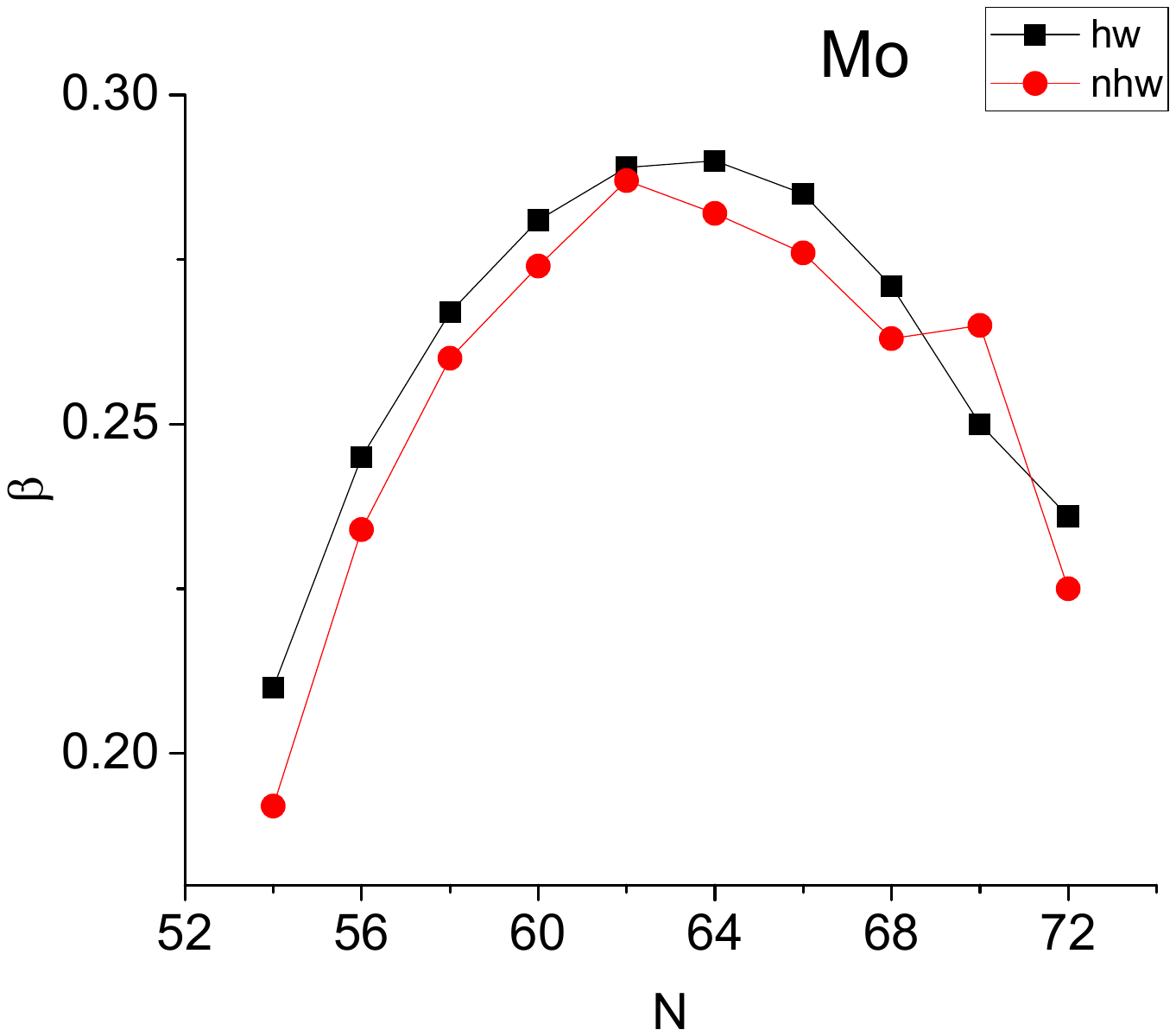} }
    {\includegraphics[width=75mm]{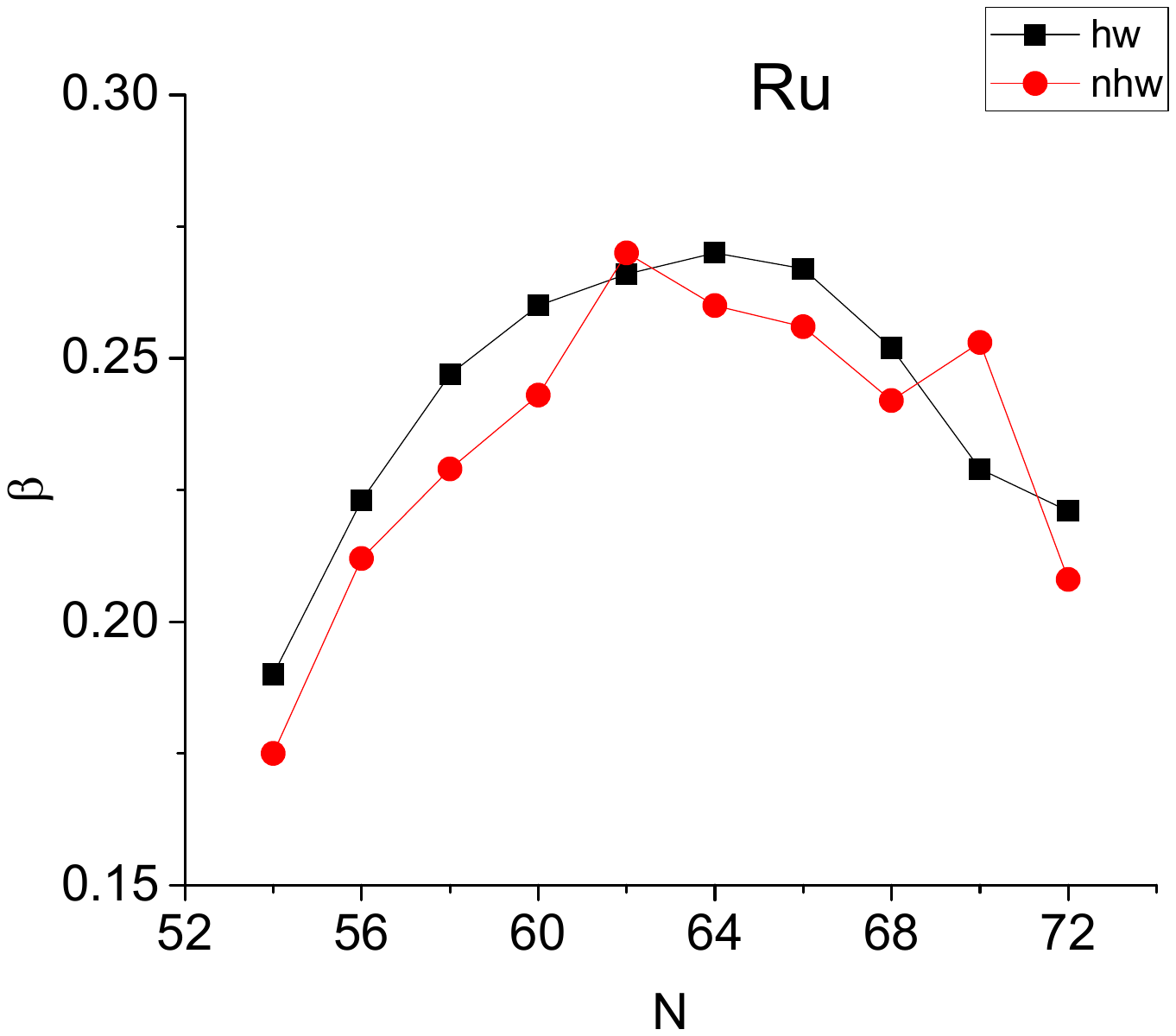} \hspace{5mm}    \includegraphics[width=75mm]{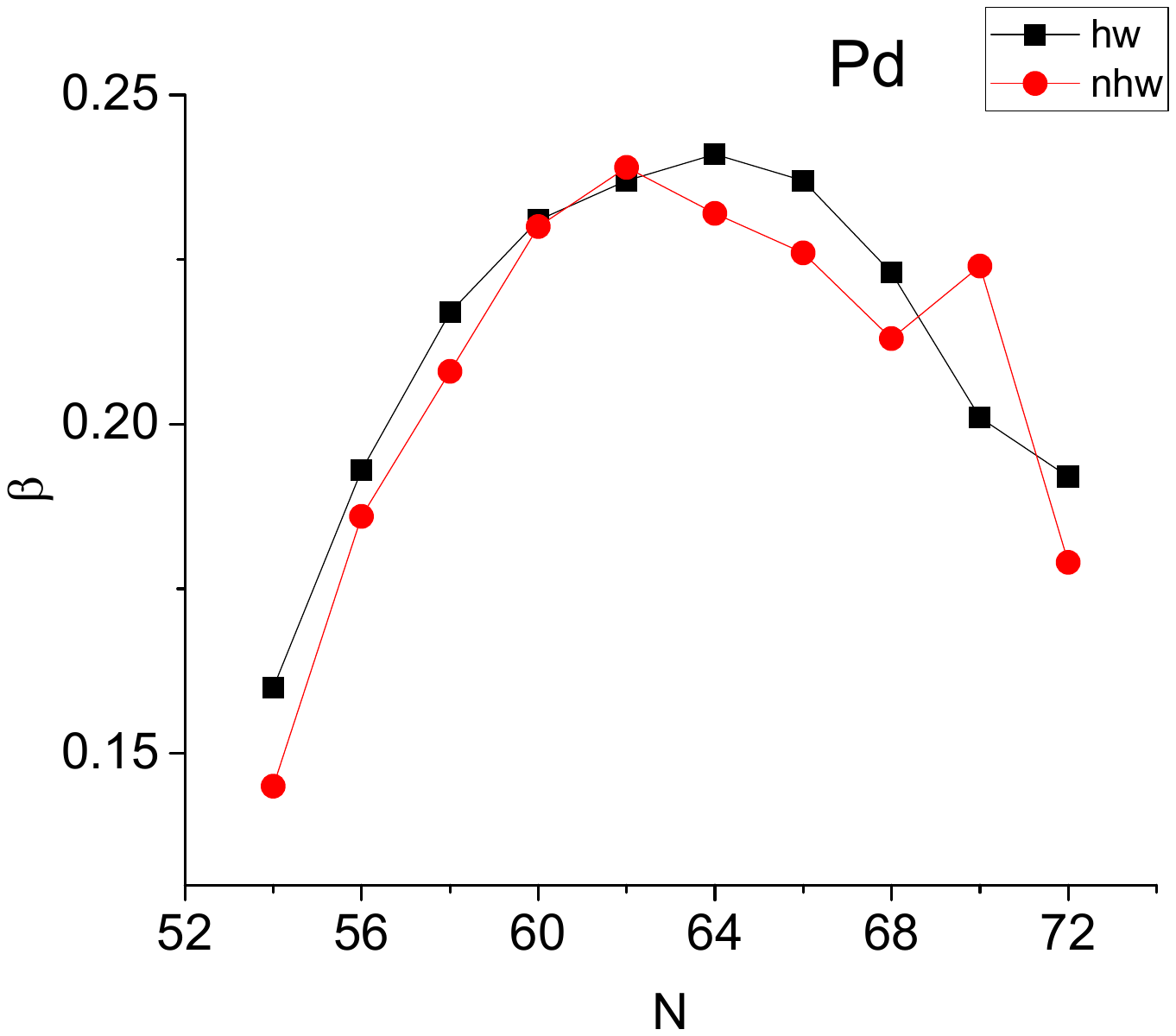} }

    \caption{Same as Fig. 3, but for the $Z=36$-46, $N=54$-72 region, with predictions taken from Table V. See Sec. \ref{nhw} for further discussion.} 
    
\end{figure*}


\begin{figure*} [htb]

    {\includegraphics[width=75mm]{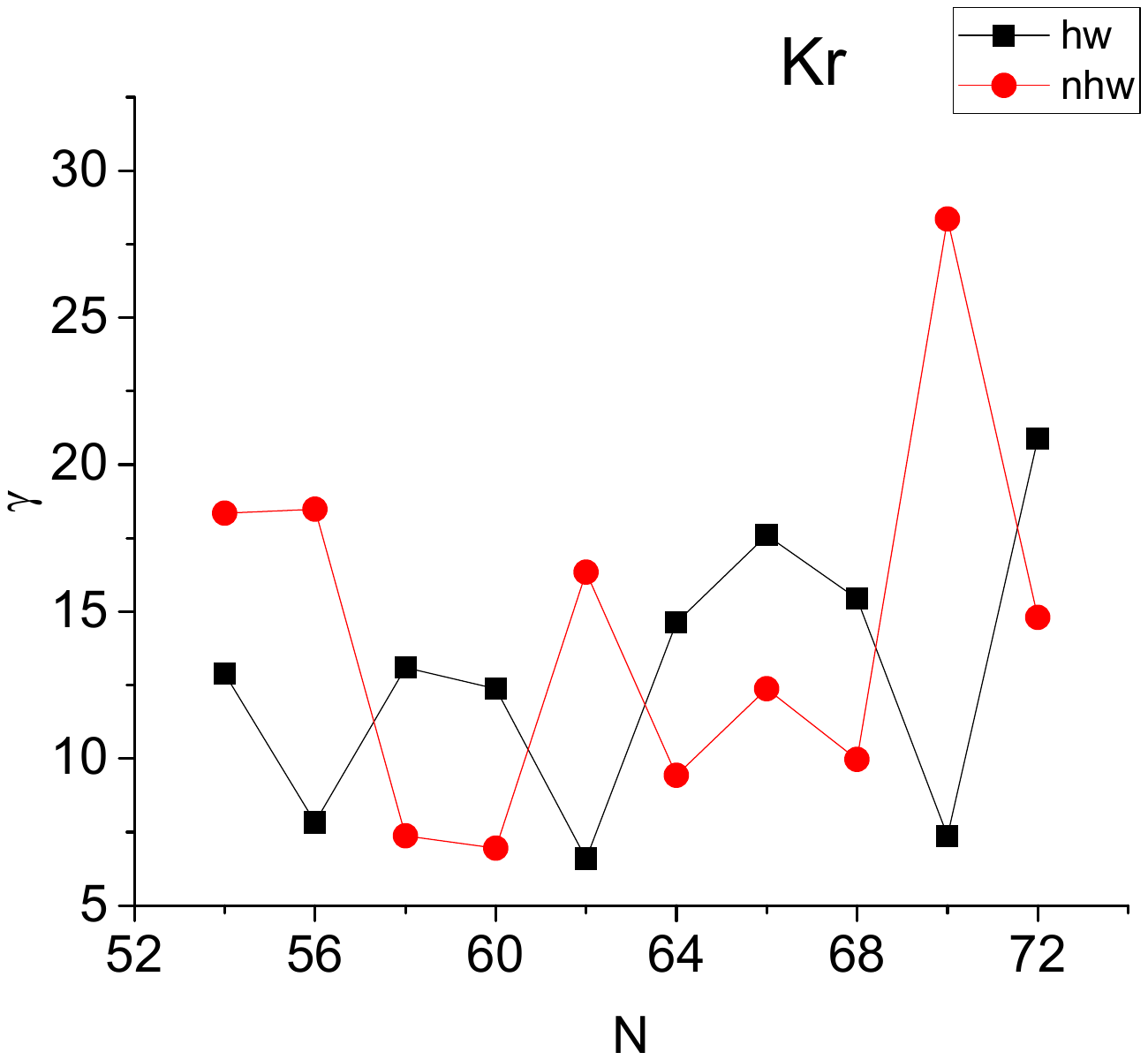}  \hspace{5mm}   \includegraphics[width=75mm]{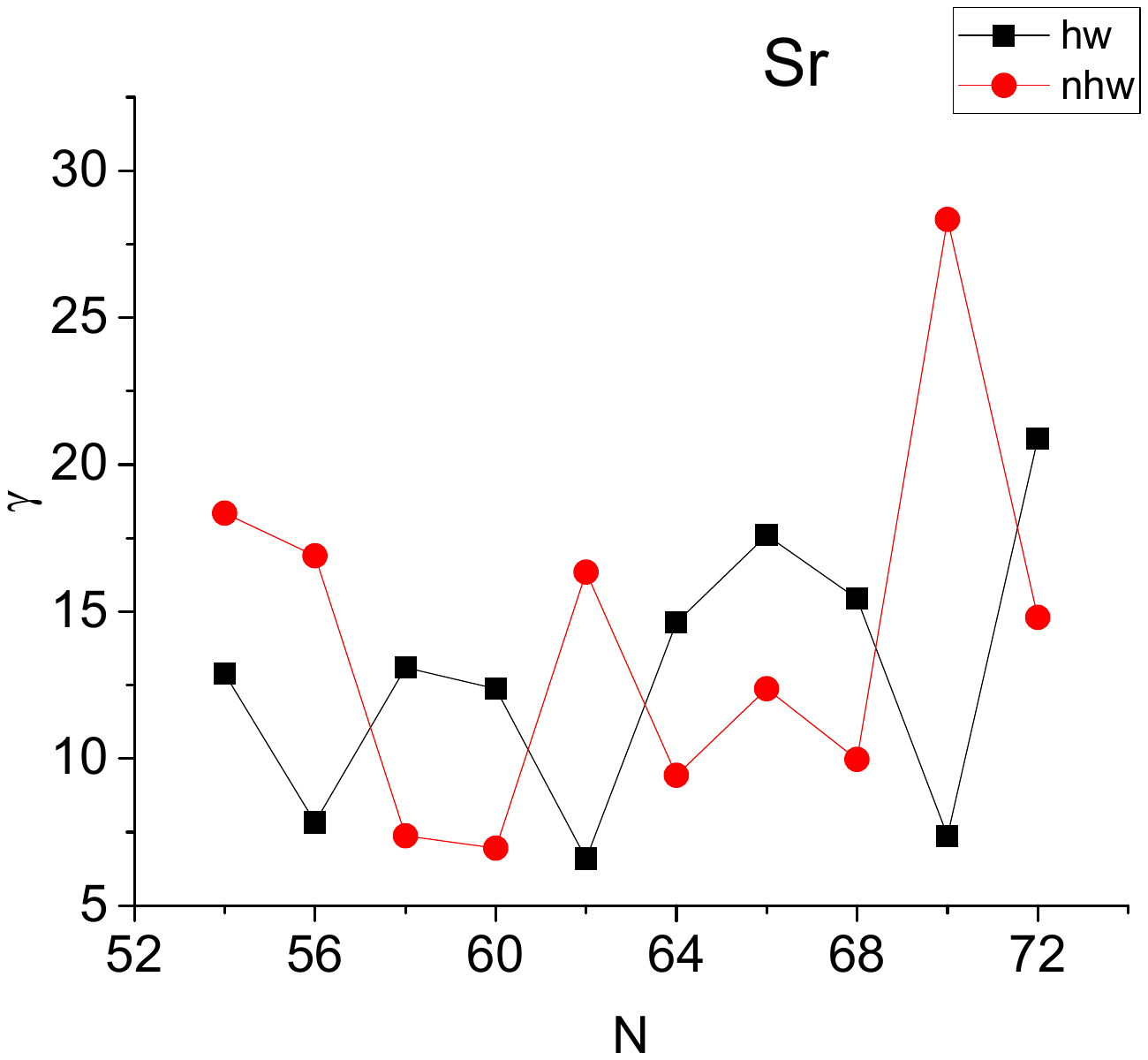} }
    {\includegraphics[width=75mm]{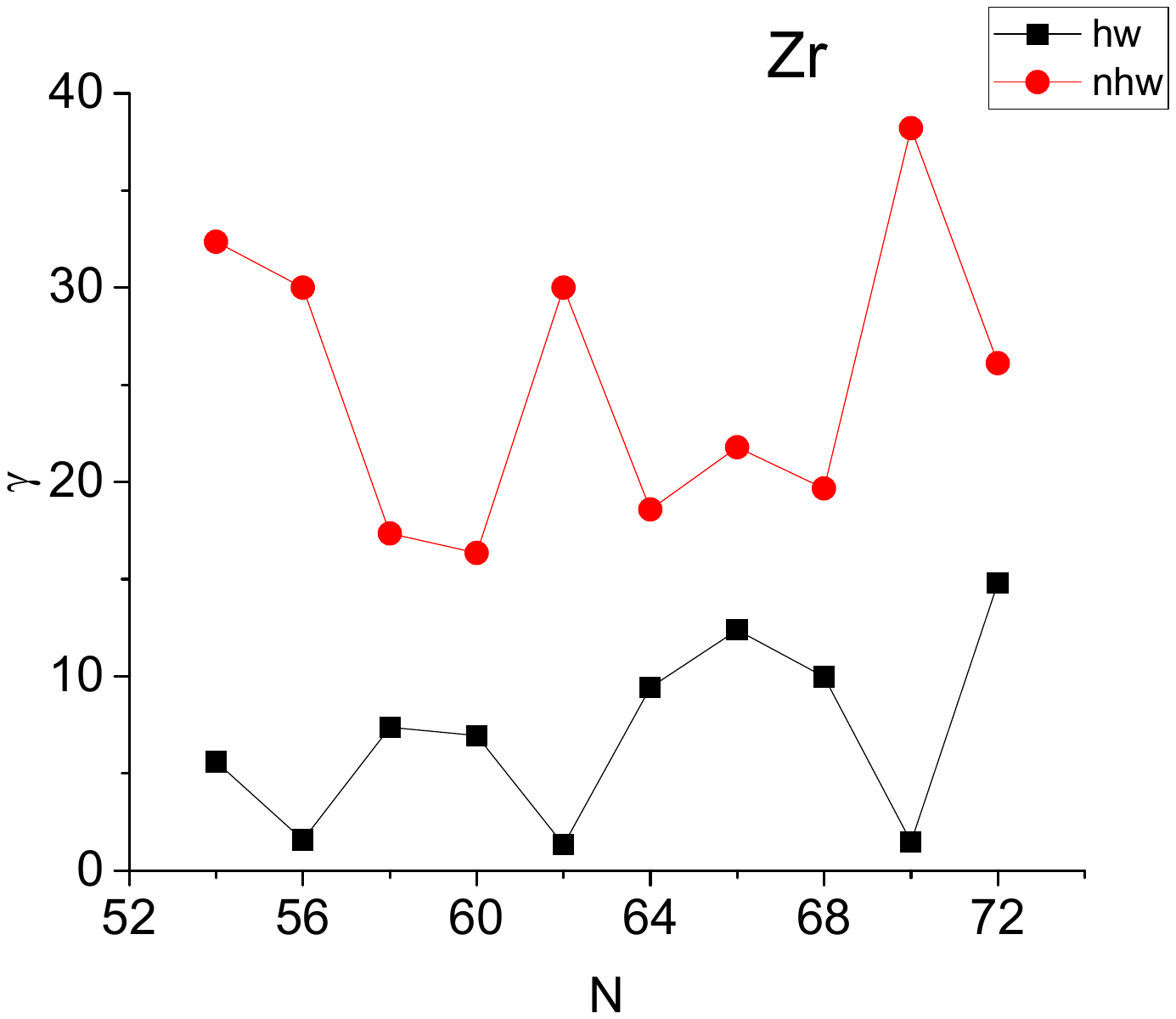} \hspace{5mm}   \includegraphics[width=75mm]{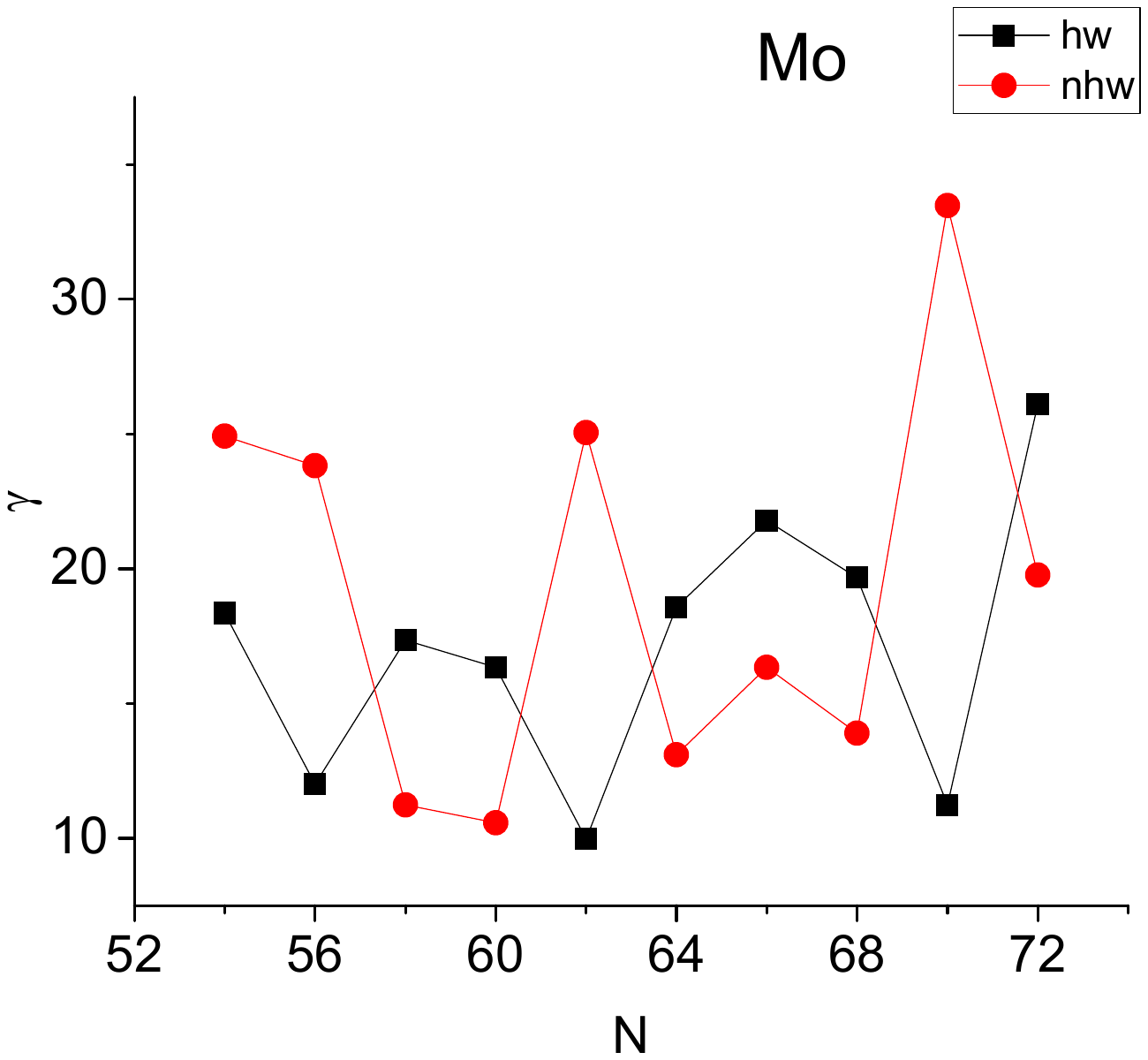} }
    {\includegraphics[width=75mm]{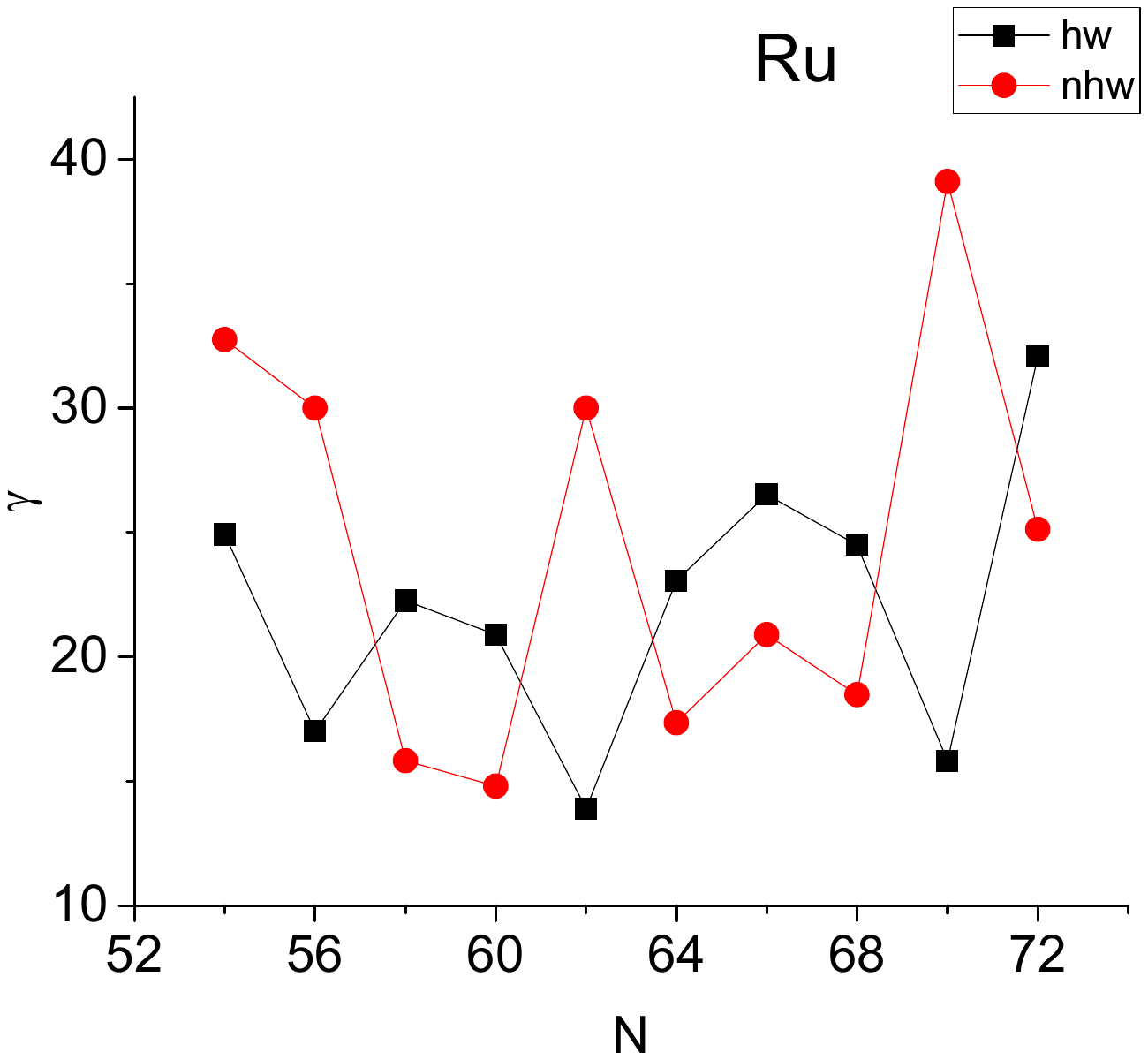} \hspace{5mm}    \includegraphics[width=75mm]{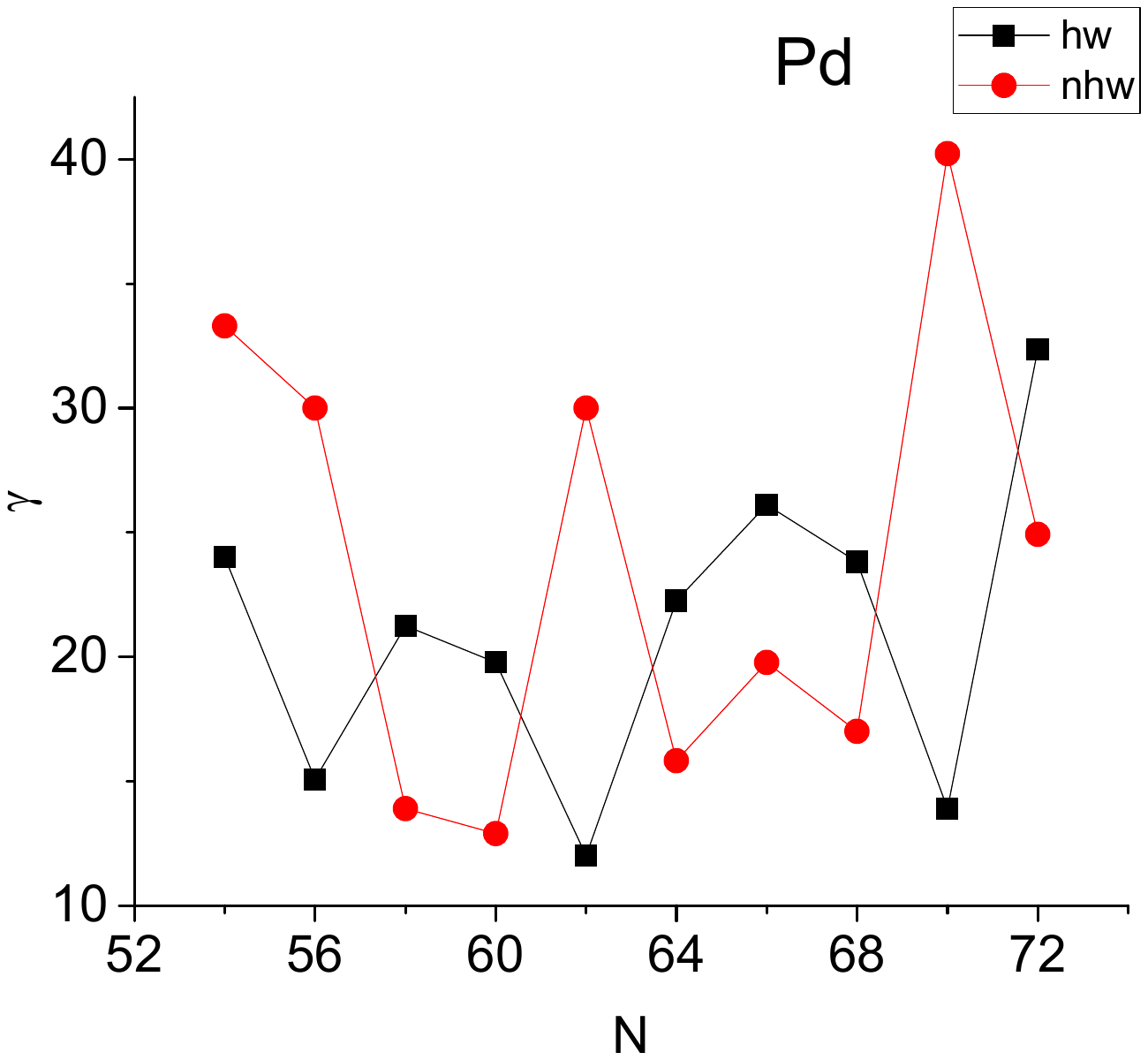} }

    \caption{Same as Fig. 4, but for the $Z=36$-46, $N=54$-72 region, with predictions taken from Table V. See Sec. \ref{nhw} for further discussion.} 
    
\end{figure*}


\begin{figure*} [htb]

  {\includegraphics[width=75mm]{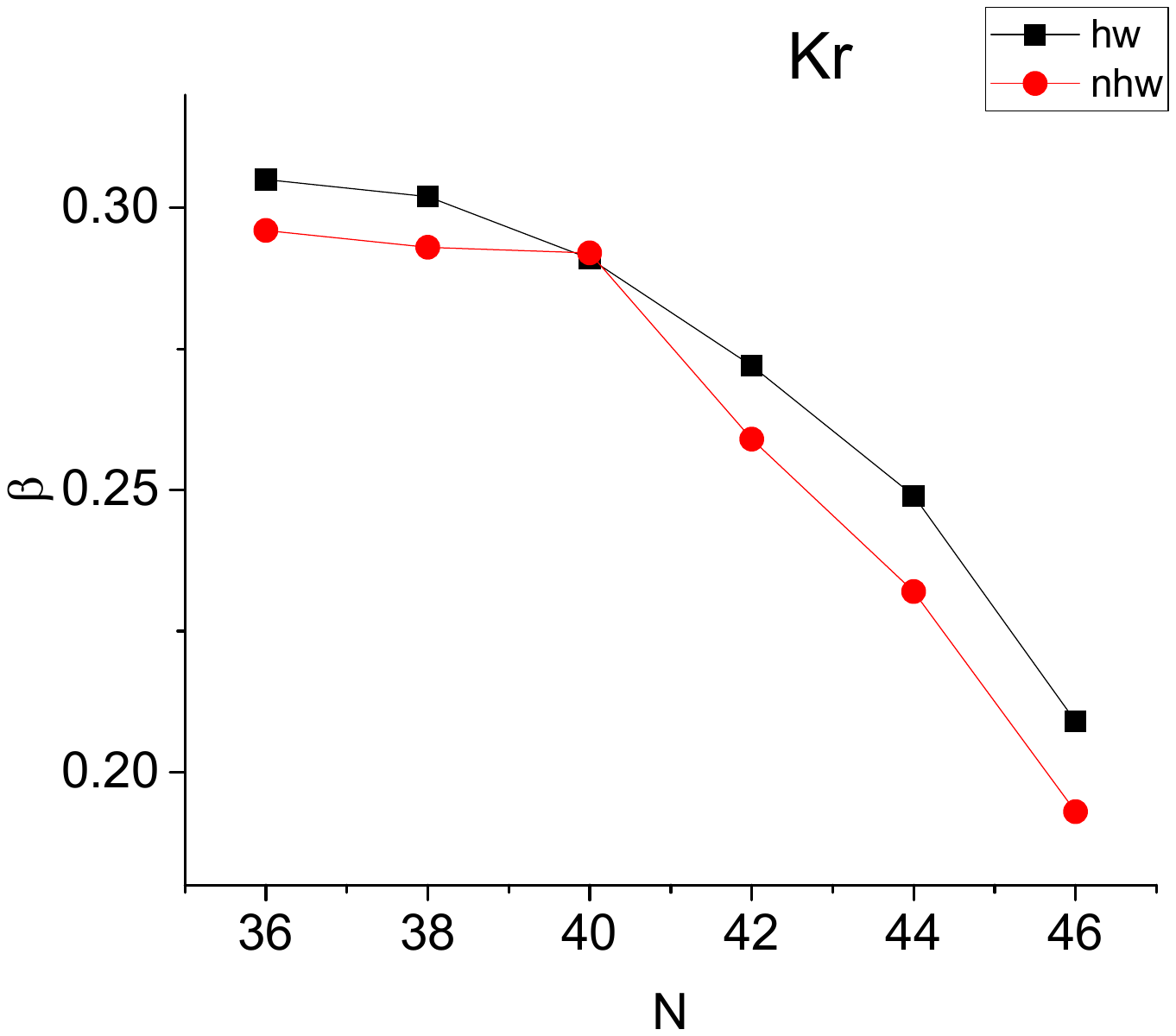}  \hspace{5mm}   \includegraphics[width=75mm]{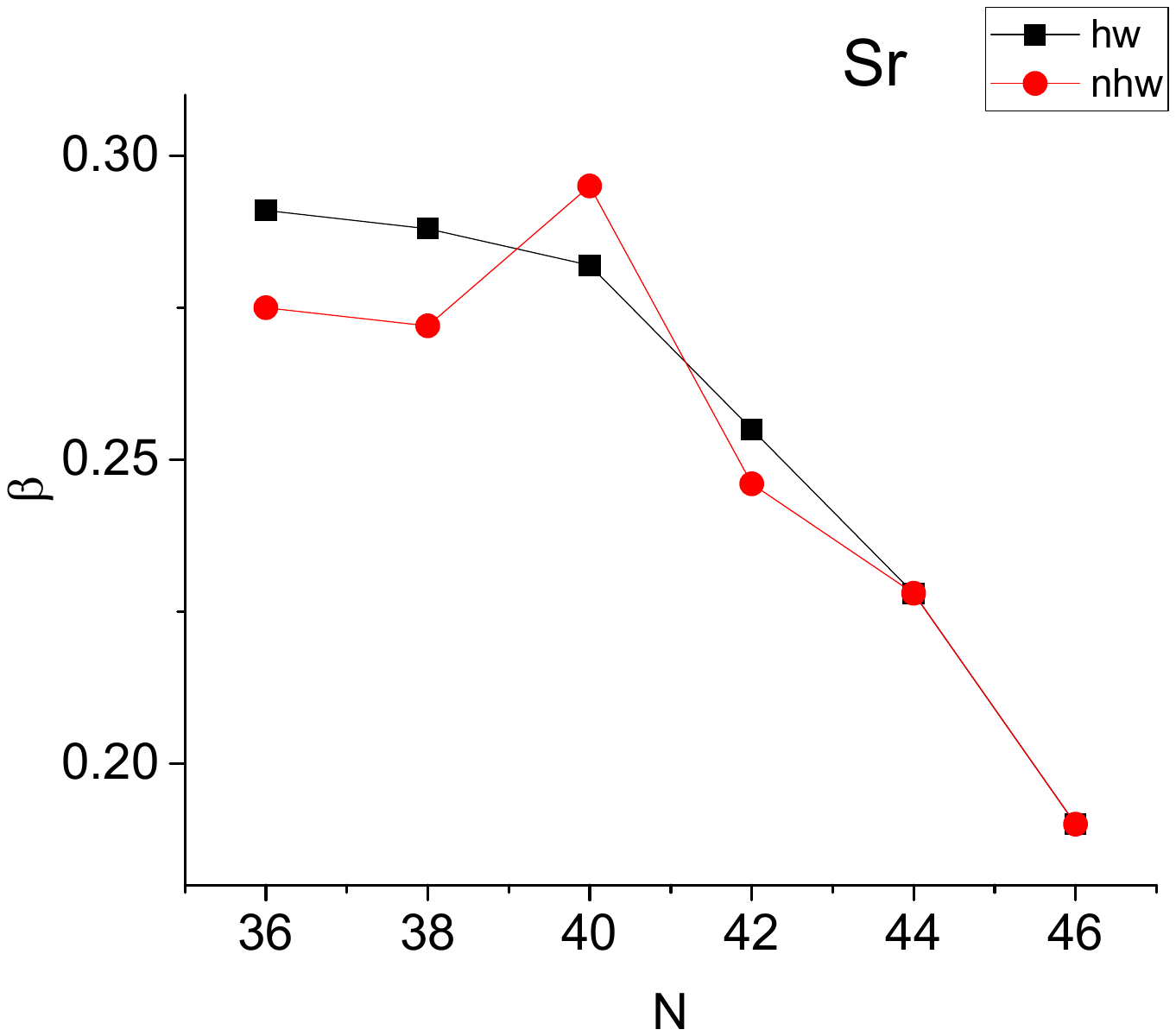} }
    {\includegraphics[width=75mm]{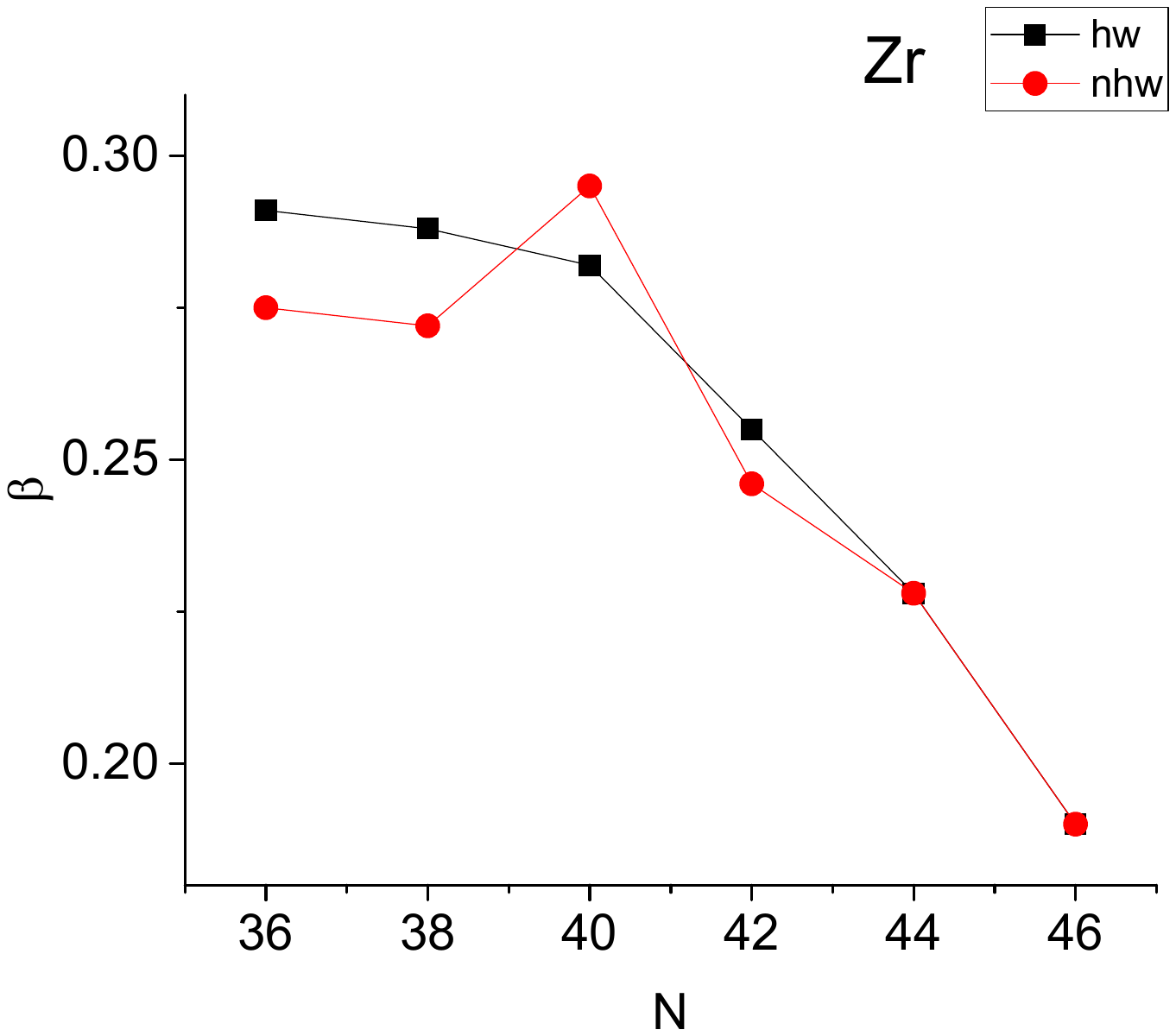} \hspace{5mm}   \includegraphics[width=75mm]{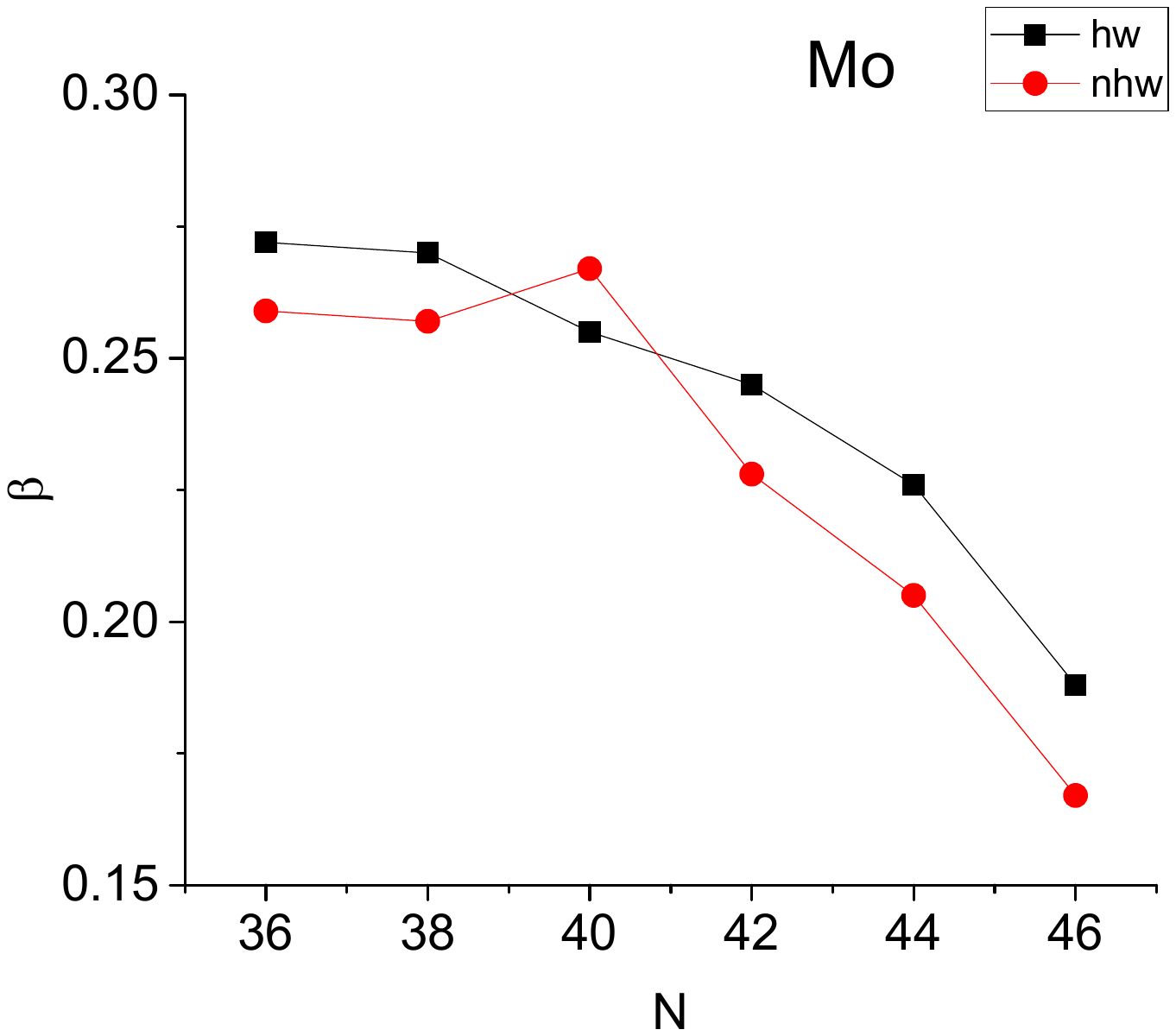} }
    {\includegraphics[width=75mm]{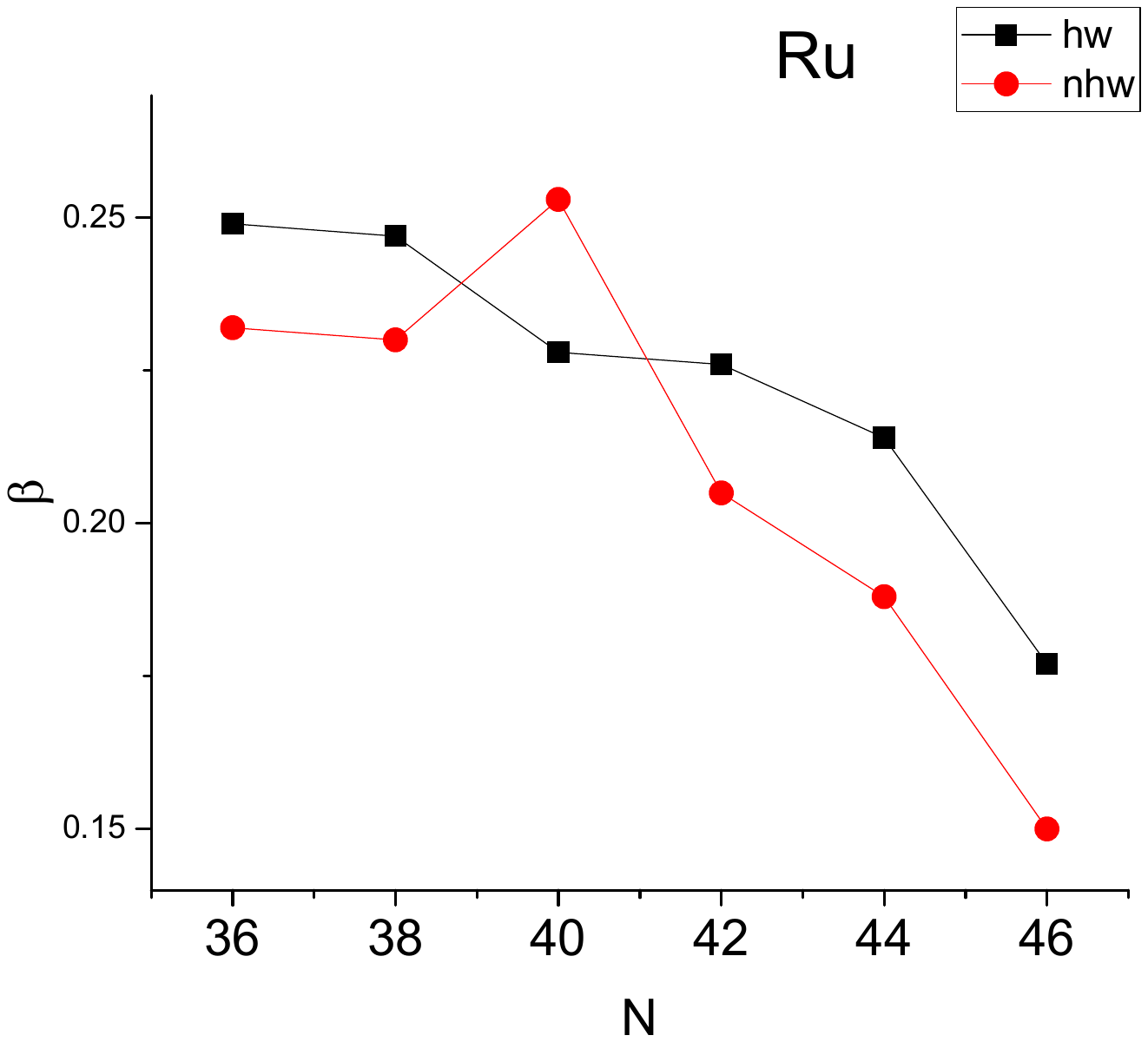} \hspace{5mm}    \includegraphics[width=75mm]{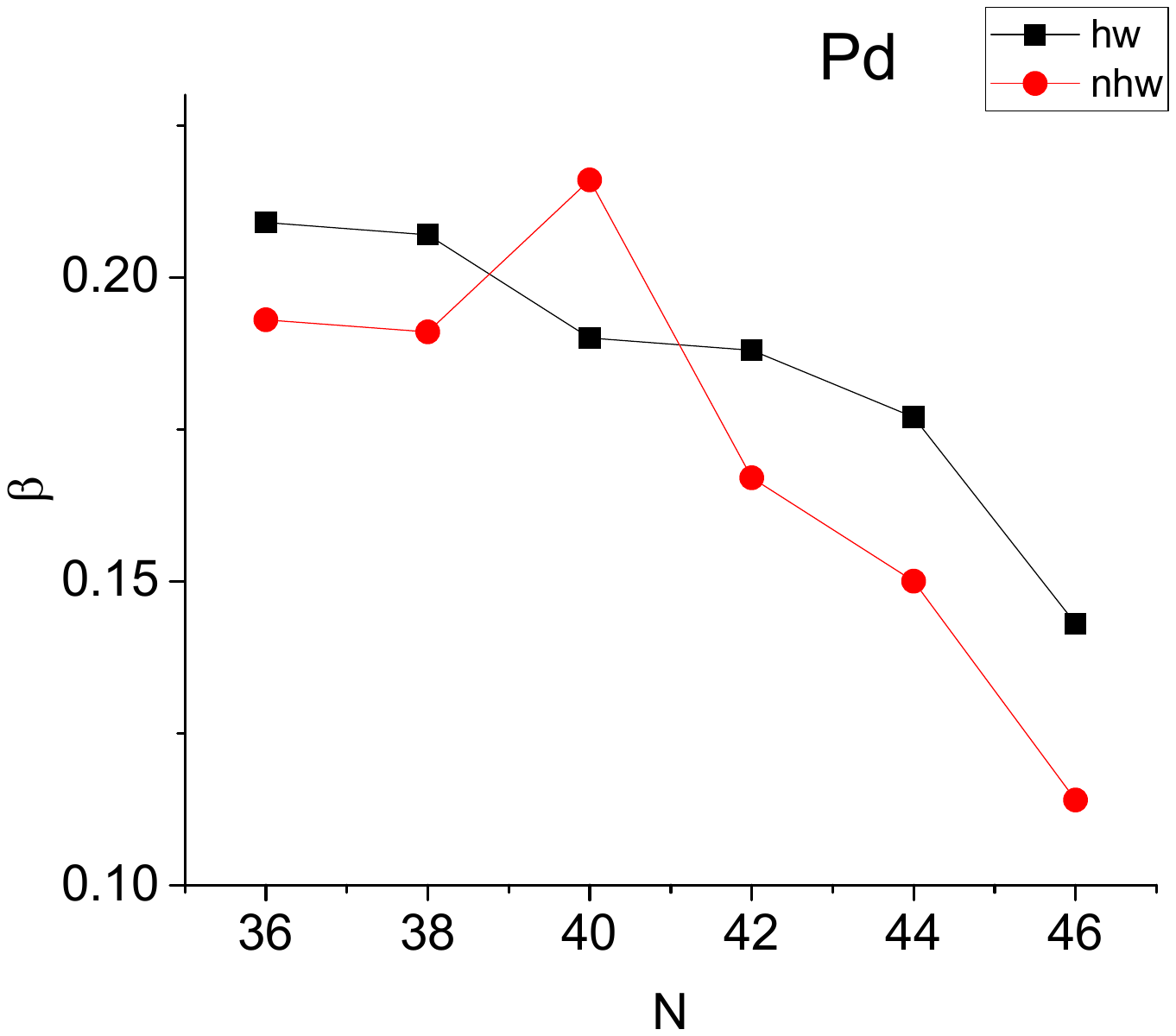} }

    \caption{Same as Fig. 3, but for the $Z=36$-46, $N=36$-46 region, with predictions taken from Table VI. See Sec. \ref{nhw} for further discussion.} 
    
\end{figure*}


\begin{figure*} [htb]

    {\includegraphics[width=75mm]{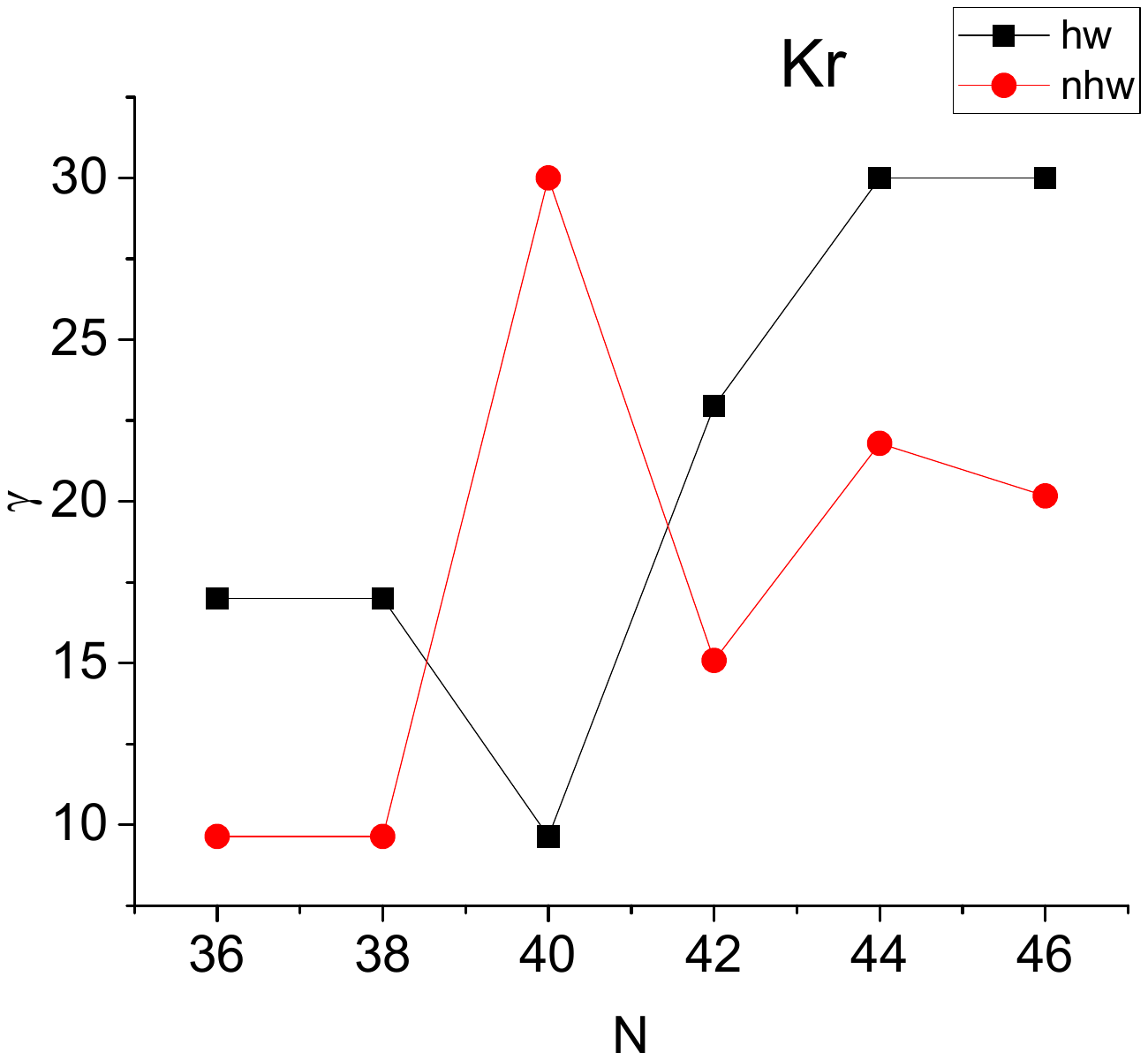}  \hspace{5mm}   \includegraphics[width=75mm]{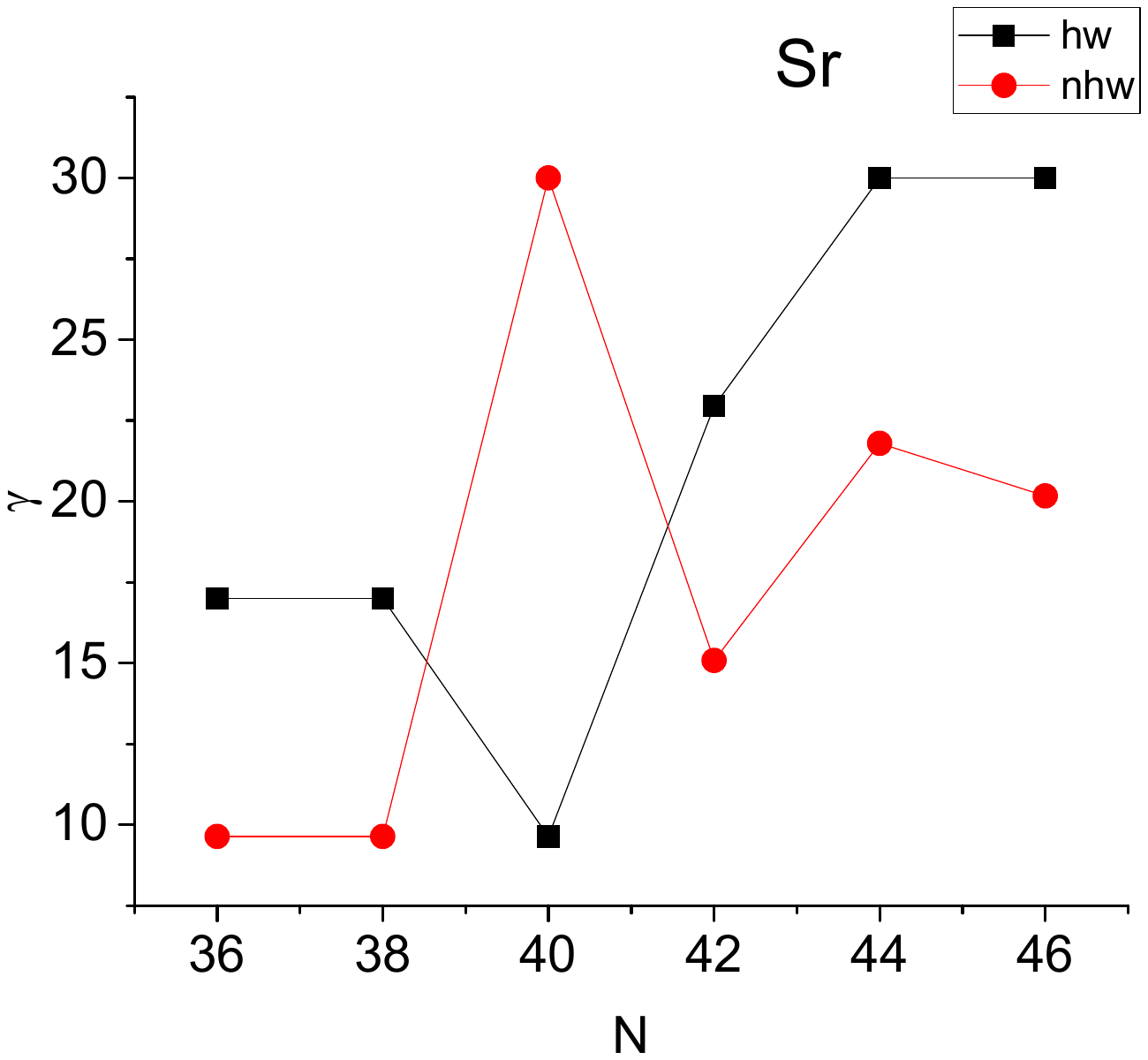} }
    {\includegraphics[width=75mm]{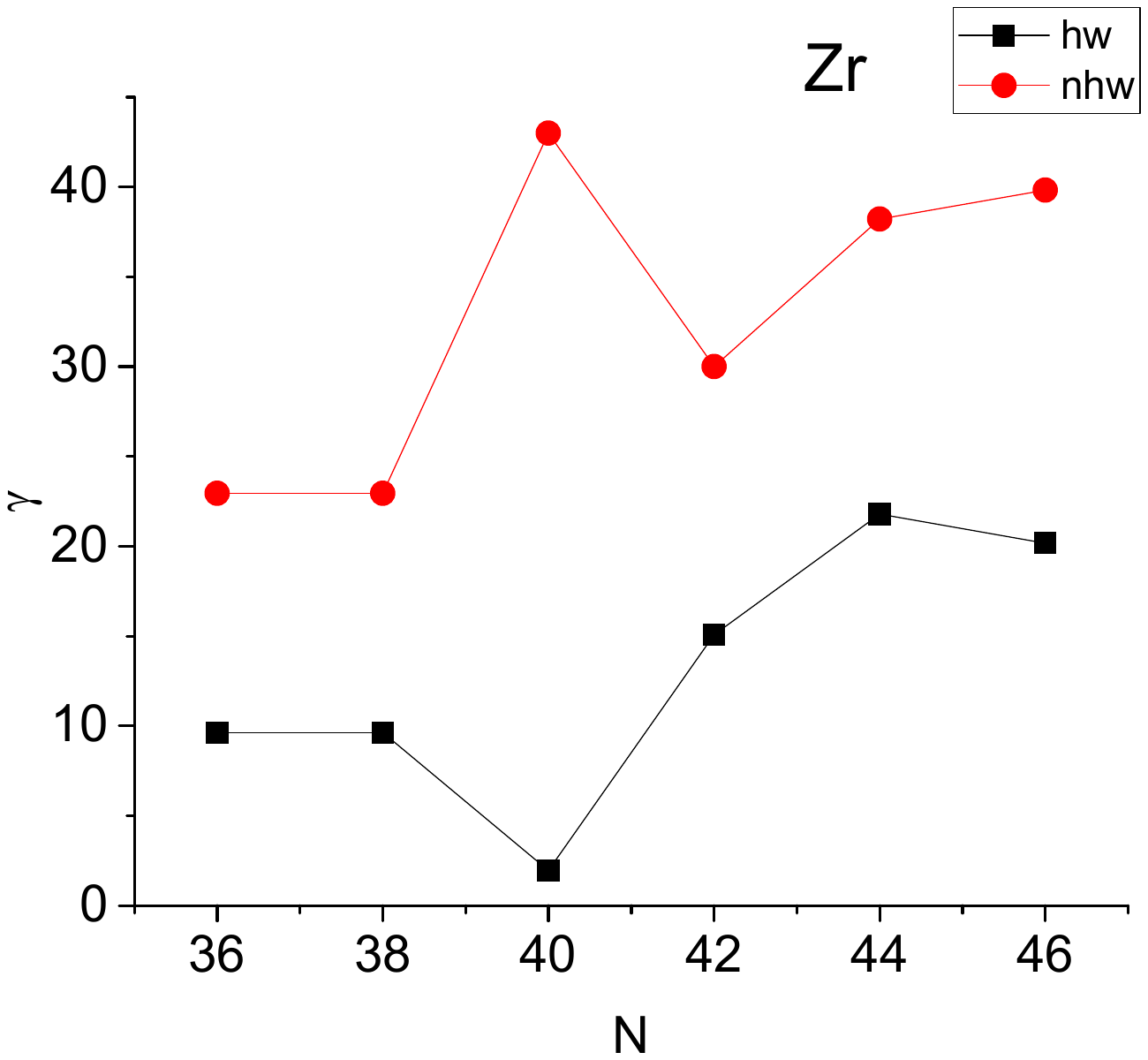} \hspace{5mm}   \includegraphics[width=75mm]{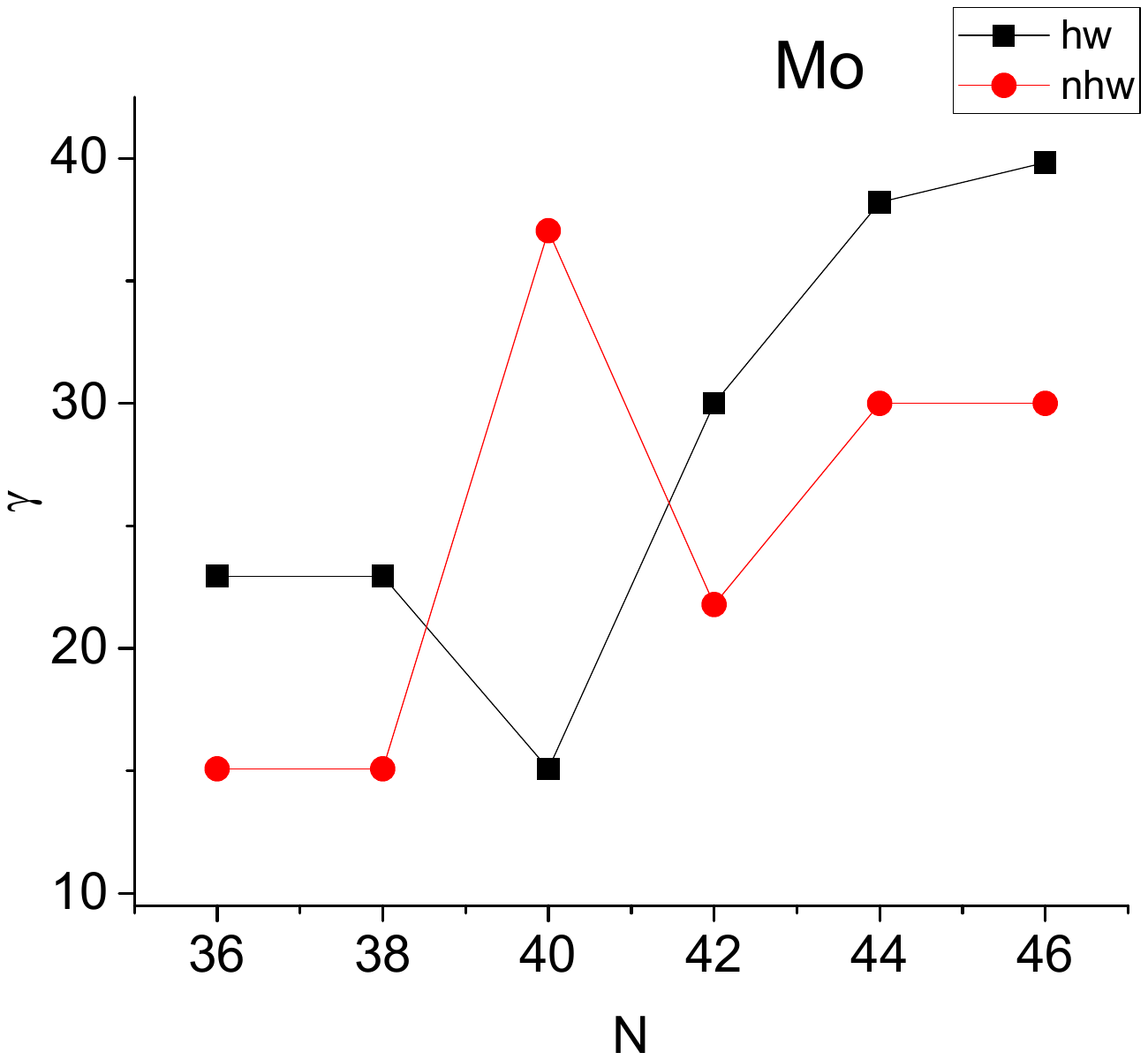} }
    {\includegraphics[width=75mm]{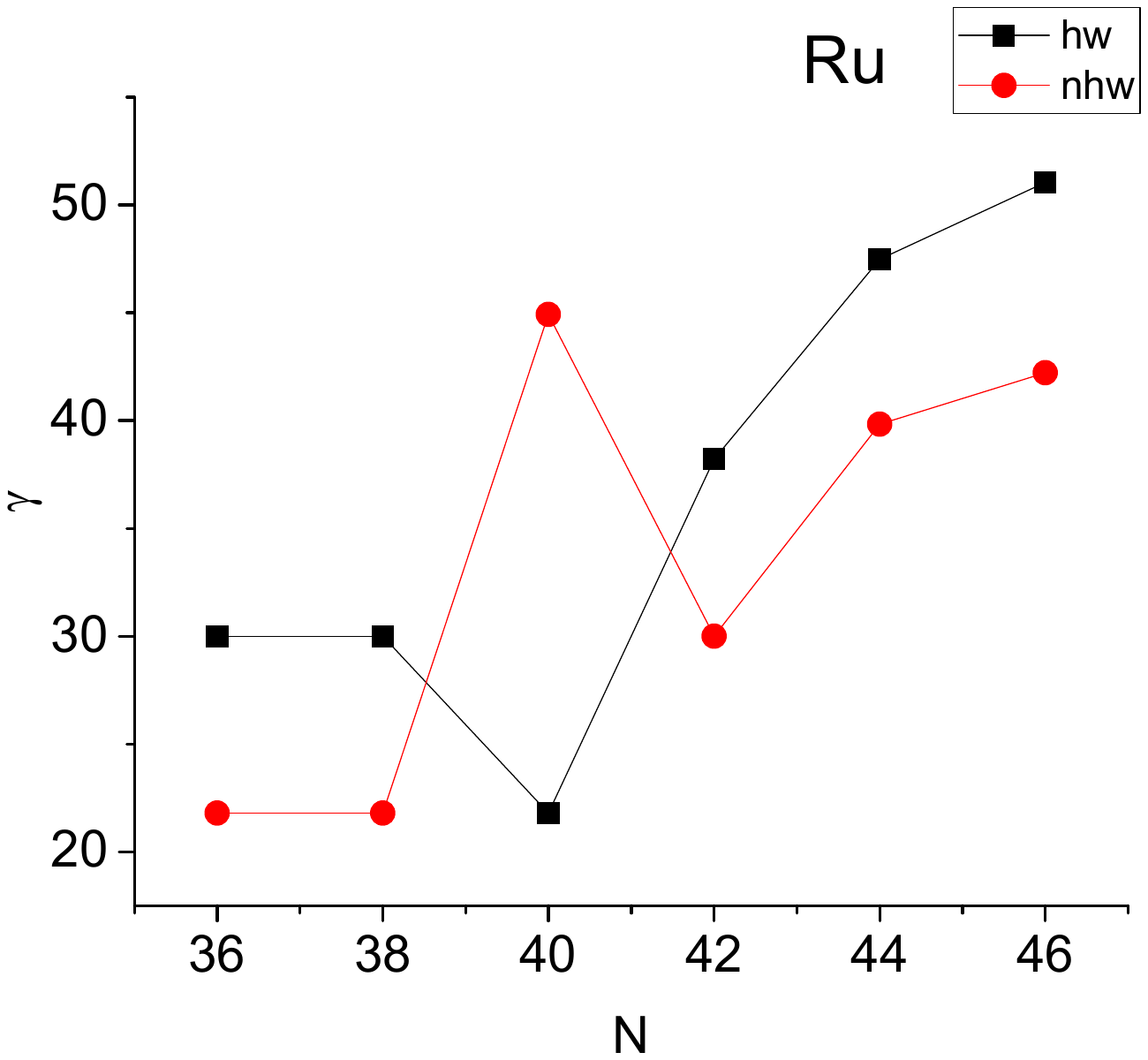} \hspace{5mm}    \includegraphics[width=75mm]{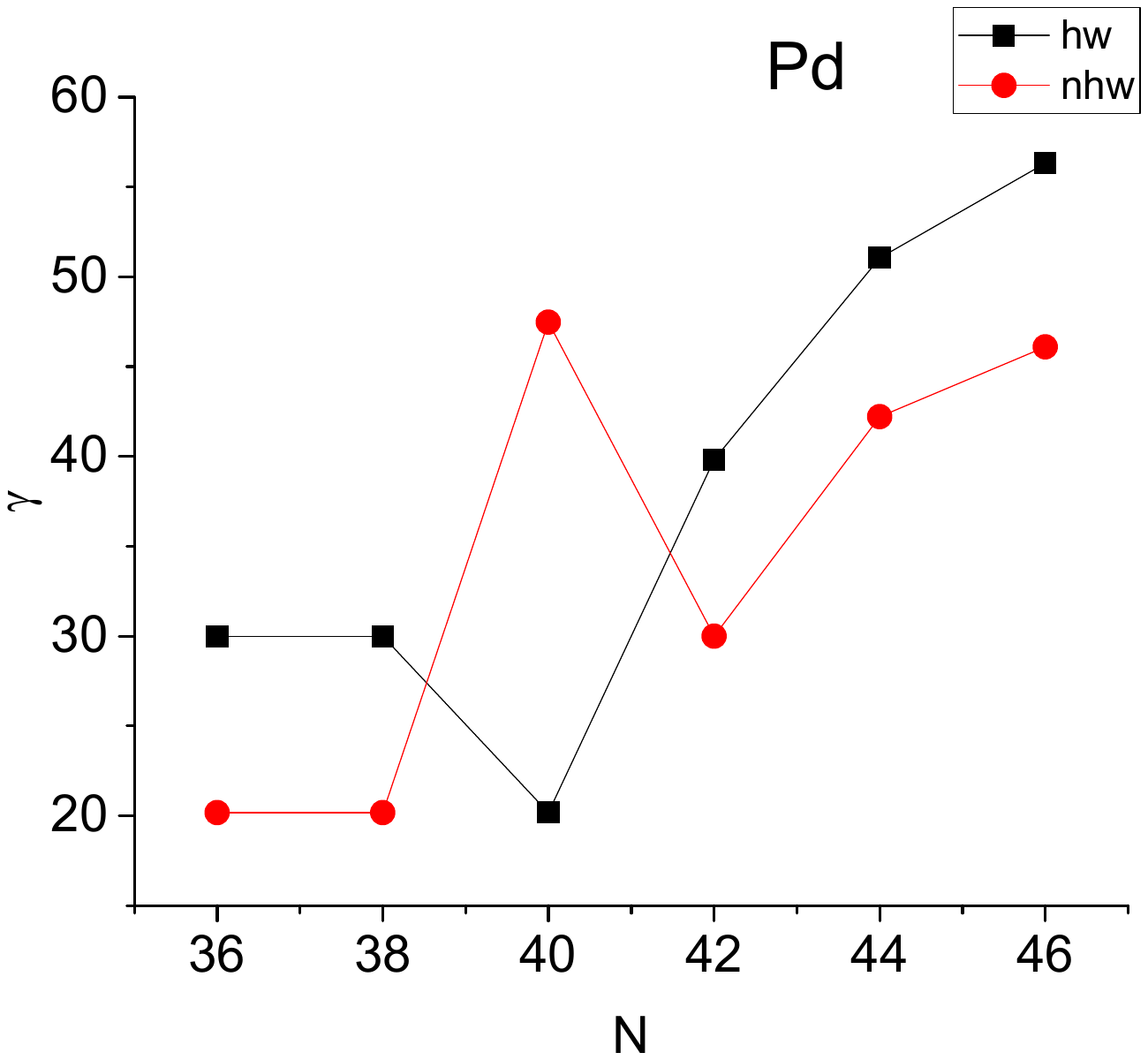} }

    \caption{Same as Fig. 4, but for the $Z=36$-46, $N=36$-46 region, with predictions taken from Table VI. See Sec. \ref{nhw} for further discussion.} 
    
\end{figure*}


\begin{thebibliography}{999}

\bibitem{Kota2020}
Kota, V. K. B.
\textit{SU(3) symmetry in atomic nuclei};  Springer Nature: Singapore, 2020. 

\bibitem{Mayer1948}
Mayer, M. G. 
On Closed Shells in Nuclei.
\textit{Phys. Rev.} \textbf{1948}, \textit{74}, 235.  

\bibitem{Mayer1949}
Mayer, M. G.
On Closed Shells in Nuclei. II.
\textit{Phys. Rev.} \textbf{1949}, \textit{75}, 1969.

\bibitem{Haxel1949}
Haxel, O.; Jensen, J. H. D.; Suess, H. E. 
On the "Magic Numbers" in Nuclear Structure.
\textit{Phys. Rev.} \textbf{1949}, \textit{75}, 1766. 

\bibitem{Mayer1955}
Mayer, M. G.; Jensen, J. H. D. 
\textit{Elementary Theory of Nuclear Shell Structure}; Wiley: New York, 1955. 

\bibitem{Heyde1990}
Heyde, K. L. G. \textit{The Nuclear Shell Model}; Springer: Berlin, 1990. 

\bibitem{Talmi1993}
Talmi, I. \textit{Simple Models of Complex Nuclei: The Shell Model and the Interacting Boson Model}; Harwood: Chur, 1993.

\bibitem{Wybourne1974}
Wybourne, B. G. \textit{Classical Groups for Physicists}; Wiley: New York, NY, USA, 1974. 

\bibitem{Moshinsky1996}
Moshinsky, M.;  Smirnov, Yu. F.
\textit{The Harmonic Oscillator in Modern Physics}; Harwood: Amsterdam, 1996. 

\bibitem{Iachello2006}
Iachello, F. \textit{Lie Algebras and Applications};  Springer: Berlin, 2006.

\bibitem{Bonatsos1986}
Bonatsos, D.; Klein, A.
Exact boson mappings for nuclear neutron (proton) shell-model algebras having SU(3) subalgebras.
\textit{Ann. Phys. (NY)} \textbf{1986}, \textit{169}, 61. 

\bibitem{Elliott1958a}
Elliott, J. P. 
Collective motion in the nuclear shell model. I. Classification schemes for states of mixed configurations.
\textit{Proc. Roy. Soc. A Ser. A} \textbf{1958}, \textit{245}, 128. 

\bibitem{Elliott1958b}
Elliott, J. P. 
Collective motion in the nuclear shell model II. The introduction of intrinsic wave-functions.
\textit{Proc. Roy. Soc. A Ser. A} \textbf{1958}, \textit{245}, 562.

\bibitem{Elliott1963}
Elliott, J. P.; Harvey, M.
Collective motion in the nuclear shell model III. The calculation of spectra.
\textit{Proc. Roy. Soc. A Ser. A} \textbf{1963}, \textit{272}, 557.

\bibitem{Elliott1968}
Elliott , J. P.; Wilsdon, C. E. 
Collective motion in the nuclear shell model IV. Odd-mass nuclei in the sd shell.
\textit{Proc. Roy. Soc. A Ser. A} \textbf{1968}, \textit{302}, 509.

\bibitem{Harvey1968}
Harvey, M. 
The nuclear SU$_3$ model.
\textit{Adv. Nucl. Phys.} \textbf{1968}, \textit{1}, 67; M. Baranger, M., Vogt, E., Eds.; Plenum: New York.    

\bibitem{Bohr1952}
Bohr, A. The coupling of nuclear surface oscillations to the motion of individual nucleons. 
{Dan. Mat. Fys. Medd.} \textbf{1952},  {26}, no. 14. 

\bibitem{Bohr1953}
Bohr, A.; Mottelson, B. R.
Collective and individual-particle aspects of nuclear structure.
\textit{Dan. Mat. Fys. Medd.} \textbf{1953},  \textit{27}, no. 16.

\bibitem{Bohr1998a}
Bohr, A.; Mottelson, B. R. 
\textit{Nuclear Structure Vol. I: Single-Particle Motion}; World Scientific: Singapore, 1998. 

\bibitem{Bohr1998b}
Bohr, A.; Mottelson, B. R. 
\textit{Nuclear Structure Vol. II: Nuclear Deformations}; World Scientific: Singapore, 1998. 

\bibitem{Arima1969}
Arima, A.; Harvey, M.; Shimizu, K.
Pseudo LS coupling and pseudo SU3 coupling schemes. 
\textit{Phys. Lett. B} \textbf{1969},  \textit{30}, 517.

\bibitem{Hecht1969}
Hecht, K. T.;  Adler, A.
Generalized seniority for favored $J\neq 0$ pairs in mixed configurations.
\textit{Nucl. Phys. A} \textbf{1969}, \textit{137}, 129. 

\bibitem{RatnaRaju1973}
Ratna Raju,  R. D.; Draayer, J. P.;  Hecht, K. T. 
Search for a coupling scheme in heavy deformed nuclei: The pseudo SU(3) model.
\textit{Nucl. Phys. A} \textbf{1973},  \textit{202}, 433.  

\bibitem{Draayer1982}
Draayer, J. P.; Weeks, K. J.; Hecht, K. T.
Strength of the $Q_\pi \cdot Q_\nu$  interaction and the strong-coupled pseudo-SU(3) limit.
\textit{Nucl. Phys. A} \textbf{1982}, \textit{381}, 1.

\bibitem{Draayer1983}
Draayer, J. P.; Weeks, K. J.
Shell-Model Description of the Low-Energy Structure of Strongly Deformed Nuclei.
\textit{Phys. Rev. Lett.} \textbf{1983}, \textit{51}, 1422. 

\bibitem{Draayer1984}
Draayer, J. P.; Weeks, K. J.
Towards a shell model description of the low-energy structure of deformed nuclei I. Even-even systems.
\textit{Ann. Phys. (NY)} \textbf{1984}, \textit{156}, 41. 

\bibitem{Ginocchio1997}
Ginocchio, J. N.
Pseudospin as a Relativistic Symmetry.
\textit{Phys. Rev. Lett.} \textbf{1997}, \textit{78}, 436. 

\bibitem{Bahri1992} 
Bahri, C.; Draayer, J. P.; Moszkowski, S. A.
Pseudospin symmetry in nuclear physics.
\textit{Phys. Rev. Lett.} \textbf{1992},  \textit{68}, 2133.  

\bibitem{Zuker1995}
Zuker, A. P.; Retamosa, J.; Poves, A.; Caurier, E. 
Spherical shell model description of rotational motion.
\textit{Phys. Rev. C} \textbf{1995}, \textit{52}, R1741(R). 

\bibitem{Zuker2015}
Zuker, A. P.;  Poves, A.; Nowacki, F.; Lenzi, S. M. 
Nilsson-SU3 self-consistency in heavy N=Z nuclei.
\textit{Phys. Rev. C} \textbf{2015},  \textit{92}, 024320. 

\bibitem{Bonatsos2017a}
Bonatsos, D.; Assimakis, I. E.; Minkov, N.; Martinou, A.; Cakirli, R. B.; Casten, R. F.; Blaum, K. 
Proxy-SU(3) symmetry in heavy deformed nuclei.
\textit{Phys. Rev. C} \textbf{2017}, \textit{95}, 064325.

\bibitem{Bonatsos2017b}
Bonatsos, D.; Assimakis, I. E.; Minkov, N.; Martinou, A.; Sarantopoulou, S.; Cakirli, R. B.; Casten, R. F.; Blaum, K. 
Analytic predictions for nuclear shapes, prolate dominance, and the prolate-oblate shape transition in the proxy-SU(3) model.
\textit{Phys. Rev. C} \textbf{2017},  \textit{95}, 064326.

\bibitem{Bonatsos2023}
Bonatsos, D.; Martinou, A.; Peroulis, S. K.; Mertzimekis, T. J.; Minkov, N.
The Proxy-SU(3) Symmetry in Atomic Nuclei.
\textit{Symmetry} \textbf{2023}, \textit{ 15}, 169. 

\bibitem{Rosensteel1977}
Rosensteel, G.; Rowe, D. J. 
Nuclear Sp(3,R) Model.
\textit{Phys. Rev. Lett.} \textbf{1977}, \textit{38}, 10. 

\bibitem{Rosensteel1980}
Rosensteel, G.; Rowe, D. J.; 
On the algebraic formulation of collective models III. The symplectic shell model of collective motion.
\textit{Ann. Phys. (NY)} \textbf{1980}, \textit{126}, 343.  

\bibitem{Rowe1985}
Rowe, D. J.
Microscopic theory of the nuclear collective model.
\textit{Rep. Prog. Phys.} \textbf{1985}, \textit{48}, 1419. 

\bibitem{Dytrych2008}
Dytrych, T.; Sviratcheva,  K. D.; Draayer, J. P.; Bahri, C.; Vary, J. P.
Ab initio symplectic no-core shell model.
\textit{J. Phys. G: Nucl. Part. Phys.} \textbf{2008},  \textit{35}, 123101.  

\bibitem{Launey2015}
Launey, K. D.; Draayer, J. P.; Dytrych, T.; Sun, G.-H.; Dong, S.-H.
Approximate symmetries in atomic nuclei from a large-scale shell-model perspective.
\textit{Int. J. Mod. Phys. E} \textbf{2015}, \textit{24}, 1530005. 

\bibitem{Launey2016}
Launey, K. D.; Dytrych, T.; Draayer, J. P. 
Symmetry-guided large-scale shell-model theory.
\textit{Prog. Part. Nucl. Phys.} \textbf{2016}, \textit{89}, 101. 

\bibitem{Launey2020}
Launey, K. D.; Dytrych, T.;  Sargsyan, G. H.; Baker, R. B.; Draayer, J. P. 
Emergent symplectic symmetry in atomic nuclei: Ab initio symmetry-adapted no-core shell model.
\textit{Eur. Phys. J. Special Topics} \textbf{2020}, \textit{229}, 2429. 

\bibitem{Launey2021}
Launey, K. D.; Mercenne, A.; Dytrych, T.
Nuclear Dynamics and Reactions in the Ab Initio Symmetry-Adapted Framework.
\textit{Annu. Rev. Nucl. Part. Sci.} \textbf{2021}, \textit{71},  253. 

\bibitem{Arima1975}
Arima, A.; Iachello, F.
Collective Nuclear States as Representations of a SU(6) Group.
\textit{Phys. Rev. Lett.} \textbf{1975},  \textit{35}, 1069. 

\bibitem{Iachello1987}
Iachello, F.; Arima, A. \textit{The Interacting Boson Model}; Cambridge U. Press: Cambridge, 1987. 

\bibitem{Iachello1991}
Iachello, F.; Van Isacker, P. \textit{The Interacting Boson-Fermion Model}; Cambridge U. Press: Cambridge, 1991.

\bibitem{Casten1993} 
Casten, R. F., ed., 
\textit{Algebraic Approaches to Nuclear Structure: Interacting Boson and Fermion Models};  Harwood: Chur, 1993. 

\bibitem{Frank2005}
Frank, A.; Van Isacker, P. 
\textit{Symmetry Methods in Molecules and Nuclei}; S y G Editores: Mexico, D.F., 2005. 

\bibitem{Arima1978}
Arima, A.; Iachello, F. 
Interacting boson model of collective nuclear states II. The rotational limit.
\textit{Ann. Phys. (NY)} \textbf{1978}, \textit{111},  201.

\bibitem{Casten1988}
Casten, R. F.; Warner, D. D. 
The interacting boson approximation.
\textit{Rev. Mod. Phys.}  \textbf{1988}, \textit{60}, 389.  

\bibitem{Casten1990}
Casten, R. F.
\textit{Nuclear Structure from a Simple Perspective}; Oxford University Press: Oxford, 1990. 

\bibitem{Minkov1997}
Minkov, N.; Drenska, S. B.; Raychev, P. P.; Roussev, R. P.; Bonatsos, D.
Broken SU(3) symmetry in deformed even-even nuclei.
\textit{Phys. Rev. C} \textbf{1997},  \textit{55}, 2345. 

\bibitem{Martinou2020}
Martinou, A.; Bonatsos, D.; Minkov, N.; Assimakis, I. E.; Peroulis, S. K.; Sarantopoulou, S.; Cseh, J. 
Proxy-SU(3) symmetry in the shell model basis.
\textit{Eur. Phys. J. A} \textbf{2020},  \textit{56}, 239.

\bibitem{Martinou2021b}
Martinou, A.; Bonatsos, D.; Karakatsanis, K. E.; Sarantopoulou, S.; Assimakis, I. E.; Peroulis, S. K.; Minkov, N.
Why nuclear forces favor the highest weight irreducible representations of the fermionic SU(3) symmetry.
\textit{Eur. Phys. J. A} \textbf{2021},  \textit{57}, 83.

\bibitem{Ring1980}
Ring, P.; Schuck, P.
\textit{The Nuclear Many-Body Problem}; Springer: Berlin, 1980. 

\bibitem{Bonatsos2017c}
Bonatsos, D.
Prolate over oblate dominance in deformed nuclei as a consequence of the SU(3) symmetry and the Pauli principle.
\textit{Eur. Phys. J. A} \textbf{2017},  \textit{53}, 148. 

\bibitem{Martinou2021}
Martinou, A.; Bonatsos, D.; Metzimekis, T. J.; Karakatsanis, K. E.; Assimakis, I. E.; Peroulis, S. K.; Sarantopoulou, S.; Minkov, N.
The islands of shape coexistence within the Elliott and the proxy-SU(3) Models.
\textit{Eur. Phys. J. A} \textbf{2021}, \textit{57}, 84.

\bibitem{Martinou2023}
Martinou,  A.; Bonatsos,  D.; Peroulis, S. K.; Karakatsanis,  K. E.; Mertzimekis, T. J.; Minkov, N.
Islands of Shape Coexistence: Theoretical Predictions and Experimental Evidence.
\textit{Symmetry} \textbf{2023}, \textit{15}, 29. 

\bibitem{Bonatsos2023c}
Bonatsos, D,; Andriana Martinou, A.; Peroulis, S. K.; Mertzimekis, T. J.;  Minkov, N.
Shape Coexistence in Even–Even Nuclei: A Theoretical Overview.
\textit{Atoms} \textbf{2023}, \textit{11}, 117.

\bibitem{Martinou2017}
Martinou, A.; Bonatsos, D.; Assimakis, I. E.; Minkov, N.; Sarantopoulou, S.; Cakirli, R. B.; Casten, R. F.; Blaum, K.
Parameter Free Predictions within the Proxy-SU(3) Model.
\textit{Bulg. J. Phys.} \textbf{2017}, \textit{44}, 407.

\bibitem{Martinou2017b} 
Martinou, A.; Peroulis, S.; Bonatsos, D.; Assimakis, I. E.;  Sarantopoulou, S.; Minkov, N.; Cakirli, R. B.; Casten, R. F.; Blaum, K.
Parameter-independent predictions for nuclear shapes and B(E2) transition rates in the Proxy-SU(3) Model.
\textit{HNPS: Adv. Nucl. Phys.} \textbf{2017}, \textit{25}, 21.

\bibitem{Draayer1989a}
Draayer, J. P.; Leschber, Y.; Park, S. C.; R. Lopez, R.
Representations of U(3) in U(N). 
\textit{Comput. Phys. Commun.} \textbf{1989}, \textit{56}, 279.

\bibitem{Langr2019}
Langr, D,;  Dytrych, T.;  Draayer, J. P.;  Launey, K. D.; Tvrd\'{\i}k, P.
Efficient algorithm for representations of U(3) in U(N).
\textit{Comput. Phys. Commun.} \textbf{2019}, \textit{244}, 442.

\bibitem{Kota2018}
Kota, V. K. B.
Simple formula for leading SU(3) irreducible representation for nucleons in an oscillator shell.
\textit{arXiv} \textbf{2018}, 1812.01810 [nucl-th]. 

\bibitem{Bonatsos2024}
Bonatsos, D.; Martinou, A.; Peroulis, S. K.; Petrellis, D.; Vasileiou, P.; Mertzimekis, T. J.; Minkov, N. 
Preponderance of triaxial shapes in atomic nuclei predicted by the proxy-SU(3) symmetry.
\textit{J. Phys. G: Nucl. Part. Phys.} \textbf{2024}, in print. https://doi.org/10.1088/1361-6471/ad903a

\bibitem{ensdf}
ENSDF database https://www.nndc.bnl.gov/ensdf

\bibitem{Borner1991}
B\"{o}rner, H. G.; Jolie, J.; Robinson, S. J.; Krusche, B.; Piepenbring, R.; Casten, R. F.; Aprahamian, A.;  Draayer, J. P. 
Evidence for the existence of two-phonon collective excitations in deformed nuclei. 
\textit{Phys. Rev. Lett.} \textbf{1991}, \textit{66}, 691.

\bibitem{Wu1994}
Wu, X.; Aprahamian, A.; Fischer, S. M.; Reviol, W.; Liu, G.; Saladin, J. X.
Multiphonon vibrational states in deformed nuclei.
\textit{Phys. Rev. C} \textbf{1994}, \textit{49}, 1837.

\bibitem{Burke1994}
Burke, D. G.
Hexadecapole-Phonon versus Double-$\gamma$-Phonon Interpretation for $K^{\pi}=4^+$  Bands in Deformed Even-Even Nuclei. 
\textit{Phys. Rev. Lett.} \textbf{1994}, \textit{73}, 1899.

\bibitem{Garrett2005}
Garrett, P. E.; Kulp, W. D.;  Wood,J. L.;  Bandyopadhyay, D.;  Christen, S.; Choudry, S.; Dewald, A.; Fitzler, A.; Fransen, C.; Jessen, K.; Jolie, J.; Kloezer, A.;  Kudejova, P.; Kumar, A.;  Lesher, S. R.; Linnemann, A.; Lisetskiy, A.;  Martin, D.; Masur, M.;  McEllistrem, M. T.;  M\"{o}ller, O.; Mynk, M.;  Orce, J. N.;  Pejovic, P.; Pissulla,T.;  Regis, J. M.; Schiller,A.;  D Tonev, D.;  Yates, S. W. 
Octupole and hexadecapole bands in $^{152}$Sm.
\textit{J. Phys. G: Nucl. Part. Phys.} \textbf{2005}, \textit{31}, S1855.

\bibitem{Jolos2006}
Jolos, R. V.; von Brentano, P.
Mass coefficient and Grodzins relation for the ground-state band and $\gamma$ band.
\textit{Phys. Rev. C} \textbf{2006}, \textit{74}, 064307.

\bibitem{Jolos2007}
Jolos, R. V.; von Brentano, P.
Bohr Hamiltonian with different mass coefficients for the ground- and $\gamma$ bands from experimental data.
\textit{Phys. Rev. C} \textbf{2007}, \textit{76}, 024309. 

\bibitem{Bonatsos2021}
Bonatsos, D.; Assimakis, I. E.;  Martinou, A.;  Sarantopoulou, S.;  Peroulis, S. K.;  Minkov, N.
Energy differences of ground state and $\gamma_1$ bands as a hallmark of collective behavior.
\textit{Nucl. Phys. A} \textbf{2021}, \textit{1009}, 122158. 

\bibitem{Castanos1988}
Casta\~{n}os, O.; Draayer, J. P.; Leschber, Y.
Shape variables and the shell model. 
\textit{Z. Phys. A} \textbf{1988}, \textit{329}, 33. 

\bibitem{Draayer1989}
Draayer, J. P.; Park, S. C.; Casta\~{n}os, O.
Shell-Model Interpretation of the Collective-Model Potential-Energy Surface.
\textit{Phys. Rev. Lett.} \textbf{1989},  \textit{62}, 20. 

\bibitem{DeVries1987}
De Vries, H.; De Jager, C. W.;  De Vries, C.
Nuclear charge-density-distribution parameters from elastic electron scattering.
\textit{At. Data Nucl. Data Tables} \textbf{1987},  \textit{36}, 495.

\bibitem{Stone2014}
Stone, J. R.; Stone, N. J.; Moszkowski, S. A.
Incompressibility in finite nuclei and nuclear matter.
\textit{Phys. Rev. C} \textbf{2014}, \textit{89}, 044316. 

\bibitem{Pritychenko2016}
Pritychenko, B.; Birch, M.;  Singh, B.; Horoi, M. 
Tables of E2 transition probabilities from the first $2^+$  states in even-even nuclei.
\textit{At. Data Nucl. Data Tables} \textbf{2016},  \textit{107}, 1.

\bibitem{Davydov1958}
Davydov, A. S.; Filippov, G. F. 
Rotational states in even atomic nuclei.
\textit{Nucl. Phys.} \textbf{1958}, \textit{8}, 237. 

\bibitem{Davydov1959}
Davydov, A. S.;  Rostovsky, V. S.
Relative transition probabilities between rotational levels of non-axial nuclei.
\textit{Nucl. Phys.} \textbf{1959}, \textit{12}, 58. 

\bibitem{Minkov1999}
Minkov, N.; Drenska, S. B.; Raychev, P. P.; Roussev, R. P.; Bonatsos, D.
Ground-$\gamma$ band coupling in heavy deformed nuclei and SU(3) contraction limit.
\textit{Phys. Rev. C} \textbf{1999}, \textit{60}, 034305. 

\bibitem{Minkov2000}
Minkov, N.; Drenska, S. B.; Raychev, P. P.; Roussev, R. P.; Dennis Bonatsos, D. 
Ground-$\gamma$ band mixing and odd-even staggering in heavy deformed nuclei
\textit{Phys. Rev. C} \textbf{2000}, \textit{61}, 064301.

\bibitem{Kota2024} 
Kota, V. K. B.; Sahu, R.
Proxy-SU(4) symmetry in A = 60-90 region. 
\textit{Phys. Scr.} \textbf{2024}, \textit{99}, 065306.

\bibitem{Sarantopoulou2017} 
Sarantopoulou, S.; Bonatsos, D.; Assimakis, I. E.; Minkov, N.; Martinou, A.; Cakirli, R. B.; Casten, R. F.; Blaum, K.
Proxy-SU(3) Symmetry in Heavy Nuclei: Prolate Dominance and Prolate-Oblate Shape Transition.
\textit{Bulg. J. Phys.} \textbf{2017}, \textit{44}, 417.

\bibitem{Bhatt2000}
Bhatt, K. H.; Kahane, S.; Raman, S. 
Collective properties of nucleons in the abnormal-parity states. 
\textit{Phys. Rev. C} \textbf{2000}, \textit{61}, 034317.

\bibitem{Bhatt1992}
Bhatt, K. H.; Nestor, Jr., C. W.; Raman, S. 
Do nucleons in abnormal-parity states contribute to deformation?
\textit{Phys. Rev. C} \textbf{1992}, \textit{46}, 164.

\bibitem{Bhatt1994}
Bhatt, K. H.; Raman, S.; Nestor, Jr., C. W. 
Symplectic pseudo-SU(3) model and $B(E2; 0_1^+\to 2_1^+)$ value of  $^{238}$U.  
\textit{Phys. Rev. C} \textbf{1994}, \textit{49}, 808. 

\bibitem{Kahane1997}
Kahane, S.;  Raman, S.; Bhatt, K. H. 
Angular-momentum structure of the yrast bands of deformed nuclei.
\textit{Phys. Rev. C} \textbf{1997}, \textit{55}, 2885.

\bibitem{Raman1988}
Raman, S.; Nestor, Jr., C. W.; Bhatt, K. H.
Systematics of $B(E2; 0_1^+ \to 2_1^+)$ values for even-even nuclei.
\textit{Phys. Rev. C} \textbf{1988}, \textit{37}, 805. 

\bibitem{Alex2011}
Alex, A.; Kalus, M.; Huckleberry, A.; von Delft, J.
A numerical algorithm for the explicit calculation of SU(N) and SL(N,C) Clebsch–Gordan coefficients.  
\textit{J. Math. Phys.} \textbf{2011}, \textit{52}, 023507. 

\bibitem{vonDelft2010}
http://homepages.physik.uni-muenchen.de/~vondelft/Papers/ClebschGordan/ \textbf{2010}.

\bibitem{Dytrych2021}
Dytrych, T.;  Langr, D.; Draayer, J. P.; Launey, K. D.; Gazda, D.
SU3lib: A C++ library for accurate computation of Wigner and Racah coefficients of SU(3). 
\textit{Comput. Phys. Commun.} \textbf{2021}, \textit{269}, 108137.

\bibitem{Bahri2004}
Bahri, C.; Rowe, D. J.; Draayer, J. P.
Programs for generating Clebsch–Gordan coefficients of SU(3) in SU(2) and SO(3) bases
\textit{Comput. Phys. Commun.} \textbf{2004}, \textit{159}, 121.

\bibitem{Brink2005}
Brink, D. M.; Broglia, R. A.
 \textit{Nuclear Superfluidity: Pairing in Finite Systems}; Cambridge U. Press: Cambridge, 2005. 

\bibitem{Rakavy1957}
Rakavy, G.
The classification of states of surface vibrations.
\textit{Nucl. Phys.} \textbf{1957}, \textit{4}, 289.

\bibitem{Troltenier1995a}
Troltenier, D.; Bahri, C.; Draayer, J. P. 
Generalized pseudo-SU(3) model and pairing.
\textit{Nucl. Phys. A} \textbf{1995}, \textit{586}, 53. 

\bibitem{Troltenier1995b}
Troltenier, D.; Bahri, C.; Draayer, J. P. 
Effects of pairing in the pseudo-SU(3) model.
\textit{Nucl. Phys. A} \textbf{1995}, \textit{589}, 75. 

\bibitem{Bahri1994}
Bahri, C.
Pairing and quadrupole interactions in the pseudo-SU(3) shell model.
 Ph.D. thesis, Louisiana State University \textbf{1994}. 

\bibitem{Bahri1995}
Bahri, C.; Escher, J.; J.P. Draayer, J. P.
Monopole-pairing and deformation in atomic nuclei.
\textit{Nucl. Phys. A} \textbf{1995}, \textit{592}, 171. 

\bibitem{Escher1998}
Escher, J.;  Bahri, C.; Troltenier, D.;  Draayer, J. P. 
Pairing-plus-quadrupole model and nuclear deformation: A look at the spin-orbit interaction.
\textit{Nucl. Phys. A} \textbf{1998}, \textit{633}, 662. 

\bibitem{Bonatsos2020}
Bonatsos, D.; Martinou, A.; Sarantopoulou, S.; Assimakis, I. E.; Peroulis, S.; Minkov, N.
Parameter-free predictions for the collective deformation variables $\beta$ and $\gamma$ within the pseudo-SU(3) scheme.
\textit{Eur. Phys. J. Special Topics} \textbf{2020},  \textit{229}, 2367. 

\bibitem{Castanos1992a}
Casta\~{n}os, O.; Moshinsky, M.; Quesne, C.
Transformation from U(3) to pseudo U(3) basis. 
In \textit{Group Theory and Special Symmetries in Nuclear Physics (AnnArbor, 1991)}; Draayer, J. P., J\"{a}necke, J., Eds.;  World Scientific: Singapore, 1992; p. 80.   

\bibitem{Castanos1992b}
Casta\~{n}os, O.; Moshinsky, M.; Quesne, C.
Transformation to pseudo-SU(3) in heavy deformed nuclei.
\textit{Phys. Lett. B} \textbf{1992}, \textit{277}, 238. 

\bibitem{Castanos1994}
Casta\~{n}os, O.; Vel\'{a}zquez A., V.; Hess, P. O.; Hirsch, J. G.
Transformation to pseudo-spin-symmetry of a deformed Nilsson hamiltonian.
\textit{Phys. Lett. B} \textbf{1994},  \textit{321}, 303. 


\end{thebibliography}
\end{document}